\documentclass[floatfix,a4paper,tightenlines,amsmath,amssymb,nofootinbib,showpacs,notitlepage,showkeys,prd,twocolumn,10pt,aps]{revtex4-1}

\usepackage[utf8]{inputenc}
\usepackage{amsmath}
\usepackage{amssymb}
\usepackage{graphicx}
\usepackage{hyperref}
\usepackage[export]{adjustbox}
\usepackage[update]{epstopdf}

\newcommand{\de}{\delta}
\newcommand{\eref}[1]{Eq. (\ref{#1})}
\newcommand{\fref}[1]{Fig. \ref{#1}}
\newcommand{\nnnl}{\nonumber\\}	

\DeclareMathOperator{\Tr}{Tr}

\pdfoutput=1

\allowdisplaybreaks 

\begin{document}

\title{Correlation functions of Landau gauge Yang-Mills theory}

\author{Markus Q.~Huber}
\email{markus.huber@physik.jlug.de}
\affiliation{Institut f\"ur Theoretische Physik, Justus-Liebig--Universit\"at Giessen, 35392 Giessen, Germany}

\date{\today}

\begin{abstract}
Correlation functions of Yang-Mills theory in the Landau gauge are calculated from their equations of motion.
The employed setup is completely parameter free and leads, within errors, to good quantitative agreement with corresponding lattice results for the ghost and gluon propagators as well as the ghost-gluon and three-gluon vertices.
Also the four-gluon vertex is calculated.
The present setup allows for the first time for a unique subtraction of quadratic divergences in the gluon propagator Dyson-Schwinger equation.
Thus, there are no ambiguities which can arise due to the use of models or auxiliary workarounds.
In addition, several self-tests of the results are described that allow assessing the truncation error in a self-consistent way.
This enables a new perspective on how to identify limitations of the present setup and develop future improvements.
\end{abstract}



\maketitle

\section{Introduction}
\label{sec:introduction}
 
Quantum chromodynamics (QCD) describes the strongly interacting part of the Standard Model and exhibits a rich phenomenology.
Describing it from first principles is a challenge for contemporary physics.
Functional methods are one possible approach for this.
They are formulated in terms of the correlation functions of quarks and gluons and constitute a system of nonlinear equations.
To solve them, some form of approximation is required, and often models are used to fill the gaps.
This has led to a successful description of many bound states and allowed exploring the phase structure of QCD; see \cite{Cloet:2013jya,Eichmann:2016yit} and \cite{Pawlowski:2010ht,Fischer:2018sdj} for reviews, respectively.
In recent years, though, some of these gaps have been closed, and a coherent picture is emerging of how functional equations can lead to a self-contained description without any modeling.
Thus, it seems within reach to establish a direct link between the QCD action and its phenomenology with functional equations.

The aim of this work is to combine several of the recent individual advances using equations of motion and provide a state-of-the-art solution for the correlation functions of Yang-Mills theory.
Understanding Yang-Mills theory, which describes only the gluonic part of QCD, is a cornerstone for the treatment of full QCD, but solving the corresponding functional equations faces a few challenges which were overcome only recently.
Some of these challenges were of a technical nature which required to adopt and develop the correspond tools.
Others are related to problems which can be alleviated by adjustments of the employed models or other modifications of the equations.
However, for a self-contained calculation this freedom is lost, as by definition no free parameters are allowed.
Consequently, these problems have to be dealt with properly.

Functional equations are used with various gauges, but here solely the Landau gauge is considered.
It has many advantageous features which make it a very convenient choice.
Propagators and vertices were calculated with different types of functional equations using a variety of different approximations.
Other methods, like lattice simulations or effective models, have been used as well.
However, to test the robustness and reliability of the presented results, not only comparisons with other methods will be performed, but also some self-tests.
Such tests are crucial to make the method independent from the availability of other results.
In addition to elucidating structures and mechanisms qualitatively, the method then becomes also quantitatively predictive starting from first principles.

The final test of results for gauge-dependent quantities consists in calculating a gauge-invariant quantity.
When, as is the case here, no free parameters exist, the self-consistency of the whole setup is decisive, as any inconsistencies cannot be covered up anymore by tuning a model.
For Yang-Mills theory a natural, though not an easy example for such a gauge-invariant quantity is glueballs.
While fore real glueballs the effects of quarks still need to be added, we can for pure Yang-Mills theory settle with a comparison to corresponding lattice calculations.
The results presented here were used to this end in Ref.~\cite{Huber:2020ngt} where the masses of scalar and pseudoscalar glueballs were calculated.
The good quantitative agreement obtained is an important indication that the results from the employed truncation have reached the necessary quantitative precision, at least for this task.

The encouraging agreement with other methods and the successful calculation of gauge-invariant quantities should not hide the fact that there is still room for improvement.
Having reached the present level of accuracy, one should try to identify the remaining shortcomings as a next step.
They might be small, but depending on the specific task, they might still play a relevant role.
Thus, it is important for the future improvement of truncations to understand their sources and how to get rid of them.
In this respect, it should be stressed that qualitatively this process is now different than in smaller truncations because there are no ambiguities left.
A prime example of this are quadratic divergences which are discussed in detail in Sec.~\ref{sec:renormalization}.
In the past, their removal in the gluon propagator equation of motion had a quantitative influence on the result which made it difficult to disentangle the effects of different parts of the truncation.
Hence, using the present truncation as a starting point, the impact of various improvements can be studied in a much clearer way.
To this end, detailed comparisons and analyses are presented in Secs.~\ref{sec:results} and \ref{sec:comparison}.

The remainder of the paper is organized as follows.
Yang-Mills theory in the Landau gauge and the correlation functions calculated in this work are introduced in Sec.~\ref{sec:corrFuncs}.
The employed functional equations are discussed in Sec.~\ref{sec:eoms} including details on the truncation.
Their renormalization is explained in Sec.~\ref{sec:renormalization}.
Results are presented in Sec.~\ref{sec:results} and compared to results from other methods in Sec.~\ref{sec:comparison}.
Sec.~\ref{sec:discussion} contains a discussion of the results and Sec.~\ref{sec:conclusions} the conclusions.
Results for the system using an alternative equation for the ghost-gluon vertex, the renormalization procedure for the scaling case and numerical details are deferred to appendices.

\section{Correlation functions of Yang-Mills theory}
\label{sec:corrFuncs}

The Lagrangian density of Yang-Mills theory with the gauge group $SU(N)$ and the gauge fixed to linear covariant gauges is\footnote{The conventions in this paper follow those of Ref.~\cite{Huber:2018ned}.}
\begin{align}\label{eq:Lagrangian-YM}
 \mathcal{L}_{YM}&=\frac{1}{4} F_{\mu \nu }^rF^r_{\mu \nu}+\frac{1}{2\xi}(\partial A)^2-\bar{c}^r\partial_\mu D_\mu^{rs} c^s,
\end{align}
where the field strength tensor and the covariant derivative are given by
\begin{align}
 F_{\mu \nu }^r&=\partial_\mu A_\nu^r-\partial_\nu A_\mu^r-g\,f^{rst}A_\mu^s A_\nu^t,\\
 D_\mu^{rs}&=\delta^{rs}\partial_\mu+g\,f^{rst} A_\mu^t.
\end{align}
respectively.
The gluon field $A$ and the ghost (antighost) field $c$ ($\bar c$) live in the adjoint representation.
The fundamental generators obey the relations
\begin{align}
[ T^r,T^s]&=i f^{rst}T^t,\\
\Tr \lbrace T^r T^s \rbrace &= \frac{1}{2} \delta^{rs}.
\end{align}

\subsection{Propagators}

The gluon propagator is given by
\begin{align}
 D^{ab}_{\mu\nu}(p)&=\de^{ab}D_{\mu\nu}(p)=\de^{ab}(D^T_{\mu\nu}(p)+D^L_{\mu\nu}(p)),\\
 D^T_{\mu\nu}(p)&=\left(g_{\mu\nu}-\frac{p_\mu p_\nu}{p^2}\right)D(p^2), \quad D(p^2)=\frac{Z(p^2)}{p^2},\\
 D^L_{\mu\nu}(p)&=\frac{p_\mu p_\nu}{p^2}\frac{\xi}{p^2}.
\end{align}
In the Landau gauge, the propagator is completely transverse, viz., $\xi=0$.
The gluon propagator is then completely described by the scalar part $D(p^2)=Z(p^2)/p^2$.

The propagator of the ghost field is
\begin{align}
D^{ab}_G(p)=-\de^{ab}\frac{G(p^2)}{p^2}.
\end{align}

\subsection{Three-point functions}

Landau gauge Yang-Mills theory has two three-point functions, the ghost-gluon and the three-gluon vertex.
The former has two dressing functions, of which in Landau gauge only one is relevant.
Thus, the vertex is written as
\begin{align}
 \Gamma^{A\bar cc,abc,T}_\mu(p_2;p_1,p_2)=i\,g\,f^{abc} D^{A\bar cc}(p_2^2;p_1^2,p_2^2)P_{\mu\nu}(p_2)p_{1\nu}.
\end{align}
Note that although the transverse projector is put explicitly, this would not be necessary since the vertex is always contracted with a transverse projector anyway.
The fields in the superscript indicate the order of the arguments.
Here and for other vertices, all momenta are incoming.
The bare vertex is
\begin{align}
 \Gamma^{(0),A\bar cc,abc}_\mu(p_2;p_1,p_2)=i\,g\,f^{abc} p_{1\nu}.
\end{align}

The three-gluon vertex has 14 dressing functions of which four are transverse:
\begin{align}
 &\Gamma^{abc,T}_{\mu\nu\rho}(p_1,p_2,p_3)=\nnnl
 &\quad i\,g\,f^{abc}\sum_{i=1}^{4}\tau^i_{\mu\nu\rho} C^{AAA,i}(p_1^2,p_2^2,p_3^2).
\end{align}
The bare vertex is given by
\begin{align}\label{eq:bare_three-gluon_vertex}
 &\Gamma_{\mu\nu\rho}^{(0),abc}(p_1,p_2,p_3) =\nnnl
   &-i g f^{abc} \left[ (p_1-p_2)_\rho \delta_{\mu\nu} + (p_2-p_3)_\mu \delta_{\nu\rho} + (p_3-p_1)_\nu \delta_{\mu\rho} \right].
\end{align}
In the following, only a dressing function $C^{AAA}(p_1^2,p_2^2,p_3^2)$ of the tree-level tensor is considered, viz., for the full vertex
\begin{align}
 \Gamma^{abc}_{\mu\nu\rho}(p_1,p_2,p_3)=C^{AAA}(p_1^2,p_2^2,p_3^2)\Gamma_{\mu\nu\rho}^{(0),abc}(p_1,p_2,p_3)
\end{align}
is used.

\begin{figure*}[tb]
 \includegraphics[width=0.4\textwidth]{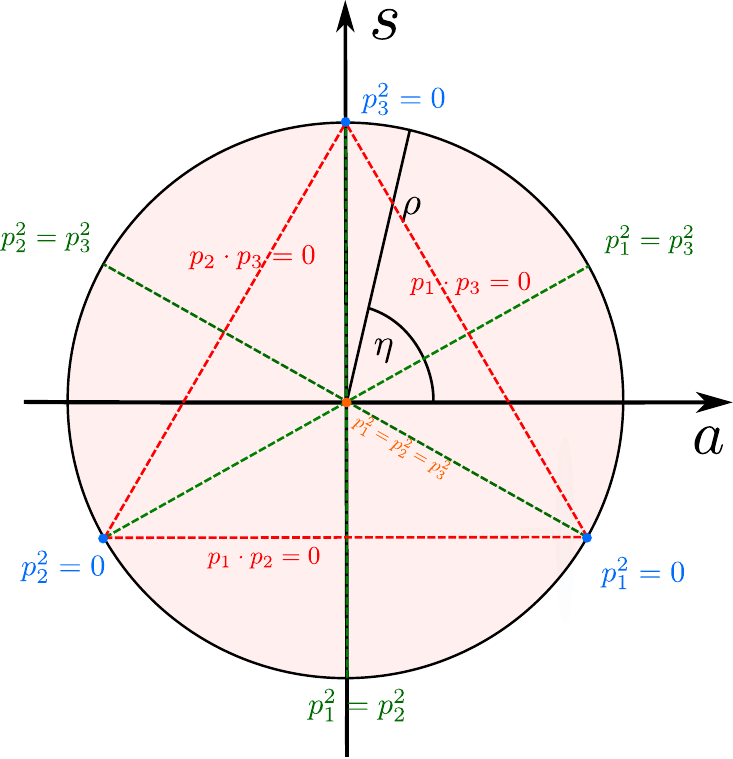}\hfill
 \includegraphics[width=0.4\textwidth]{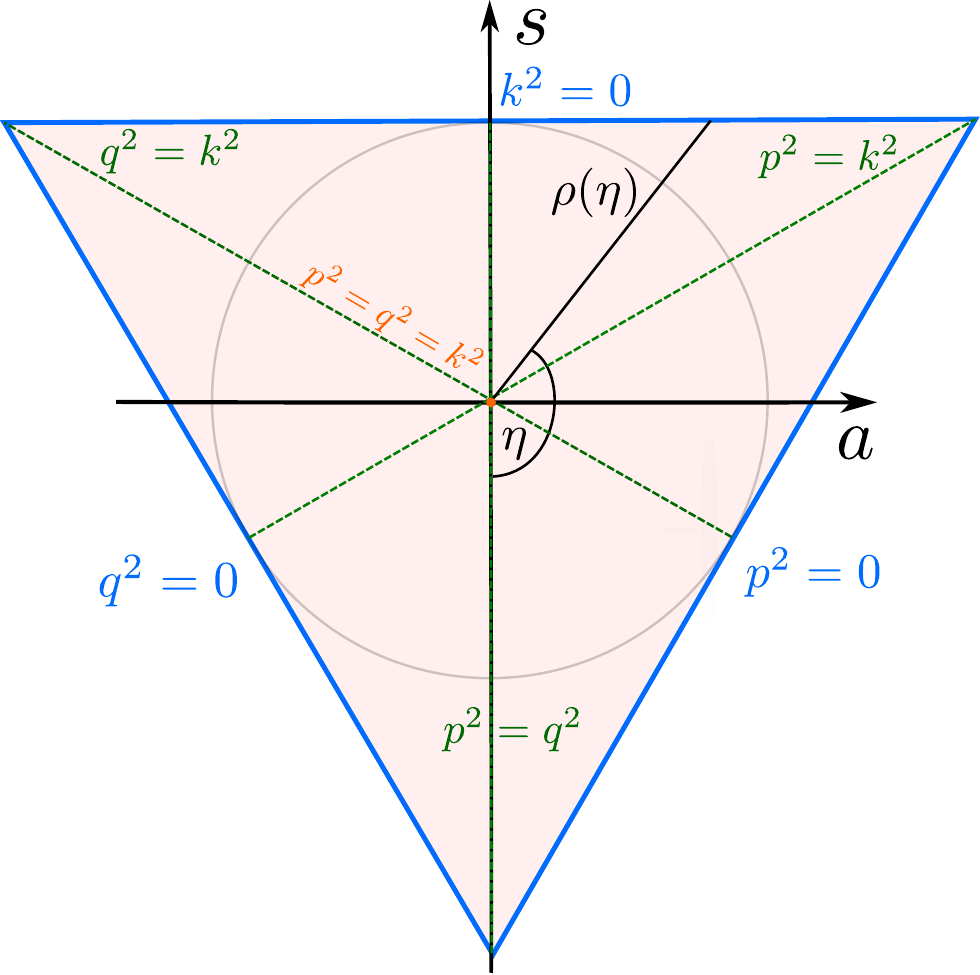}
\caption{Kinematic regions for the $S_3$ doublet of three-point functions (left) and the $S_4$ doublet of four-point functions (right); see \cite{Eichmann:2014xya} and \cite{Eichmann:2015nra}, respectively, for more details.
Kinematic points of interest for three-point functions are the symmetric point $p_1^2=p_2^2=p_3^2$, the lines of two equal momentum squares $p_i^2=p_j^2$, the lines of orthogonal momenta $p_i\cdot p_j=0$, and the points of one vanishing momentum $p_i^2=0$.
Similar points exist for four-point functions with $k=p_1+p_2$, $p=p_2+p_3$, and $q=p_3+p_1$.}
\label{fig:S3}
\end{figure*}

The color structure employed for these vertices is fixed to the antisymmetric structure constant $f^{abc}$.
In principle, $SU(N)$ also has the symmetric structure constant $d^{abc}$ for $N>2$.
However, due to the charge invariance of QCD, only the totally antisymmetric $f^{abc}$ is allowed \cite{Smolyakov:1980wq,Blum:2015lsa}.

Three-point functions depend on two independent momenta from which three kinematic variables can be constructed.
A typical choice are two momentum squares and an angle.
However, to exploit potential symmetry properties of the vertices and for reasons of technical simplifications, another set motivated by the symmetry properties of the $S_3$ permutation group is used here \cite{Eichmann:2014xya}.
This basis makes the Bose symmetry of the three-gluon vertex manifest, but it is also useful for the ghost-gluon vertex which lacks this symmetry.
From the three momenta $p_1$, $p_2$, $p_3$, one constructs the following variables \cite{Eichmann:2014xya} which are a singlet ($S_0$) and a doublet ($a$, $s$) under the permutation group $S_3$:
\begin{align}
\label{eq:S3_vars}
 S_0&=\frac{p_1^2+p_2^2+p_3^2}{6},\\
 a&=\sqrt{3}\frac{p_2^2-p_1^2}{p_1^2+p_2^2+p_3^2},\\
 s&=\frac{p_1^2+p_2^2-2p_3^2}{p_1^2+p_2^2+p_3^2}.
\end{align}
$S_0$ can take all positive values and $a$ and $s$ are restricted to a unit disk, $a^2+s^2\leq1$; see the left plot of \fref{fig:S3}.
Computationally, it is thus advantageous to use radial coordinates for $a$ and $s$,
\begin{align}
 a&=\rho \cos\eta, & s&=\rho \sin\eta\\
 \rho&=\sqrt{a^2+s^2}, &\eta&=\arctan \frac{s}{a}.
\end{align}
For clarity, the arguments of the dressing functions will be given in the following by the momenta and not the actually used kinematic variables.
The inverse transformation is
\begin{align}
p_1^2 &= S_0(2-\sqrt{3}a+s),\\
p_2^2 &= S_0(2+\sqrt{3}a+s),\\
p_3^2 &= -2S_0(s-1).
\end{align}

\subsection{Four-point functions}

Finally, the four-gluon vertex has to be discussed.
It has 136 tensors in Lorentz space, 41 of which are transverse \cite{Eichmann:2015nra}.
In color space, there are nine independent tensors for general $SU(N)$.
They are obtained from the 15 possible combinations of Kronecker deltas and the symmetric and antisymmetric structure constants.
Only nine of them are linearly independent as can be seen from various identities like the Jacobi identity listed in Ref.~\cite{Pascual:1980yu}.
This number reduces to eight for $SU(3)$ due to an additional identity \cite{Pascual:1980yu} and to three for $SU(2)$ due to the absence of the symmetric structure constant.
However, for $SU(3)$, only five color tensors are required since the other three decouple \cite{Driesen:1997wz,Huber:2017txg}.
The tree-level four-gluon vertex reads
\begin{align}
\label{eq:bare_four-gluon_vertex}
  &\Gamma^{{(0)},abcd}_{\mu\nu\rho\sigma}(p_1,p_2,p_3,p_4) =\nnnl
    &\quad -g^2  \Big[\;\,\left(f^{acn'}f^{bdn'}-f^{adn'}f^{cbn'}\right)\delta_{\mu\nu}\delta_{\rho\sigma}\nnnl
	  &\quad\quad + \left(f^{abn'}f^{cdn'}-f^{adn'}f^{bcn'}\right)\delta_{\mu\rho}\delta_{\nu\sigma} \nnnl
	  &\quad\quad +\left(f^{acn'}f^{dbn'}-f^{abn'}f^{cdn'}\right)\delta_{\mu\sigma}\delta_{\rho\nu}\Big].
\end{align}
For the full vertex only a dressing function $F^{AAAA}(p_1,p_2,p_3,p_4)$ of the tree-level tensor is taken into account:
\begin{align}\label{eq:AAAA}
 \Gamma&^{abcd,T}_{\mu\nu\rho\sigma}(p_1,p_2,p_3,p_4) = \nnnl
 &F^{AAAA}(p_1,p_2,p_3,p_4)\Gamma^{{(0)},abcd}_{\mu\nu\rho\sigma}(p_1,p_2,p_3,p_4) .
\end{align}

The dressing function $F^{AAAA}(p_1,p_2,p_3,p_4)$ depends on three independent momenta out of which six variables can be formed.
They can be organized according to the $S_4$ permutation group into a singlet, a doublet, and a triplet \cite{Eichmann:2015nra}.
From the three independent momenta $p_1$, $p_2$, and $p_3$, the following momentum combinations are constructed:
\begin{align}
 k&=p_1+p_2,& p&=p_2+p_3,& q&=p_3+p_1.
\end{align}
The momentum squares $k^2$, $p^2$ and $q^2$ are used together with the scalar products between different momenta,
\begin{align}
 \omega_1&=q\cdot k,& \omega_2 &=p\cdot k,&\omega_3&=p\cdot q,
\end{align}
to define the singlet variable\footnote{Due to the structural similarities to the case of three-point functions, the same variable names are used.
It is always clear from the context, which case is meant.}
\begin{align}
 S_0=\frac{p^2+q^2+k^2}{4},
\end{align}
the doublet variables
\begin{align}
 a&=\sqrt{3}\frac{q^2-p^2}{p^2+q^2+k^2}, & s&=\frac{p^2+q^2-2k^2}{p^2+q^2+k^2},
\end{align}
and the triplet variables
\begin{align}
 u&=-2\frac{\omega_1+\omega_2+\omega_3}{p^2+q^2+k^2},\\
 v&=-\sqrt{2}\frac{\omega_1+\omega_2-3\omega_3}{p^2+q^2+k^2},\\
 w&=\sqrt{6}\frac{\omega_1-\omega_2}{p^2+q^2+k^2}.
\end{align}
The doublet variables are similar to the case of three-point functions, except that $a$ and $s$ are restricted to the interior of a triangle \cite{Eichmann:2015nra}; see the right plot of \fref{fig:S3}.
Also, in this case, $a$ and $s$ are expressed via radial coordinates,
\begin{align}
 a&=\rho(\eta) \sin \eta, & s&=-\rho(\eta)\cos \eta,\nnnl
 \rho(\eta)&=\sqrt{a^2+s^2}, &\eta&=\arctan \frac{a}{-s }.
\end{align}
For symmetry reasons, the angle $\eta$ is measured from the negative $s$-axis and not the $a$-axis as for the three-point functions.
The different shape of the domain of $a$ and $s$ is encoded in an angular dependence of the radial coordinate $\rho(\eta)$.
It is convenient to rescale it to the interval $[0,1]$ as
\begin{align}
 \hat{\rho}(\eta)=\frac{\rho(\eta)}{\rho_\text{max}(\eta)}
\end{align}
where $\rho_\text{max}$ is given by
\begin{align}
 \rho_\text{max}(\eta)=\frac{1}{\sin\left(\eta+\frac{\pi}{6}\left(1-4\left\lfloor\frac{3\eta}{2\pi}\right\rfloor\right)\right)}.
\end{align}
$\lfloor \cdot \rfloor$ denotes the floor function.

The dependence of the four-gluon vertex on six kinematic variables is slowing down calculations considerably.
As an approximation, only the singlet and doublet are taken into account here by setting $\omega_1=\omega_2=\omega_3=0$.
This entails $p_1^2=p_2^2=p_3^2=p_4^2=S_0$.
A nontrivial angular dependence enters via the doublet variables
\begin{align}
 p_1\cdot p_2=\frac{k^2-p^2-q^2}{4},\nnnl
 p_2\cdot p_3=\frac{p^2-q^2-k^2}{4},\nnnl
 p_1\cdot p_3=\frac{q^2-k^2-p^2}{4}.
\end{align}
Even with this approximation, the calculation of the four-gluon vertex dominates the computing time.

\section{Equations of motion}
\label{sec:eoms}

In this section, the equations of motion from the one-particle irreducible (1PI) effective action, the Dyson-Schwinger equations, and from higher $n$PI effective actions, which will be used to calculate propagators and vertices, are discussed.
For simplicity, we refer with $n$PI effective action always to higher $n$PI effective actions with $n>1$ and denote their equations of motion as EOMs, while the ones from the 1PI effective action are denoted by DSEs.
The equations of both cases are similar and can be treated numerically with the same methods.
A difference consists in the structures of the complete systems of equations:
DSEs form an infinite set of equations of finite size, while $n$PI effective actions lead to a finite number of equations with (possibly) infinitely many terms.
Such differences become important when considering how to truncate the equations.
For more details beyond the short overview presented here, see, e.g., \cite{Roberts:1994dr,Alkofer:2000wg,Berges:2004pu,Alkofer:2008nt,Carrington:2010qq,Huber:2018ned}.

After a general discussion of the equations, the specific equations for the two-, three- and four-point functions are presented and how they are truncated.

\subsection{Dyson-Schwinger equations}
\label{sec:DSEs}

The 1PI effective action $\Gamma[\Phi]$ depends on the classical fields only.
In the shorthand notation employed here, $\Phi$ is the classical superfield that represents all fields, i.e., $\Phi=\{A,\bar{c},c\}$.
It is related to the underlying action $S[\phi]$ via a Legendre transformation of the generating functional of connected correlation functions $W[J]$, depending on the corresponding sources $J$ of the quantum fields $\phi$, with $\Phi=\delta W/\delta J=\langle \phi \rangle_J$,
\begin{align}
 \Gamma[\Phi]&=-W[J]+\Phi_i J_i,\label{eq:effAct}\\
Z[J]&=\int D[\phi] e^{-S[\phi] + \phi_i J_i}=e^{W[J]}.
\end{align}
Summation as well as integration over repeated indices, which represent field-types and all internal and space/momentum variables, are understood.
DSEs can be derived from the invariance of the path integral under translations of the fields which lead to (see, e.g., \cite{Roberts:1994dr,Alkofer:2000wg,Alkofer:2008nt,Huber:2018ned} for details)
\begin{align}\label{eq:DSE-master}
  \frac{\delta \Gamma[\Phi]}{\delta \Phi_i}=\frac{\delta S[\phi]}{\delta
  \phi_i}\Bigg\vert_{\phi_i=\Phi_i+D^{ij}_J  \, \delta/\delta \Phi_j}.
\end{align}
The field-dependent (as indicated by the index $J$) propagator $D_J^{ij}$ is given by
\begin{align}\label{eq:prop}
 D_J^{ij}:=\frac{\de W[J]}{\de J_i \de J_j}
=\left[\left(\frac{\delta^{2}\Gamma}{\delta\Phi^2}\right)^{-1}\right]^{ij}.
\end{align}
The physical propagators are obtained by setting the sources to zero, $D=D_{J=0}$.
From the master equation \eref{eq:DSE-master}, DSEs for all correlation functions can be derived by differentiating with respect to the appropriate fields and setting the sources $J$ to zero.
This leads to infinitely many coupled equations.
The equation for an $n$-point function does not only depend on lower correlation functions but also on $(n+1)$- and, due to the presence of a four-point function in the action, possibly on $(n+2)$-point functions.
However, each equation has a finite number of terms, as can be directly inferred from \eref{eq:DSE-master} and the fact that the bare action $S[\phi]$ has a finite number of terms.

For the actual derivation of the equations employed here, the \textit{Mathematica} \cite{Wolfram:2004} package \textit{DoFun} \cite{Alkofer:2008nt,Huber:2011qr,Huber:2019dkb} is used.
The resulting expressions are optimized for numerical use with \textit{FORM} \cite{vanRitbergen:1998pn,Vermaseren:2000nd,Kuipers:2012rf,Kuipers:2013pba,Ruijl:2017dtg} and exported to \textit{C++} code using \textit{CrasyDSE} \cite{Huber:2011xc}, which is also used for solving the equations numerically.

\subsection{Equations of motion from $n$PI effective actions}
\label{sec:eoms_nPI}

\begin{figure*}
 \includegraphics[height=1.6cm]{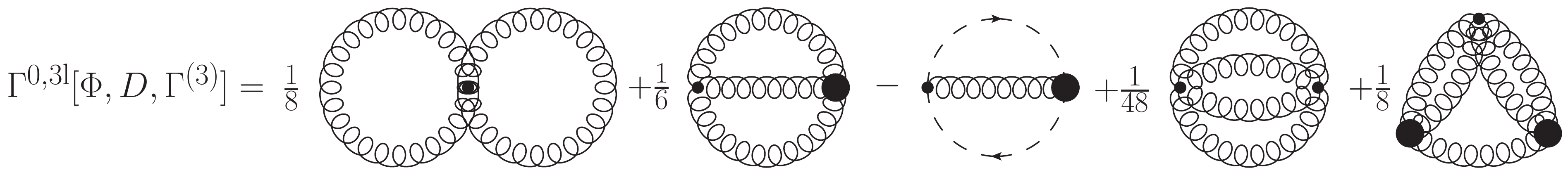}\\
 \vskip5mm
 \hskip4mm\includegraphics[height=1.6cm]{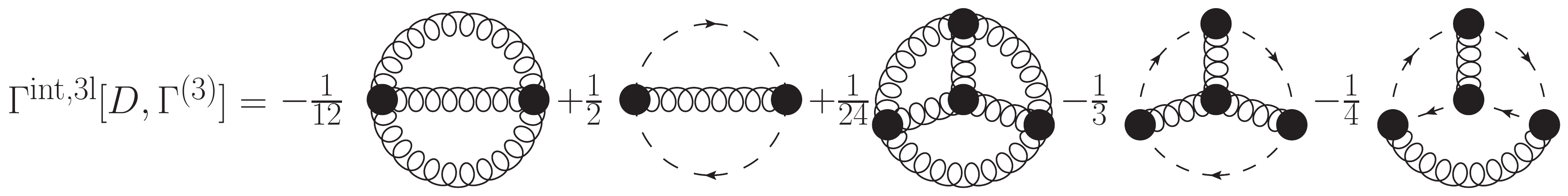}\hspace{1.55cm}
\caption{$\Gamma^0[\Phi, D, \Gamma^{(3)}]$ and $\Gamma^\text{int}[D, \Gamma^{(3)}]$ of the 3PI effective action truncated to three loops (as indicated by the superscript $\text{3l}$).
Here and in other figures, internal propagators are dressed, black disks denote dressed vertices, dots bare vertices, wiggly lines gluons, and dashed lines ghosts.}
\label{fig:3PI-3L}
\end{figure*}

The 1PI effective action is formulated as an expansion in the classical field $\Phi$ with the fully dressed 1PI correlation functions of the theory, denoted by $\Gamma^{(i)}$, as expansion coefficients.
$n$PI effective actions differ insofar as propagators and vertices up to order $n$ are treated on the same footing as the fields and are taken into account via additional Legendre transformations.
Vertices with $i>n$ only appear bare.
In analogy to the field variable, additional sources $R^{(i)}$, $2\leq i\leq n$, are introduced,
\begin{align}
 &e^{W[J,R^{(2)},R^{(3)},\ldots]} = Z[J,R^{(2)},R^{(3)},\ldots]\nnnl
 &\quad =\int D[\phi] e^{-S + \phi_i J_i + \frac{1}{2} R^{(2)}_{ij}\phi_i \phi_j + \frac{1}{3!} R^{(3)}_{ijk}\phi_i \phi_j \phi_k + \ldots}.
\end{align}
The $n$PI effective action is then obtained from corresponding Legendre transformations and depends in addition to the classic field $\Phi$ also on the propagator $D$ and potentially vertices $\Gamma^{(i)}$,
\begin{align}\label{eq:nPI-action}
 &\Gamma[\Phi, D, \Gamma^{(3)},\ldots] =\nnnl
 &- W + \frac{\de W}{\de J_i} J_i + \frac{\de W}{\de R^{(2)}_{ij}} R^{(2)}_{ij} + \frac{\de W}{\de R^{(3)}_{ijk}} R^{(3)}_{ijk} + \ldots,
\end{align}
where arguments have been suppressed on the right-hand side.
Differentiating the effective action and setting all sources to zero leads to stationarity conditions which correspond to the EOMs,
\begin{align}\label{eq:stationarity_conds}
 \frac{\delta \Gamma[\Phi, D, \Gamma^{(3)},\ldots]}{\de \Phi_i}&=0, \qquad \frac{\delta \Gamma[\Phi, D, \Gamma^{(3)},\ldots]}{\de D_{ij}}=0, \nnnl
 \frac{\delta \Gamma[\Phi, D, \Gamma^{(3)},\ldots]}{\de \Gamma^{(3)}_{ijk}}&=0,\qquad \ldots
\end{align}
There are as many equations as there are Legendre transformations with respect to sources $J$ and $R^{(i)}$.
In contradistinction to DSEs, these equations have infinitely many terms.
In practice, a loop expansion of the $n$PI effective action is performed.
Truncating it renders the numbers of terms finite.
In addition, using the EOMs of higher correlation functions, cancellations in an equation can be realized \cite{Berges:2004pu,Carrington:2010qq} that lead to simplifications.
An $n$PI effective action for Yang-Mills\footnote{Due to the same structure of quark-gluon and ghost-gluon interactions, one can directly extend this to QCD by replicating the ghost terms.} theory can be written as
\begin{widetext}
\begin{align}
\label{eq:3PI}
 \Gamma[\Phi, D, \Gamma^{(3)},\ldots] &= S[\Phi] + \frac{1}{2}\ln \left[ D^{AA}_{ii}\right]^{-1} - \ln \left[ D^{\bar c c}_{ii}\right]^{-1}
 + \frac{1}{2} S^{AA}_{ij}[\Phi] D^{AA}_{ij} -S^{\bar c c}_{ij}[\Phi] D^{\bar c c}_{ij} - \widetilde{\Gamma}[\Phi,D, \Gamma^{(3)},\ldots],\\
 \widetilde{\Gamma}[\Phi,D, \Gamma^{(3)},\ldots] &= \Gamma^0[\Phi,D, \Gamma^{(3)},\ldots] + \Gamma^\text{int}[D, \Gamma^{(3)},\ldots].
\end{align}
\end{widetext}
$S_{ij}$ is the field-dependent inverse bare propagator, defined as
\begin{subequations}
\begin{align}
S^{AA}_{ij}&=\frac{\de^2 S[\Phi]}{\de A_i\de A_j},\\
S^{\bar c c}_{ij}&=\frac{\de^2 S[\Phi]}{\de c_i\de \bar{c}_j}.
\end{align}
\end{subequations}
For the full propagators $D_{ij}$, their field content was made explicit with the superscripts.
$\widetilde{\Gamma}$ is split into parts with ($\Gamma^\text{int}$) and without ($\Gamma^0$) bare vertices.
Here, we will consider the 3PI effective action at three-loop level which is required for a self-consistent truncation including three-point functions.
The corresponding $\widetilde{\Gamma}$ is given in \fref{fig:3PI-3L}.

The structures of the EOMs in \eref{eq:stationarity_conds} are, for the three-loop truncation employed here, very similar to those of DSEs.
Thus, it is straightforward to modify the DSE expressions within \textit{DoFun} and calculate them with \textit{CrasyDSE} as well.

\subsection{Equations for two-, three-, and four-point functions and their truncations}
\label{sec:equations}

The truncation of a system of DSEs specifies how to reduce the infinite system of equations to a finite set of equations.
This can be done by setting certain correlation functions to zero or by providing models for them.
Also EOMs can be truncated in this way, but $n$PI effective actions offer a more systematic way via a loop expansion.
One then obtains a concrete form for the $n$PI effective action from which all EOMs are derived.
Changing the truncation leads then to consistent changes in all EOMs and manifests the relation between different terms.
Here, we will discuss truncations of DSEs and EOMs and describe the system of equations that was used for the calculations.
It should be noted that for the self-consistent treatment of an $m$-point function, an $l$-loop expansion of the $n$PI effective action with $l\geq n$ (and trivially $m\geq n$) is needed.

The goal is to calculate all primitively divergent correlation functions of Yang-Mills theory: the gluon and ghost propagators, the ghost-gluon vertex, the three-gluon vertex, and the four-gluon vertex.
The lowest nonprimitive ones, the two-gluon-two-ghost and four-ghost vertices, were analyzed in Refs.~\cite{Huber:2017txg,Gracey:2017yfi}.
Within a kinematic approximation, their impact on the DSEs of primitively divergent correlation functions was found to be negligible \cite{Huber:2017txg}.
Thus, they are neglected here as are five- and higher $n$-point functions.
In the following, the equations for individual correlation functions and details of the employed truncations are discussed.

\begin{figure}
 \includegraphics[width=0.48\textwidth]{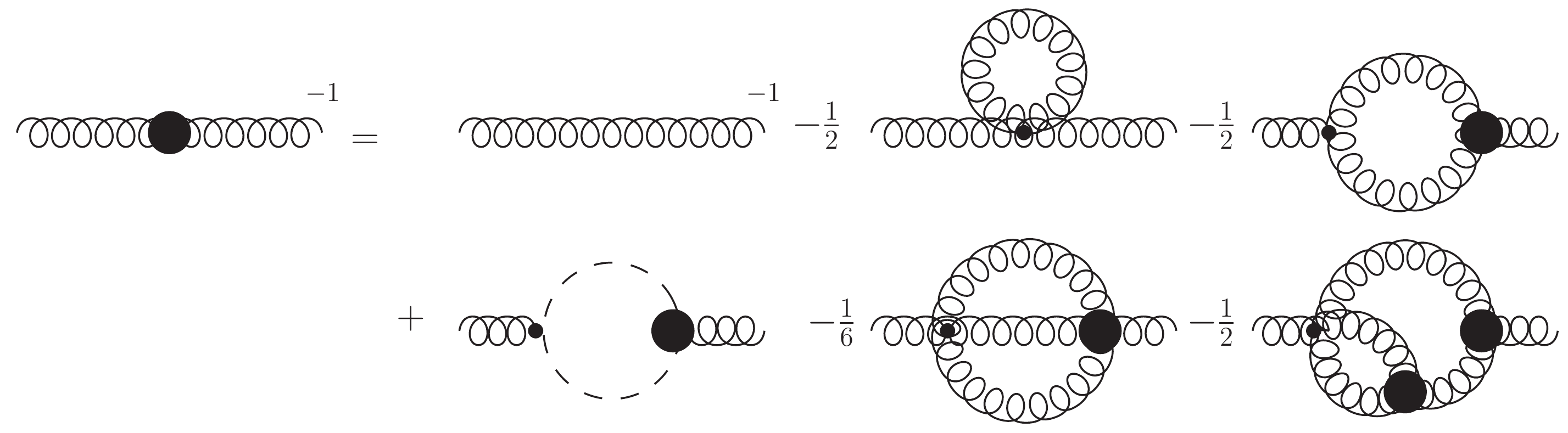}\\
 \vskip5mm
 \includegraphics[width=0.36\textwidth]{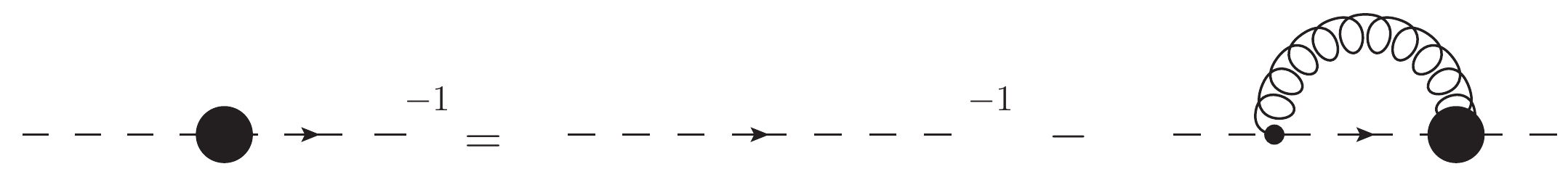}\hspace{1.85cm}
\caption{The gluon (top) and ghost (bottom) propagator DSEs.
In the former, the loop diagrams are referred to as tadpole, gluon loop, ghost loop, sunset, and squint diagrams.}
\label{fig:dse_props}
\end{figure}

The simplest equation is that for the ghost propagator.
Its full DSE is depicted in \fref{fig:dse_props}.
For $n$PI effective actions with $n\geq 3$, some diagrams in the corresponding EOM can be summed up such that the EOM equals the DSE \cite{Berges:2004pu,Carrington:2010qq}.
In all calculations, the full ghost DSE is used.

\begin{figure*}
 \includegraphics[height=3.6cm,left]{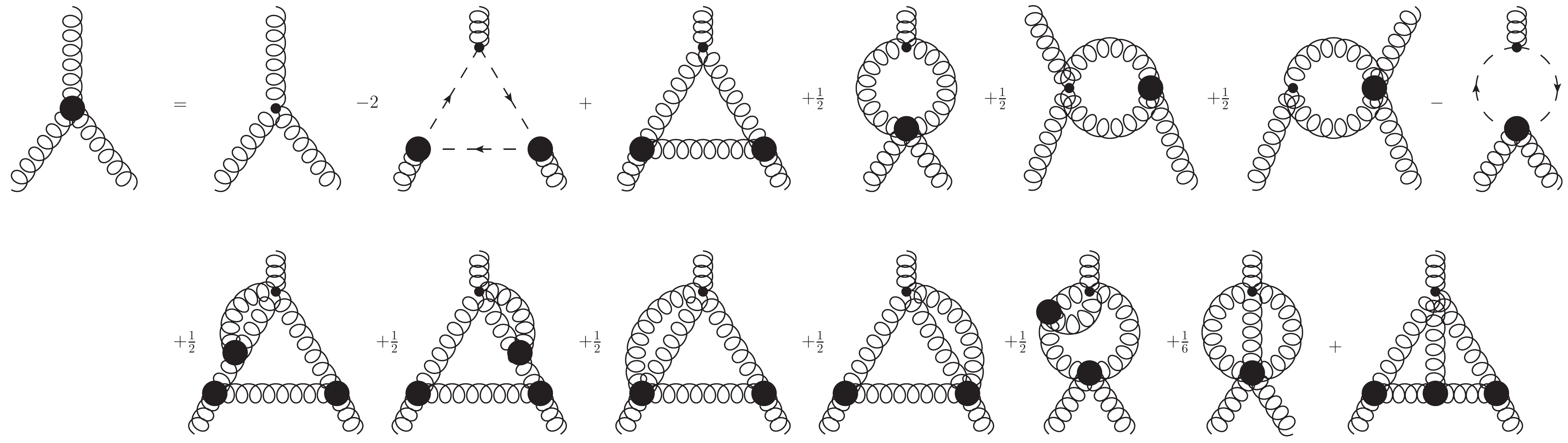}\\
 \vskip5mm
 \includegraphics[height=1.75cm,left]{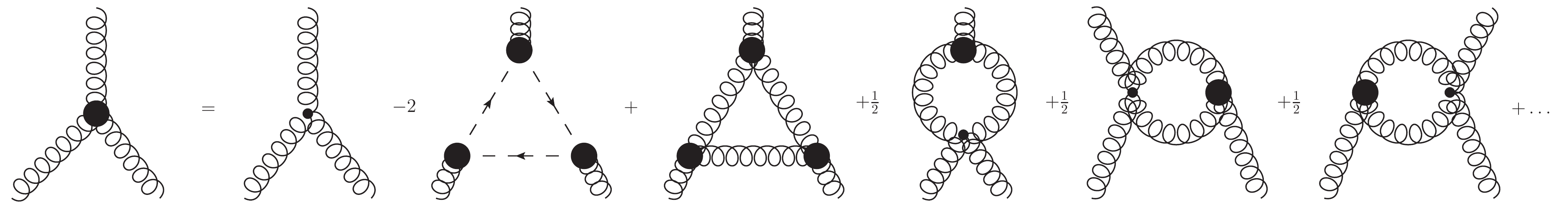}
\caption{DSE (top) and EOM from the three-loop expanded 3PI effective action (bottom) for the three-gluon vertex.
The dots represent discarded higher loop contributions in the EOM from the 3PI effective action.}
\label{fig:eqs_3g}
\end{figure*}

Also for the gluon propagator resummations can be performed and its EOM looks similar to the DSE.
The latter is depicted in \fref{fig:dse_props}.
The EOM from an $n$PI effective action with $n\geq 4$ has the same form \cite{Carrington:2010qq}.
For the three-loop truncated 3PI effective action, the only difference is the replacement of the dressed four-gluon vertex in the sunset diagram by a bare one.
Although within a certain approximation \cite{Mader:2013ru} or for three-dimensional Yang-Mills theory \cite{Huber:2016tvc} it was found that the sunset is subleading in the gluon propagator DSE, this is not the case here.
On the contrary, it is found that a dressed vertex is required for the stability of the solution.
It should be stressed, though, that this statement is made within the given truncation.
If the truncation was changed, for example, by including all tensors of the three-gluon vertex, this would need to be tested again.
For now, the dressed four-gluon vertex is always included in the calculations.

The inclusion of two-loop diagrams in the calculation of the gluon propagator DSE goes beyond most previous calculations with a few exceptions, e.g., \cite{Bloch:2003yu,Mader:2013ru,Meyers:2014iwa,Huber:2016tvc,Huber:2017txg}.
It is also noteworthy that in a perturbative treatment of the Curci-Ferrari model, where the gluon mass is treated as an effective parameter, the two-loop contributions lead to a considerable improvement compared to the one-loop order \cite{Gracey:2019xom}.
In summary, at the two-point level, all equations are untruncated DSEs and thus exact.

The complete DSE of the three-gluon vertex, depicted in \fref{fig:eqs_3g}, contains 14 diagrams.
At leading order of perturbation theory, only five of the one-loop diagrams contribute, while the sixth one-loop diagram, the ghost swordfish diagram, is perturbatively suppressed by $g^2$, because the two-ghost-two-gluon vertex starts at order $g^4$ and not $g^2$ as the four-gluon vertex.
As mentioned above, it is neglected here due to the negligible impact it has also nonperturbatively \cite{Huber:2017txg} and for simplicity this truncated equation is called one-loop truncated DSE.
Before the question of the impact of the nonperturbative two-loop diagrams is discussed, we compare the one-loop truncated DSE to the EOM of the three-loop truncated 3PI effective action which is also depicted in \fref{fig:eqs_3g}.
This equation is very similar to the one-loop truncated DSE (without ghost swordfish) with the difference that all three-/four-point functions are dressed/bare.
Obviously, the one-loop truncated DSE is not Bose symmetric, while the EOM is.
This can be remedied for the DSE by symmetrizing the result \cite{Blum:2014gna,Eichmann:2014xya}; see also Appendix~\ref{sec:numerics}.
However, it is convenient to symmetrize the result from the EOM as well to cancel subleading numeric effects arising from having to make a specific choice for the legs with momenta $p_1$ and $p_2$.
Despite the similarities of the equations, we will see below that they yield very different results and the EOM result is closer to the result from the DSE with two-loop diagrams included.

\begin{figure*}
 \includegraphics[width=0.48\textwidth]{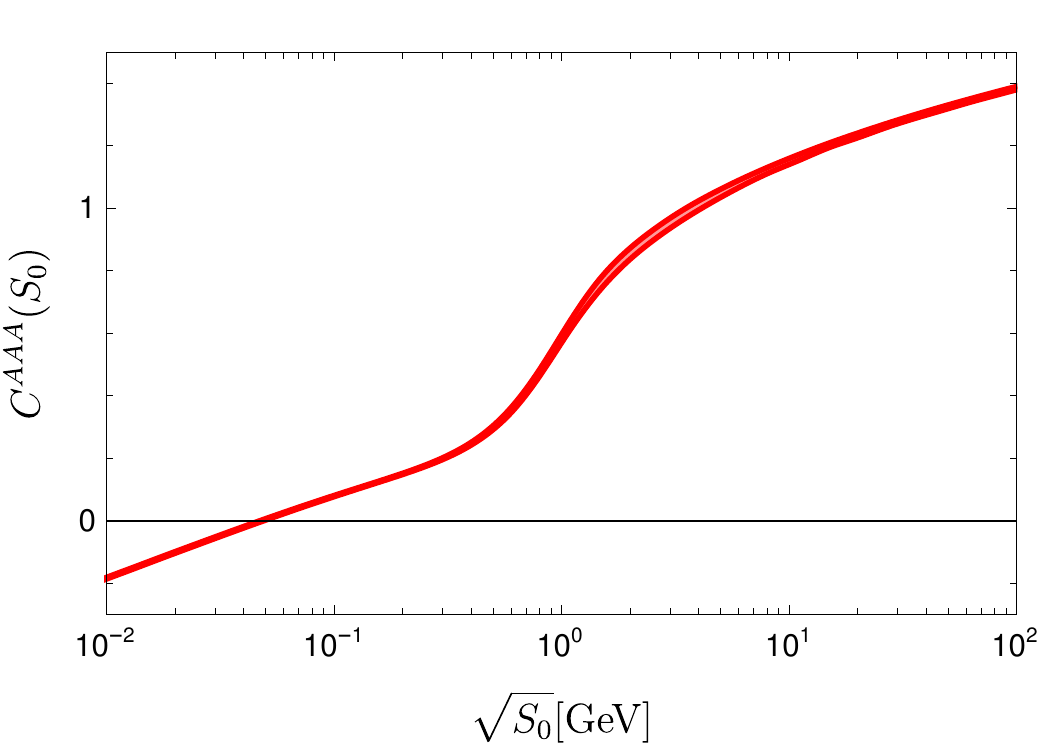}\hfill
\includegraphics[width=0.48\textwidth]{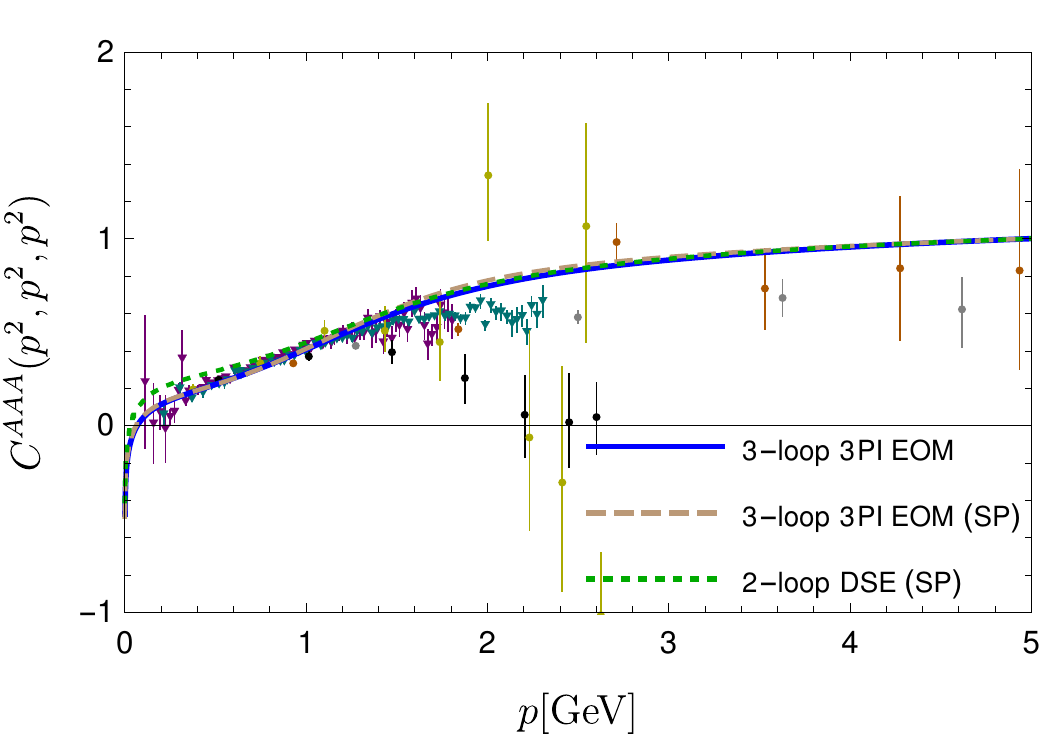}\\
\caption{Left: The three-gluon vertex dressing function for a specific solution as a function of $S_0$ with the band indicating the angular dependence.
Right: The three-gluon vertex dressing function from the 3PI effective action truncated at three loops and its DSE including two-loop diagrams.
The latter is evaluated without angular dependence at the symmetric point (SP).
For the EOM, results with and without angular approximation are shown.}
\label{fig:tg_angle_truncs}
\end{figure*}

The calculation of two-loop diagrams of three-point functions is complicated.
To assess their importance, they were calculated within a one-momentum approximation.
The diagrams with the two-ghost-two-gluon and five-gluon vertices were not included.
From previous calculations using one-loop truncations, it is known that the leading kinematic dependence comes from the momentum scale \cite{Blum:2014gna,Eichmann:2014xya,Aguilar:2019jsj}.
When using the $S_3$ variables, this is reflected in the dependence on the singlet $S_0$ and only a small dependence on the doublet variables $a$ and $s$.
In the left plot of \fref{fig:tg_angle_truncs}, this dependence is shown via the small band.
Thus, neglecting the angular dependence of the vertex should provide a reasonable first guess to test the impact of two-loop diagrams and simplifies the calculation.
Still, the computational cost increases considerably compared to a one-loop truncation, because seven instead of three integrations have to be performed numerically.

In the right plot of \fref{fig:tg_angle_truncs}, results for the three-gluon vertex from different equations and truncations are compared: from its DSE including two-loop diagrams and the EOM of the three-loop truncated 3PI effective action.
Since the former was evaluated using a kinematic approximation, the EOM was evaluated with and without the same approximation to confirm how small the effect of it is.
The equations were solved for a fixed input, for which a solution of the full system of equations was used.
It would be interesting to compare these solutions to those of the one-loop truncated DSE.
However, with the given input, no solution is obtained in that case because the gluon triangle is too strong compared to the swordfish diagrams and the three-gluon vertex equation diverges.
This imbalance depends on the details of the input and the situation can be different for variations of it.
In this respect, it is also important that no RG improvement terms are included for the bare vertices.
Such terms were found to have a substantial impact on the solution by shifting the zero crossing to lower momenta \cite{Blum:2014gna,Eichmann:2014xya}.
Conversely, without RG improvement terms, the zero crossing is at scales above $1\,\text{GeV}$ which makes such solutions very different to the ones shown in \fref{fig:tg_angle_truncs}.
For the EOM setup, solutions with and without the same kinematic approximation that was used for the two-loop DSE are shown, but the difference is marginal.
The solution of the two-loop DSE is very similar to that of the EOM, with small differences around $2\,\text{GeV}$ and below.
Within errors, both results agree with lattice results.
Thus, since it is much easier to evaluate and one does not need to resort to kinematic approximations, for the three-gluon vertex the EOM is used instead of the DSE.

\begin{figure}
 \includegraphics[height=1.5cm,left]{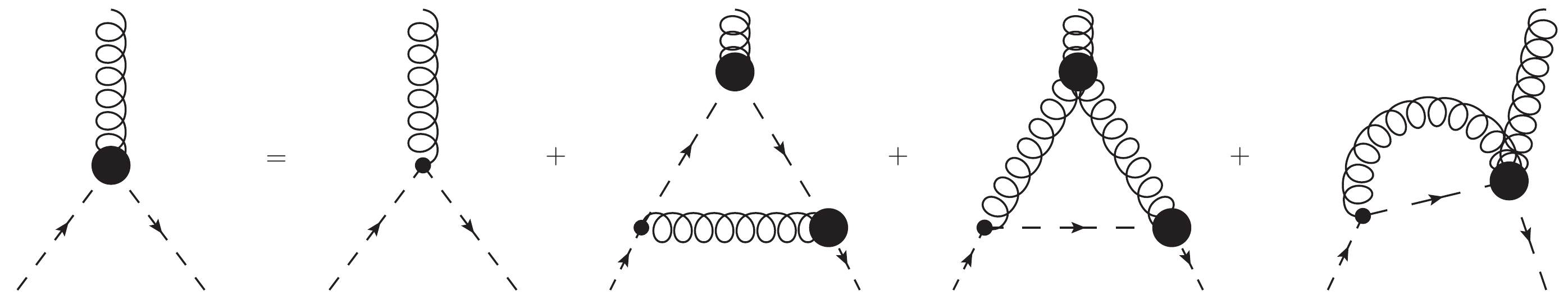}\\
 \vskip5mm
 \includegraphics[height=1.5cm,left]{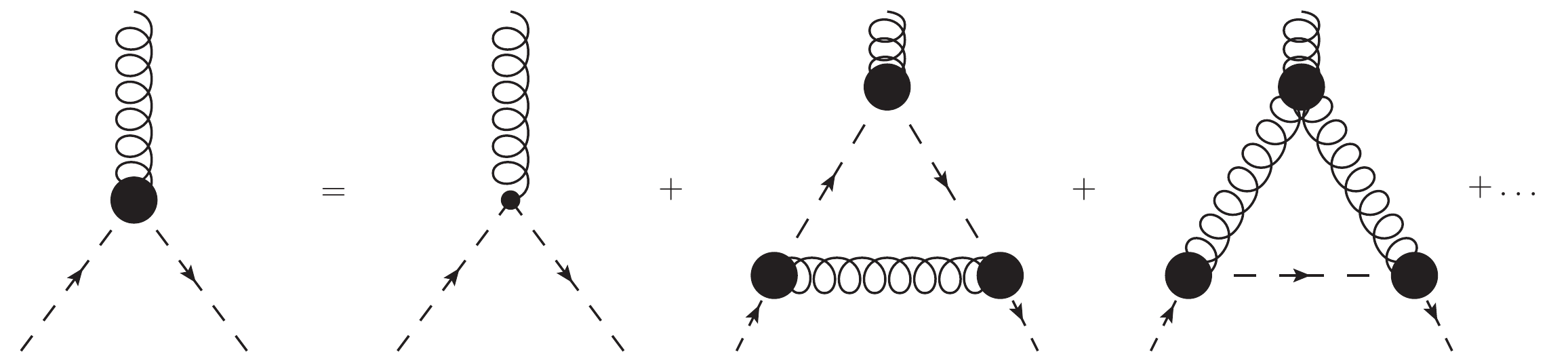}
\caption{DSE (top) and EOM from the three-loop expanded 3PI effective action (bottom) for the ghost-gluon vertex.
For the former, the equation with the bare vertex connected to the ghost leg is shown.
The triangle diagram with/without the three-gluon vertex is called non-Abelian/Abelian diagram}
\label{fig:eqs_ghg}
\end{figure}

The ghost-gluon vertex has two different DSEs that are distinguished by the leg attached to the bare vertex.
For a detailed discussion of the differences, see Refs.~\cite{Huber:2016tvc,Huber:2018ned}.
In \fref{fig:eqs_ghg}, the DSE with the ghost leg attached to the bare vertex and the EOM of the three-loop expanded 3PI effective action are shown.
The DSE where the gluon leg is connected to the bare vertex in the diagrams contains two-loop diagrams.
It is not only more difficult to calculate, one would also need the two-ghost-two-gluon and the two-ghost-three-gluon vertices which we do not include here.
The other DSE consists only of one-loop diagrams but also contains the two-ghost-two-gluon vertex.
However, the influence of this diagram was found to be very small at least under certain approximations for the four-point vertex \cite{Huber:2017txg}.
One can compare this truncation to the EOM of the three-loop truncated 3PI effective action, also shown in \fref{fig:eqs_ghg}.
It looks very similar to the truncated DSE with the exception that all vertices are dressed.
By comparing results from both equations, one can get an estimate of the truncation error.
In the following, for consistency with the three-gluon vertex, the EOM is used.
The effect of using the DSE instead is discussed in Appendix~\ref{sec:ghg_eqs}.
In summary, the vertex itself still depends on the choice of truncation, but the other correlation functions are hardly affected.

\begin{figure}
 \includegraphics[width=0.48\textwidth]{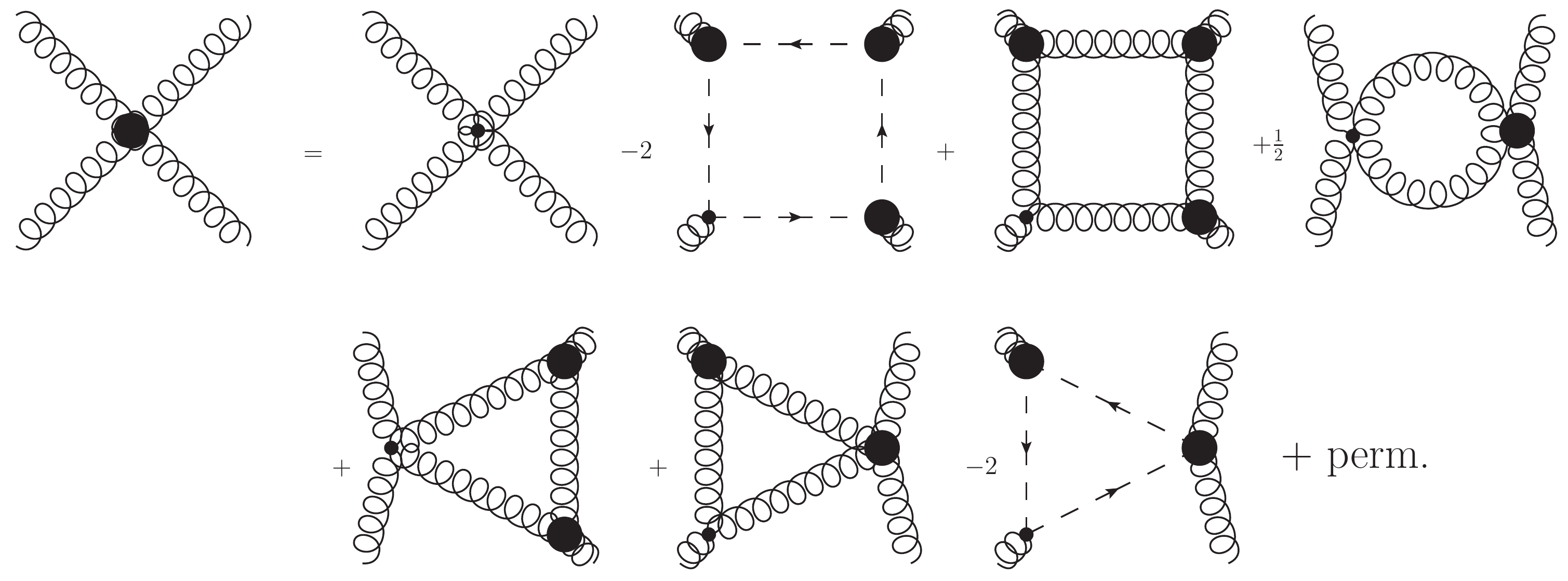}\\
\caption{The one-loop truncated four-gluon vertex DSE.}
\label{fig:dse_4g}
\end{figure}

The DSE for the four-gluon vertex truncated at one-loop level is depicted in \fref{fig:dse_4g}.
For the calculations, these diagrams without the ghost triangle, the impact of which was found to be small \cite{Huber:2017txg}, are used.
The four external legs lead to a proliferation of combinatoric possibilities of diagrams.
However, when solving the equation, each diagram type is calculated only once.
From this result, the permutated diagrams can be obtained; see Ref.~\cite{Cyrol:2014kca} and Appendix~\ref{sec:numerics} for details.
The EOM from the four-loop expanded 4PI effective action looks very similar to the one-loop part of the DSE but with all vertices dressed and an additional diagram type, a swordfish with two two-ghost-two-gluon vertices.

\section{Renormalization}
\label{sec:renormalization}

Due to their nonperturbative nature, the renormalization of DSEs can be an intricate issue.
In particular, when several correlation functions are involved, this problem becomes severe.
Also, perturbative resummation is closely related to a proper renormalization.
This becomes apparent when tracing the origins of different contributions to the perturbative series of the one-loop resummed expression.
To illustrate this, let us consider the perturbative, one-loop resummed expression
\begin{widetext}
\begin{align}
  \left(1+g^2 \beta_0\, \ln{\frac{ p^2}{\mu^2}}\right)^{\gamma} = 1+\gamma\beta_0 g^2 \ln{\frac{ p^2}{\mu^2}} +\frac{1}{2}\gamma(\gamma-1)\left(\beta_0 g^2\right)^2 \ln^2{\frac{ p^2}{\mu^2}} + \mathcal{O}(g^6).
\end{align}
\end{widetext}
$\beta_0$ is the first coefficient of the beta function and $\gamma$ the anomalous dimension, e.g., for the gluon.
The contribution proportional to $g^2 \ln{p^2/\mu^2}$ comes directly from the one-loop diagrams.
The higher order terms, however, are combinations from different sources.
In Ref.~\cite{Huber:2018ned}, it was shown explicitly for $\phi^3$ theory how the terms of order $g^4 \ln^2{p^2/\mu^2}$ are a combination of two-loop contributions and one-loop contributions including counterterms.
The same happens for QCD and the employed truncation needs to fulfill certain criteria to be able to recover the one-loop resummation.
In particular, renormalization needs to be done properly, as the counterterms contribute directly via the renormalization constants.
Thus, a naive two-loop calculation alone is not sufficient, because also the renormalization constants in front of the one-loop diagrams are crucial.
A consistent inclusion of two-loop diagrams in the gluon propagator DSE respecting this was realized for the first time in Ref.~\cite{Huber:2017txg}.
The calculation of the renormalization constants will be explained later.

In most previous calculations, the lack of perturbative self-consistency leads to problems with regard to the resummed perturbative behavior.
This problem is not specific to Yang-Mills theory but applies to full QCD as well.
Despite variations in the reasoning, all workarounds led to the introduction of artificial terms in the equations.
For example, the momentum-independent renormalization constant can be replaced by a momentum-dependent function, e.g., \cite{vonSmekal:1997vx,Fischer:2002eq,Huber:2012kd,Blum:2014gna,Eichmann:2014xya,Cyrol:2014kca,Aguilar:2019kxz}, or vertex models can be modified appropriately to cancel the renormalization constants and correct the anomalous dimensions, e.g., \cite{Maris:1997tm,Maris:1999nt,Fischer:2009wc}.
In both cases, one gets rid of the cutoff-dependent renormalization constants in a way that the RG behavior of the equation is respected.
However, there is always a model-dependent aspect to this which one would like to overcome if one aims at quantitative predictive power.
For the Yang-Mills propagators, it was demonstrated in Ref.~\cite{Huber:2017txg} how this is achieved.
Here, it is extended to all other primitively divergent correlation functions.
The crucial ingredient is the interplay of renormalization and higher-loop contributions, see Ref.~\cite{Huber:2018ned} for details.

We start the discussion of renormalization with the simplest equation, the ghost propagator DSE.
Its renormalization is realized via subtraction at vanishing momentum,
\begin{align}
 G(x)^{-1}=\widetilde{Z}_3+\Sigma_G(x)=G(0)^{-1}+\Sigma_G(x)-\Sigma_G(0),
\end{align}
where $\Sigma_G$ is the ghost self-energy, $\widetilde{Z}_3$ the ghost wave function renormalization constant, and $x=p^2$.
$G(0)$ can be varied to obtain different solutions \cite{Fischer:2008uz}.
In contrast, in functional renormalization group (FRG) calculations the initial conditions for the gluon propagator can be varied to achieve the same effect \cite{Cyrol:2016tym}.
For $G(0)\rightarrow \infty$, the so-called scaling solution is obtained \cite{vonSmekal:1997vx,vonSmekal:1997is,Zwanziger:2001kw,Lerche:2002ep,Pawlowski:2003hq} for which the dressing functions obey power laws.
The corresponding exponents are related to each other \cite{Zwanziger:2001kw,Lerche:2002ep,Alkofer:2004it,Huber:2007kc}, unique \cite{Fischer:2006vf,Alkofer:2008jy,Huber:2009wh,Fischer:2009tn,Huber:2012zj} and can be calculated analytically in terms of $\kappa=0.595$ \cite{Zwanziger:2001kw,Lerche:2002ep}.
The qualitative behavior, including the scaling relation for the propagators, is in agreement with the Gribov-Zwanziger picture \cite{Gribov:1977wm,Zwanziger:1989mf,Zwanziger:1992qr,Zwanziger:1993dh} which describes the effect of the Gribov problem on correlation functions.
Although originally analyzed in a semiperturbative way, this can be generalized to a fully nonperturbative analysis \cite{Huber:2009tx,Huber:2010cq,Huber:2010ne}.
Choosing a finite value of $G(0)$, a family of so-called decoupling solutions is obtained.
Their characteristic feature is an infrared (IR) finite gluon propagator \cite{Cornwall:1981zr,Cucchieri:2007md,Cucchieri:2008fc,Sternbeck:2007ug,Bogolubsky:2009dc,Dudal:2008sp,Boucaud:2008ji,Aguilar:2008xm,Fischer:2008uz,Alkofer:2008jy,Oliveira:2012eh,Quandt:2013wna}.
All other dressing functions are IR finite or logarithmically divergent.
This can be shown analytically \cite{Alkofer:2008jy,Fischer:2009tn} but was also seen in various numerical calculations, e.g., \cite{Huber:2012kd,Aguilar:2013xqa,Blum:2014gna,Eichmann:2014xya,Binosi:2014kka,Cyrol:2014kca,Cyrol:2016tym,Aguilar:2018csq,Aguilar:2019jsj}.
The decoupling type of solutions can be accommodated in the Gribov-Zwanziger picture by taking into account certain condensates \cite{Dudal:2008sp,Gracey:2010cg,Dudal:2011gd}.
In addition, the finite value of the gluon propagator at zero momentum has motivated effective descriptions of correlation functions using massive extensions of Yang-Mills theory \cite{Tissier:2010ts,Tissier:2011ey,Serreau:2012cg,Serreau:2013ila,Weber:2011nw,Siringo:2014lva,Machado:2016cij,Siringo2016,Gracey:2019xom}, e.g., based on the Curci-Ferrari model \cite{Curci:1976bt,Tissier:2010ts,Tissier:2011ey,Serreau:2012cg,Serreau:2013ila}.

The origin of different solutions has been attributed to the Gribov problem which refers to the fact that the perturbative definition of the Landau gauge is only an incomplete gauge fixing \cite{Gribov:1977wm,Singer:1978dk,vanBaal:1991zw}.
Indeed, lattice calculations have seen variations of the correlation functions depending on the details of the employed gauge fixing algorithm \cite{Cucchieri:1997dx,Bogolubsky:2005wf,Sternbeck:2006rd,Maas:2009se,Bornyakov:2011fn,Maas:2011se,Maas:2013vd,Bornyakov:2013ysa,Sternbeck:2012mf}.
However, a one-to-one correspondence between functional and lattice prescriptions has not been found yet; see, e.g. \cite{Sternbeck:2012mf,Maas:2019ggf}.
Here, the space of solutions is scanned by varying $G(0)$.
The scaling solution corresponds to the limit of an infinite gluon mass with $G(0)\rightarrow \infty$.
Lowering $G(0)$, also the gluon mass decreases.
As discussed in detail in Ref.~\cite{Cyrol:2016tym}, there is a critical value which separates solutions into confined  and Higgs-type classes.
The former class is characterized by a maximum in the gluon propagator at nonvanishing momenta.
For the Higgs-type solutions, the maximum is at zero.
The existence of a maximum at nonvanishing momenta does not only lead to positivity violation of the propagator \cite{Alkofer:2000wg}, it also reduces the spectral dimension, viz., the dimension felt by the propagator \cite{Kern:2019nzx}, from four to one.
The calculations in this work span solutions from scaling to the boundary of the Higgs branch.

The gluon propagator is renormalized via momentum subtraction as well.
However, there is an additional complication that haunted calculations of the gluon propagator for a long time, namely, the appearance of quadratic divergences; see, e.g., \cite{Brown:1988bm,vonSmekal:1997vx,Atkinson:1997tu,Fischer:2002eq,Fischer:2005en,Aguilar:2009ke,Huber:2012kd,Meyers:2014iwa,Huber:2014tva}.
Their origin lies in the breaking of gauge covariance by the regularization procedure.
Methods to subtract these divergences are either limited in their practical applicability, for instance, because they require an exact knowledge of the vertices, or they introduce a new parameter.
For an overview, see Ref.~\cite{Huber:2014tva}.
The new parameter is a manifestation of the fact that the regularization scheme breaks gauge covariance.
Consequently, a counterterm for the gluon mass is no longer forbidden and can be used to renormalize the quadratic divergence \cite{Collins:2008re}.
However, the corresponding renormalization condition is not fixed and one would expect that results depend on what value one chooses.
Indeed, calculations performed up to now using this renormalization scheme \cite{Meyers:2014iwa,Huber:2016hns,Huber:2017txg} saw such a behavior.
The dependence on this parameter is also studied within the present truncation scheme in Sec.~\ref{sec:glPAt0}.

The gluon propagator renormalization is realized as follows \cite{Meyers:2014iwa,Huber:2018ned}.
We write the renormalized gluon propagator DSE as
\begin{align}
 Z^{-1}(x)=Z_3+\Sigma_Z(x)-\frac{C_\text{sub}}{x},
\end{align}
where $\Sigma_{Z}$ is the self-energy, $Z_3$ the gluon wave function renormalization constant and $C_\text{sub}$ the renormalization constant for the quadratic divergences.
One can determine the renormalization constant $Z_3$ in a standard MOM scheme by choosing a fixed value for $Z(x_s)$,
\begin{align}
 Z_3 = Z^{-1}(x_s)-\Sigma_Z(x_s)+\frac{C_\text{sub}}{x_s}.
\end{align}
The term $C_\text{sub}$ is determined by demanding that the propagator $D(x)=Z(x)/x$ has a fixed value at $x_m$,
\begin{align}
 C_\text{sub}&=\frac{x_m x_s}{x_s-x_m}\left(\Sigma_Z(x_m)-\Sigma_Z(x_s)\right)\nnnl
 &+\frac{x_m x_s}{x_s-x_m} Z^{-1}(x_s)-\frac{ x_s}{x_s-x_m}D^{-1}(x_m).
\end{align}
In summary, we need to specify the renormalization conditions $D(x_m)$ and $Z(x_s)$ with their corresponding renormalization points $x_m$ and $x_s$, respectively.
In practical calculations, it turns out to be most convenient to choose $x_m$ at the lowest calculated point and $x_s$ in the perturbative regime.
It should be noted that $D(x_m)$ has a mass dimension of $-2$.
The influence of choosing different values for $D(x_m)$ is discussed in Sec.~\ref{sec:glPAt0}.

For the scaling solution, the gluon propagator does not approach a fixed value at zero momentum but vanishes as $D(x)=d_\text{gl} x^{\de_\text{gl}-1}$.
Thus, one cannot choose $D(x_m)$ freely.
On the other hand, it is known that on the right-hand side of the gluon propagator DSE, the ghost loop is responsible for the IR suppression \cite{vonSmekal:1997is,vonSmekal:1997vx}.
This does not change when two-loop diagrams are included \cite{Fischer:2006vf,Huber:2007kc,Alkofer:2008jy,Fischer:2009tn,Huber:2009wh} and leads to an analytic result for the IR fixed point of the \textit{MiniMOM} coupling, given by
\begin{align}
\label{eq:couplingMM}
 \alpha_\text{MM}(p^2)=\alpha(\mu^2)G^2(p^2) Z(p^2),
\end{align}
by comparing the IR dominant diagrams of the ghost and gluon propagator DSEs \cite{vonSmekal:1997is,vonSmekal:1997vx,Zwanziger:2001kw,Lerche:2002ep}.
Via this relation, the value for $D(x_m)$ can be determined from the ghost dressing function as explained in Appendix~\ref{sec:aud_divs_scal}.

The renormalization of the ghost-gluon vertex is trivial, as it is finite in Landau gauge \cite{Taylor:1971ff}.
Numerically, it turned out to be advantageous to increase the internally used cutoff for the vertex compared to the divergent quantities.
This reduces subleading cutoff dependencies which cannot be absorbed in the renormalization process as for other divergent quantities.

\begin{figure*}[tb]
 \includegraphics[width=0.48\textwidth]{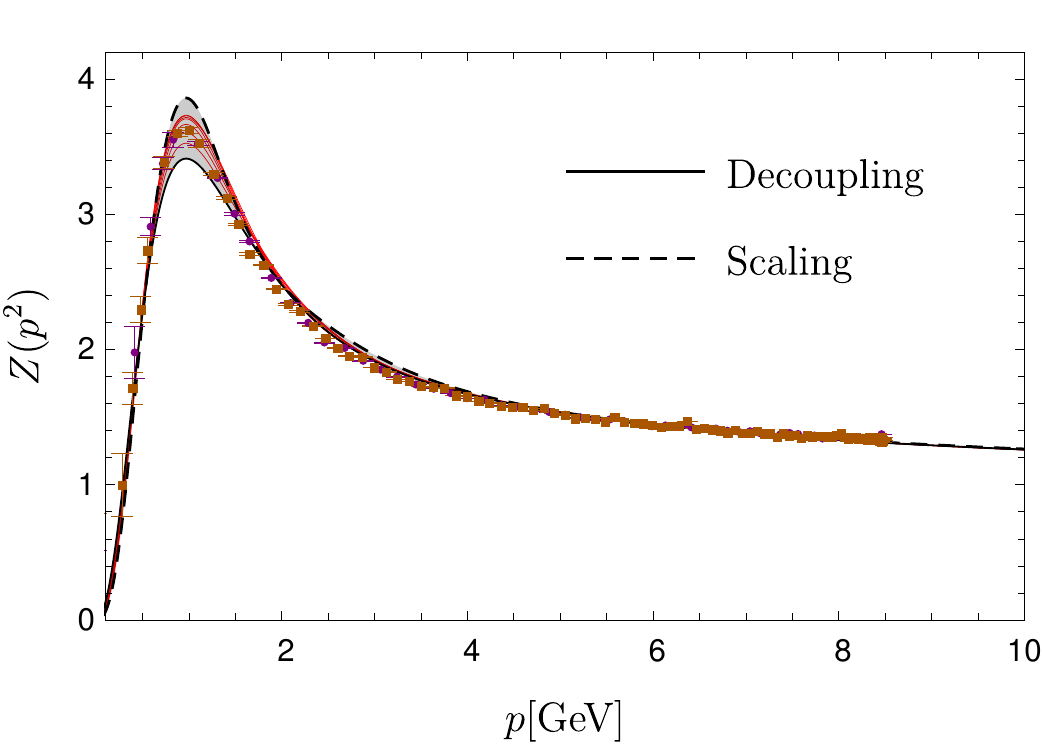}\hfill
 \includegraphics[width=0.48\textwidth]{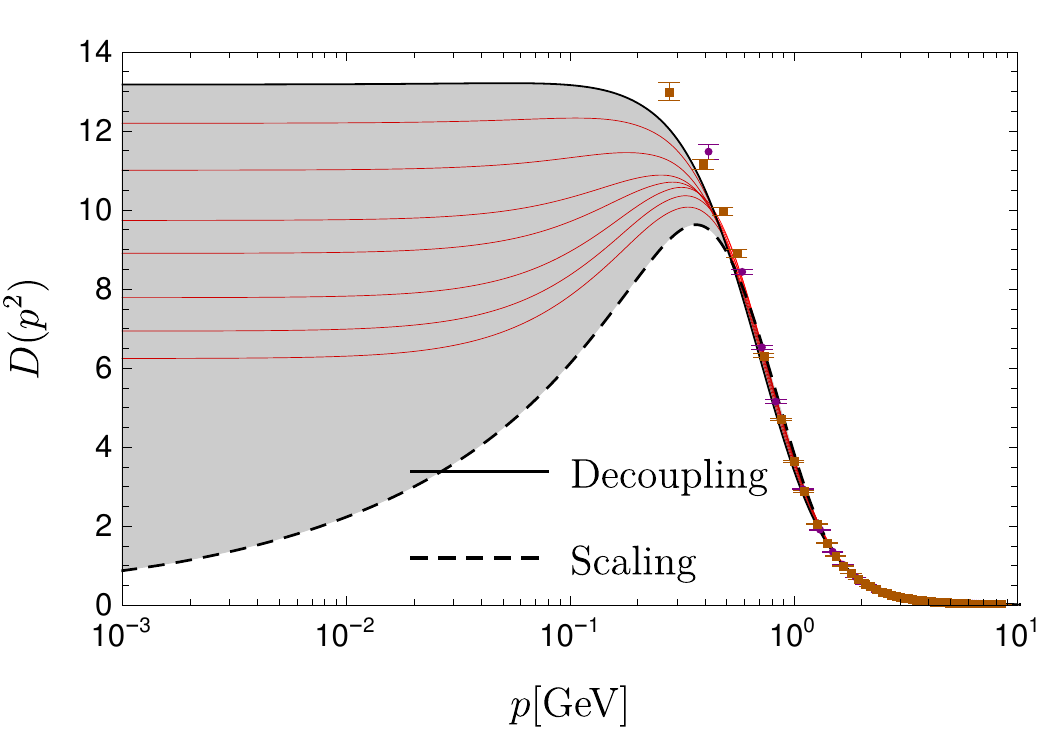}
 \caption{Gluon dressing function $Z(p^2)$ (left) and gluon propagator $D(p^2)$ (right) in comparison to lattice data \cite{Sternbeck:2006rd}.
 Different lines correspond to different solutions as explained in the text.}
 \label{fig:gl}
\end{figure*}

The employed renormalization scheme corresponds to the \textit{MiniMOM} \cite{vonSmekal:1997vx,vonSmekal:2009ae} or Taylor scheme \cite{Boucaud:2008gn} supplemented by the second renormalization condition for the gluon propagator.
This scheme fixes all the renormalization constants via momentum subtraction for the propagators and a minimal subtraction for the ghost-gluon vertex.
Due to the finiteness of this vertex in the Landau gauge \cite{Taylor:1971ff}, its renormalization constant is $\widetilde{Z}_1=1$.
The three- and four-gluon vertex renormalization constants, in turn, are fixed by Slavnov-Taylor identities (STIs).
Alternatively, one could use any other vertex instead of the ghost-gluon vertex to define a scheme; see, e.g., Refs.~\cite{Celmaster:1979km,Gracey:2014ola}.
Since the STIs are a consequence of the gauge invariance of the path integral, they are modified by the presence of the ultraviolet (UV) cutoff in the numerical calculations.
Thus, the original STIs relating the renormalization constants of the three-gluon and four-gluon vertices with those of the propagators and the ghost-gluon vertex,
\begin{align}
\label{eq:STIs}
 Z_1=\widetilde{Z}_1 \frac{Z_3}{\widetilde{Z}_3}\quad  \text{and} \quad Z_4=\widetilde{Z}_1 \frac{Z_3}{\widetilde{Z}^2_3},
\end{align}
respectively, are only fulfilled approximately and they are not used for the renormalization of the respective vertices.
Instead, a momentum-dependent form of the STIs is used which contains the longitudinal pieces of the vertices; see also Ref.~\cite{Cyrol:2016tym} for more details.
Working in Landau gauge, the longitudinal pieces are not calculated here.
The crucial point is that in the perturbative regime the longitudinal and transverse pieces agree.\footnote{For a detailed discussion on possibilities how to separate the longitudinal and transverse parts, see Ref.~\cite{Eichmann:2015nra}.}
Thus, one concludes from the STIs that the couplings of all vertices as calculated from their transverse parts must agree for high momenta.
These running couplings are defined as follows \cite{Alles:1996ka,Alkofer:2004it,Eichmann:2014xya}:
\begin{subequations}
\label{eq:couplings}
\begin{align}
 \alpha_\text{ghg}(p^2)&=\alpha(\mu^2)\left(D^{A\bar cc}(p^2)\right)^2G^2(p^2)Z(p^2),\\
 \alpha_\text{3g}(p^2)&=\alpha(\mu^2)\left(C^{AAA}(p^2)\right)^2Z^3(p^2),\\
 \alpha_\text{4g}(p^2)&=\alpha(\mu^2)F^{AAAA}(p^2)Z^2(p^2).
\end{align}
\end{subequations}
The vertex dressing functions are taken at the symmetric points.\footnote{Note that for the permutation group variables, the symmetric point is $S_0=p^2/2$ for three-point functions and $S_0=p^2$ for four-point functions with $a=s=0$.}
The perturbative equivalence of the couplings is then used to calculate a renormalization condition for the vertices from
\begin{align}
 C^{AAA}(p_r^2)&=D^{A\bar cc}(p_r^2)\frac{G(p_r^2)}{Z(p_r^2)},\nnnl
 F^{AAAA}(p_r^2)&=\left(D^{A\bar cc}(p_r^2)\right)^2\frac{G^2(p_r^2)}{Z(p_r^2)}.
\end{align}
Note that this only ensures that the couplings agree at the renormalization point $p_r$.
Agreement over a wide range of momenta, as demanded by the STIs, is not guaranteed.
This will be discussed further in Sec.~\ref{sec:couplings}.

The renormalization constants also appear in loop diagrams that feature bare vertices.
In calculations using models for dressed vertices, they are often absorbed in the models which thus become cutoff dependent; see the discussion above.
In the absence of such models as here, it is important to include the renormalization constants properly, as this is also important for the correct perturbative resummation of diagrams.
For the renormalization constants in the loop diagrams the values obtained from the STIs were used.
In the iterative process, the renormalization constants need to be updated as well, and only once the final self-consistent solution is obtained, the renormalization constants have obtained values that balance all equations.
This introduces a certain destructive element and it turned out to be most stable to use the renormalization constants from the STIs in these cases.
For the final solution, the values calculated from the two methods agree better than $1\,\%$.
In addition, the renormalization constants of the propagators also agree with the inverse dressing functions at the cutoff to the same degree.
In previous calculations, these quantities typically deviated by a few percent.

\section{Results}
\label{sec:results}

This section contains the results for the propagators and vertices.
Furthermore, self-tests of the results are discussed and performed.
All calculations were done for the gauge group $SU(3)$, but they are equivalent for all $SU(N)$ under rescaling $g^2\,Nc$.
This was explicitly checked for $SU(2)$.
For reference, plots contain also lattice data where available.
A more detailed comparison to lattice and FRG data is done in Sec.~\ref{sec:comparison}.
The results for the dressing functions can be downloaded from \url{https://github.com/markusqh/YM_data}.

\subsection{Propagators}
\label{sec:props}

\begin{figure}[tb]
 \includegraphics[width=0.48\textwidth]{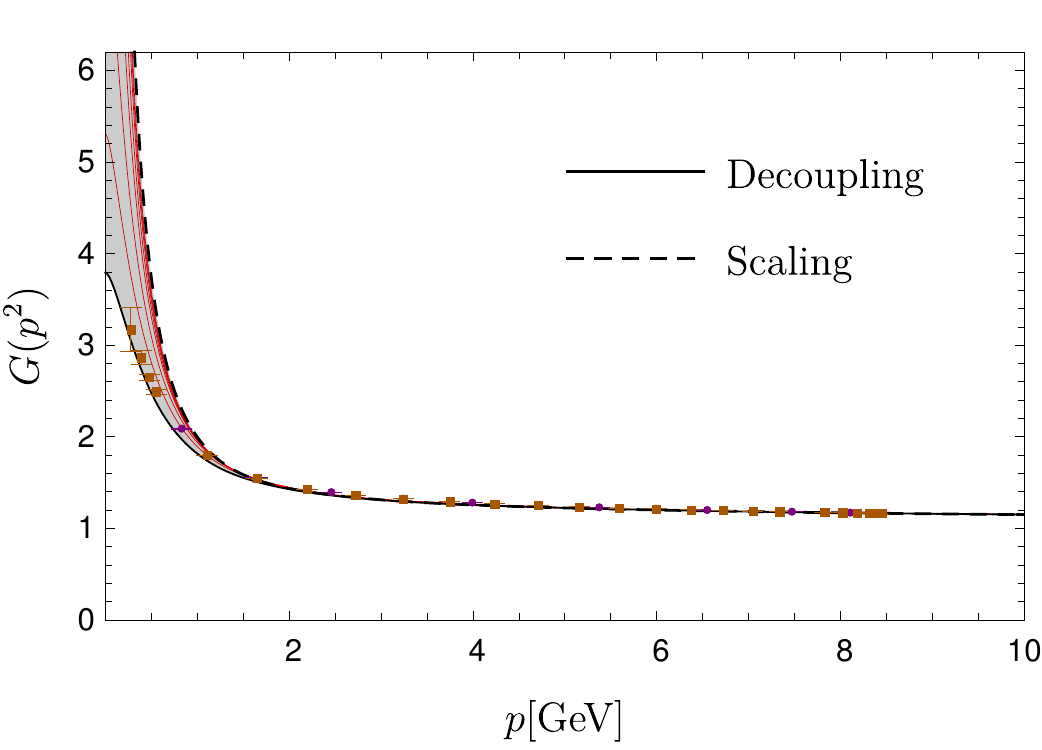}
 \caption{Ghost dressing function $G(p^2)$ in comparison to lattice data \cite{Sternbeck:2006rd}.
 Different lines correspond to different solutions as explained in the text.}
 \label{fig:gh}
\end{figure}

\begin{figure}[tb]
 \includegraphics[width=0.48\textwidth]{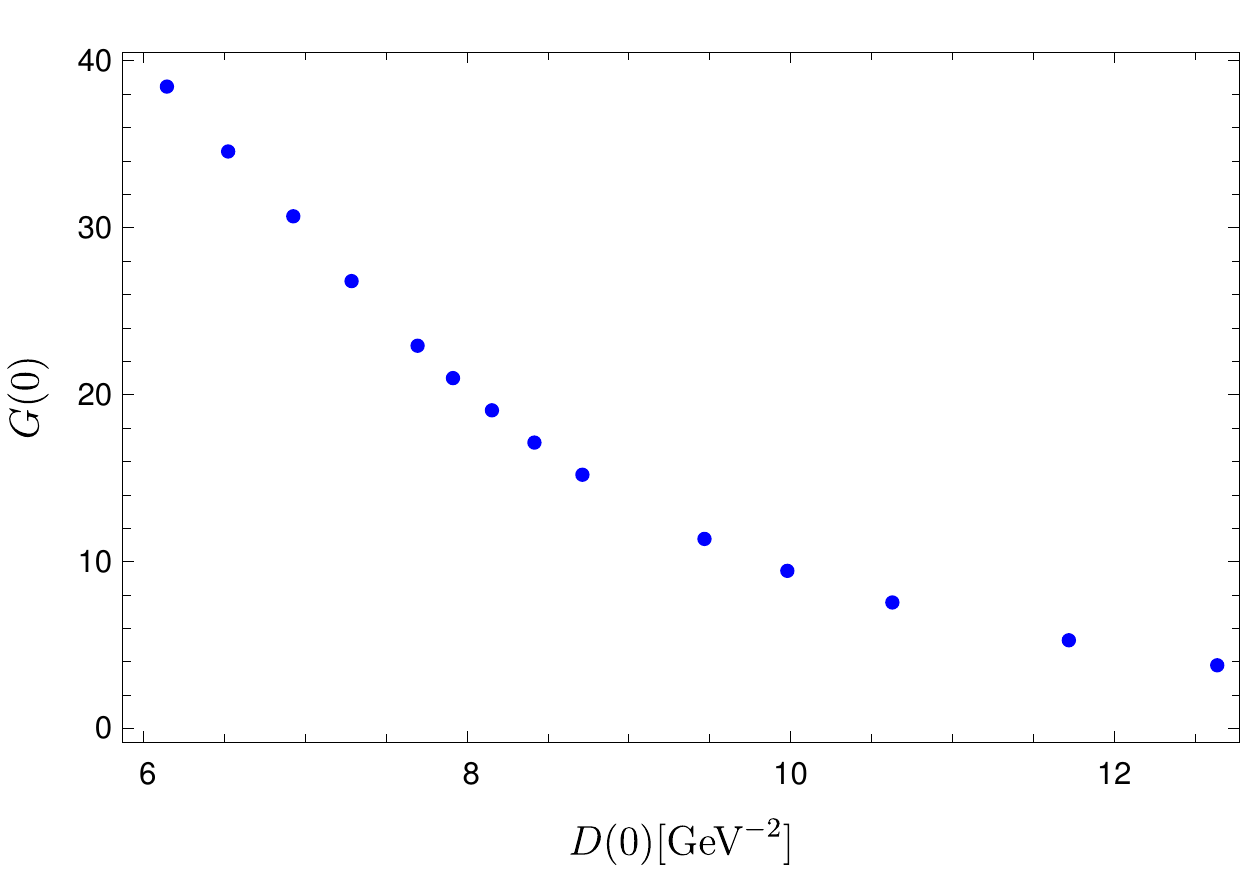}
 \caption{Relation between the ghost dressing function and the gluon propagator at vanishing momentum for various decoupling solutions.
 The scaling solution has $D(p^2\rightarrow 0)=0$ and $G(p^2\rightarrow 0)=\infty$.}
 \label{fig:glP0gh0}
\end{figure}

As discussed in Sec.~\ref{sec:renormalization}, the equations allow a continuum of solutions by choosing different values for the ghost dressing function in the IR.
For the gluon and ghost propagators, results are shown in Figs.~\ref{fig:gl} and \ref{fig:gh}, respectively.
The spectrum of solutions is represented by the gray band which is bounded by the scaling solution (dashed line).
The other bound is a decoupling solution close to the Higgs branch of solutions (black continuous line), viz. the solution with the maximum at the smallest nonvanishing momentum found in the set of calculated decoupling solutions.
As a consequence, the maximum is extremely shallow and can only be seen directly in the data and not in the plots.
Intermediate decoupling solutions are shown as thin red lines.
The relation between the ghost dressing functions and the gluon propagators at vanishing momentum is shown in \fref{fig:glP0gh0}.

All solutions agree roughly down to $2\,\text{GeV}$.
Below, the ghost dressing function, shown in \fref{fig:gh}, bends up and becomes finite for all solutions except the scaling solution for which it diverges.
The gluon dressing function, displayed in \fref{fig:gl}, looks very similar for all solutions except around the momentum scale of $1\,\text{GeV}$.
However, the propagator reveals that in the deep IR the solutions do vary as well.
For all solutions the gluon propagator settles at a nonvanishing value except for the scaling solution for which it vanishes.
The closer the solutions are to the scaling solution, the more pronounced is the maximum in the propagator.
One can also observe that the point where the propagators become basically constant is shifted further into the IR.
For both the gluon and ghost dressing functions, the IR exponents of the scaling solution agree with the analytic results, $\de_\text{gh}=-\kappa=-0.595$ and $\de_\text{gl}=1+2\kappa=1.191$, respectively \cite{Zwanziger:2001kw,Lerche:2002ep}.
Extracting momentum-dependent IR exponents, viz. calculating the power laws for small momentum intervals, one observes that the IR power laws of the scaling solution are reached also for the decoupling solutions close to the scaling solution before they turn to their final values  $\de_\text{gh}=0$ and $\de_\text{gl}=1$.

\begin{figure}[tb]
 \includegraphics[width=0.48\textwidth]{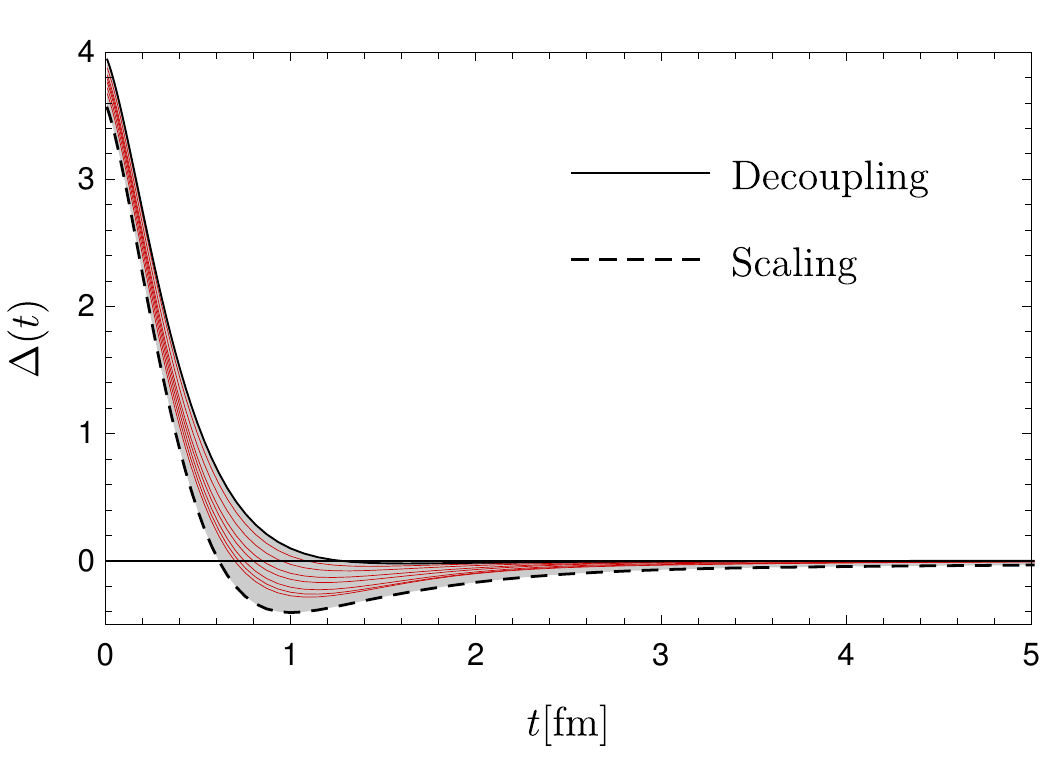}
 \caption{Schwinger function of the gluon propagator for different solutions.}
 \label{fig:Schwinger}
\end{figure}

The comparison with lattice results does not show a candidate solution that agrees well in all quantities.
The best overall agreement is obtained with the solution close to the Higgs-type branch.
However, this agreement is not perfect and one cannot establish a one-to-one correspondence between lattice and functional results.
This was also observed for the corresponding FRG solutions \cite{Cyrol:2016tym}.
Both ghost and gluon propagators show, though, that the lattice results are closer to the Higgs branch than to the scaling solution.
This also explains why the maximum in the gluon propagator is difficult to observe in lattice results for four dimensions \cite{Cucchieri:2007md,Bogolubsky:2009dc,Boucaud:2008ji}.
In general, however, there are additional arguments for the existence of a maximum from both continuum analyses, e.g., \cite{Aguilar:2013vaa,Kern:2019nzx} and from three-dimensional lattice calculations, e.g., \cite{Bornyakov:2013ysa,Maas:2014xma,Cucchieri:2016jwg}.

The observed maximum in the gluon propagator directly leads to a violation of positivity of the spectral function; see, e.g., \cite{Oehme:1979ai,Oehme:1994pv,Maas:2011se,Kern:2019nzx}.
This is reflected in the Schwinger function calculated from the propagator as
\begin{align}
 \Delta(t)&=D(t,\vec{p}=\vec{0})=\int_{-\infty}^\infty \frac{dp_0}{2\pi} e^{-i\,p_0 \,t}D(p_0, \vec{0})\nnnl
 &=\frac{1}{\pi}\int_0^\infty dp_0 \cos(p_0\,t)D(p_0,\vec{0}).
\end{align}
In \fref{fig:Schwinger}, the Schwinger function for different solutions is shown.
For the scaling solution, the violation of positivity around $1\,\text{fm}$ is most pronounced, but also the other solutions become clearly negative.
It should be noted, though, that in the Landau gauge positivity is violated already perturbatively \cite{Oehme:1994hf,Alkofer:2000wg}.

\subsection{Ghost-gluon vertex}

\begin{figure*}[tb]
 \includegraphics[width=0.48\textwidth]{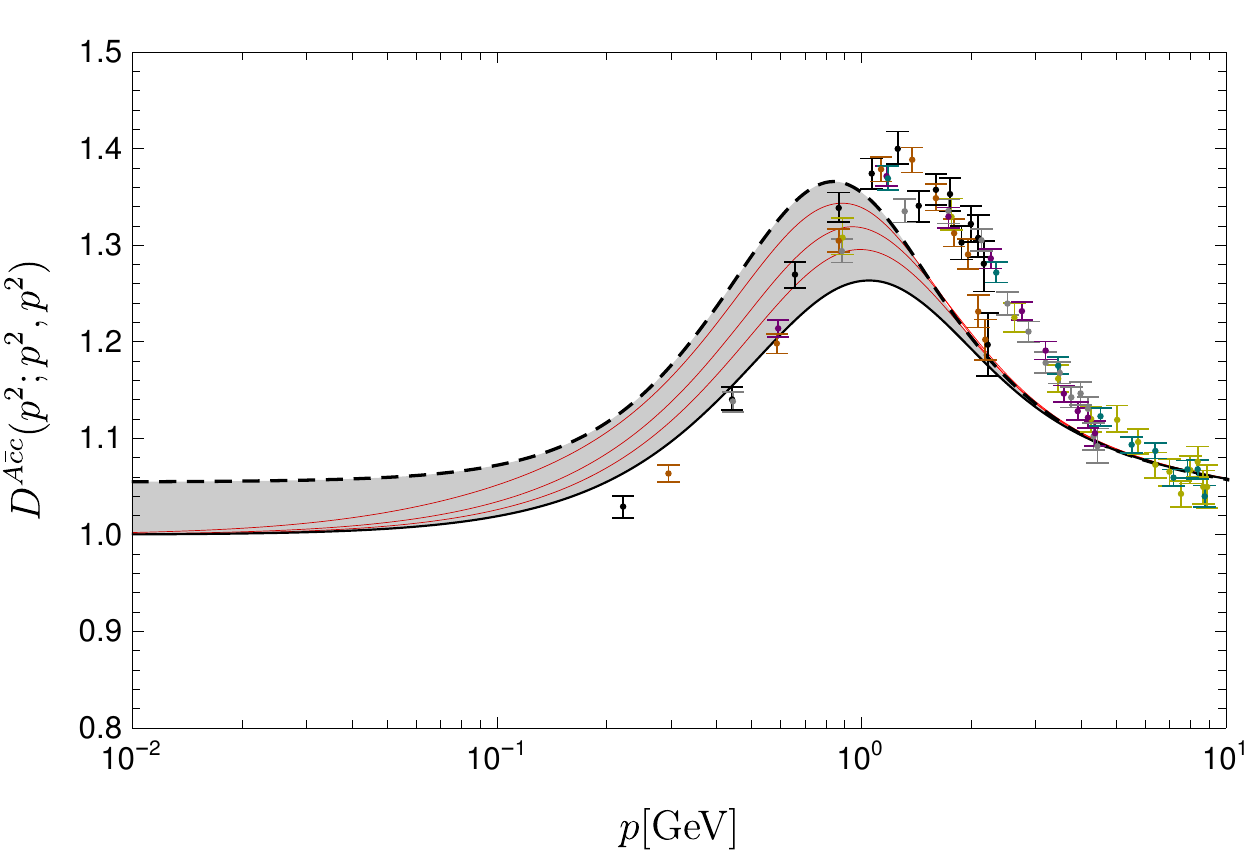}\hfill
 \includegraphics[width=0.48\textwidth]{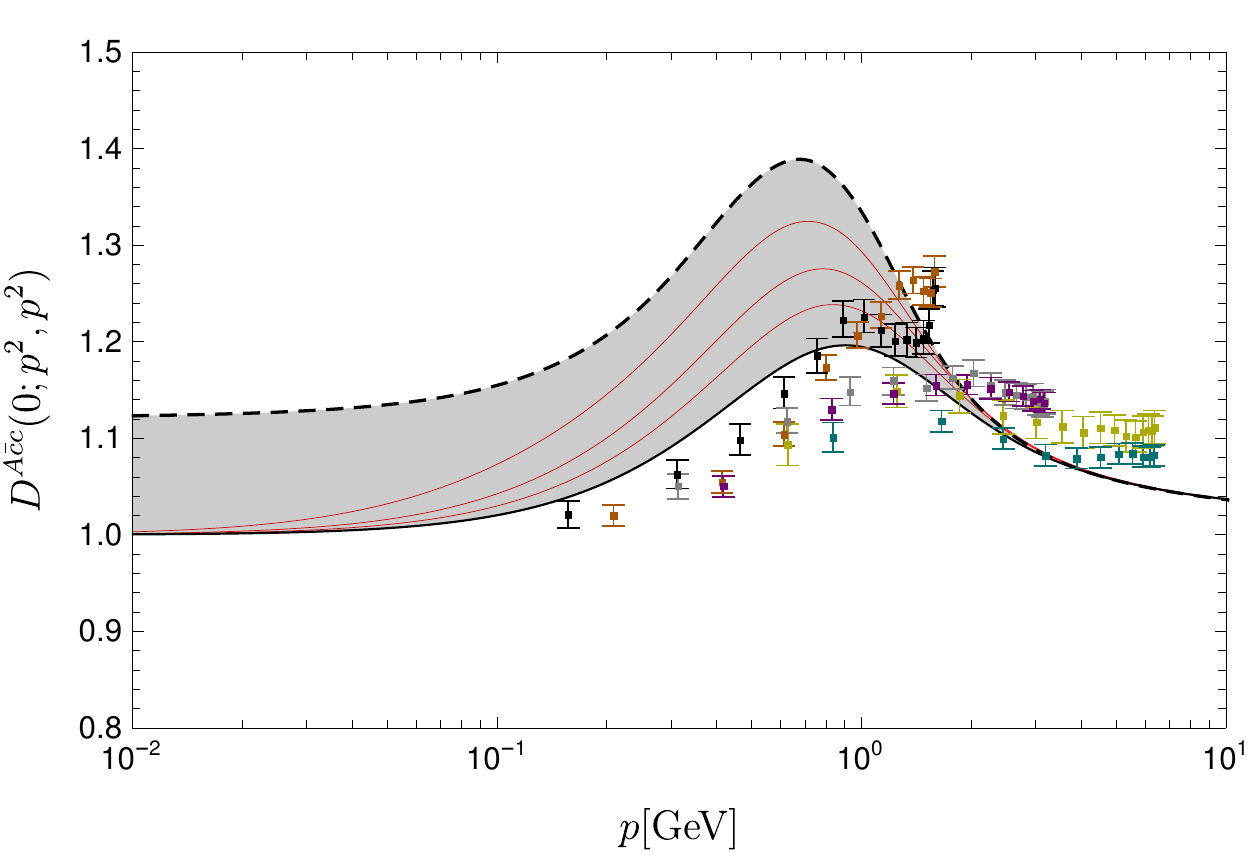}\\
 \includegraphics[width=0.48\textwidth]{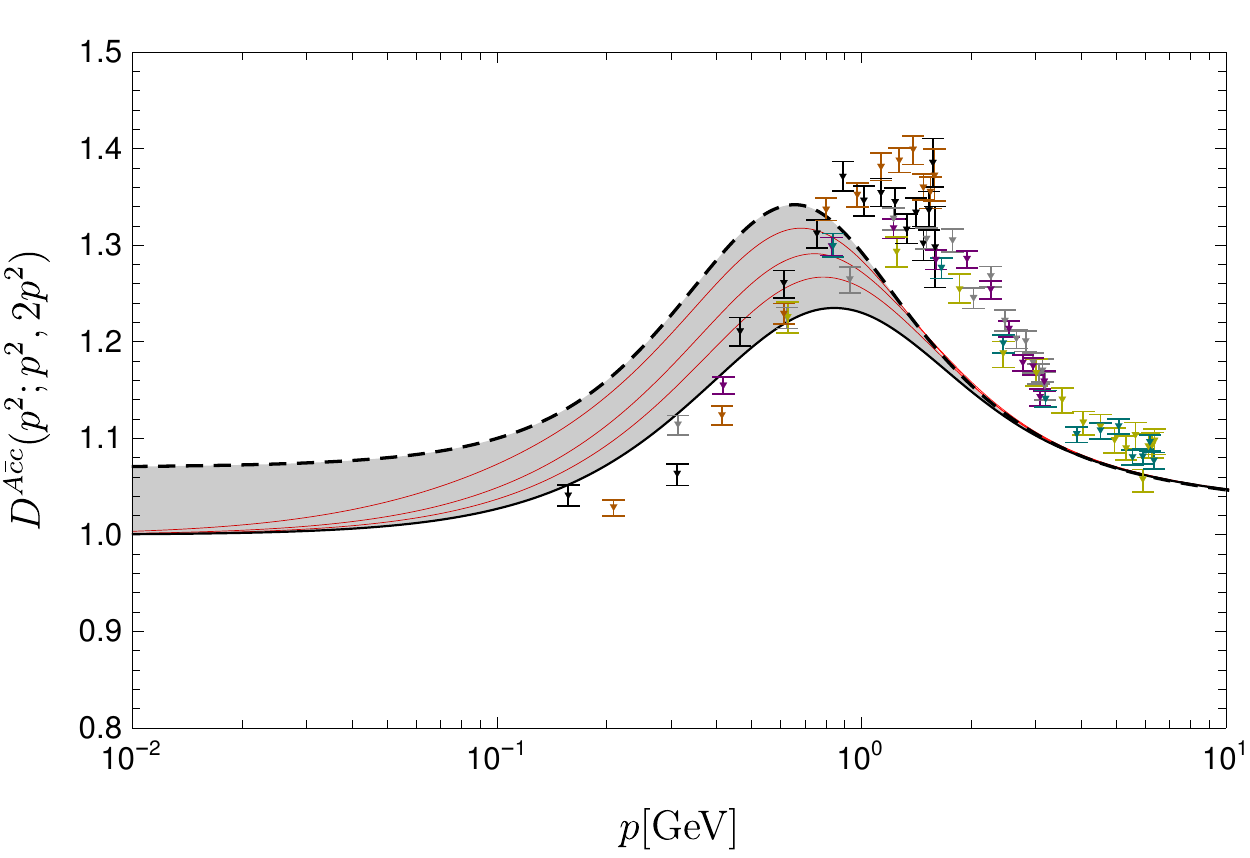}\hfill
 \includegraphics[width=0.48\textwidth]{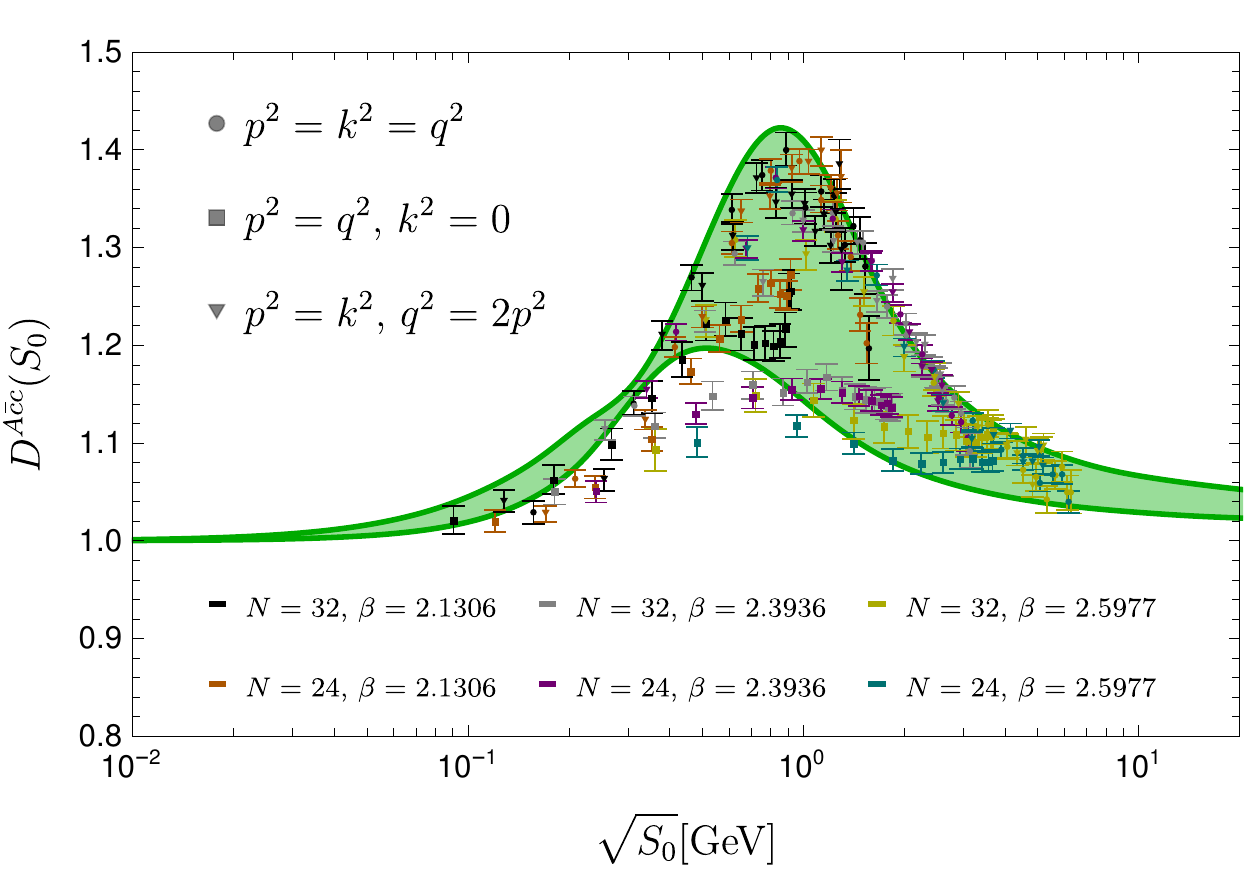}
 \caption{Ghost-gluon vertex dressing function in comparison to $SU(2)$ lattice data \cite{Maas:2019ggf}.
 The dashed/continuous black lines correspond to scaling/decoupling solutions.
 Top left/right: Symmetric/vanishing gluon momentum configuration.
 Bottom left: Ghost and gluon momenta with equal magnitude and orthogonal.
 Bottom right: Full angular dependence for the decoupling solution corresponding to the black continuous line in the other plots.}
 \label{fig:ghg}
\end{figure*}

\begin{figure}[tb]
 \includegraphics[width=0.48\textwidth]{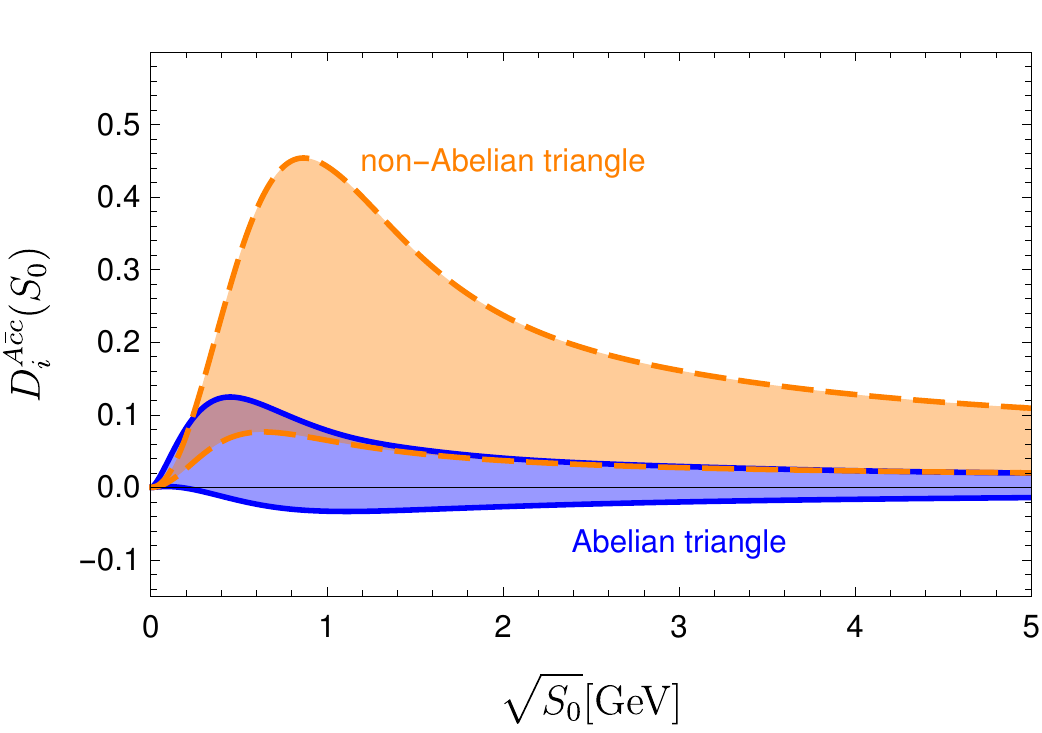}
 \caption{Contributions of individual diagrams in the ghost-gluon vertex equation for the lowest decoupling solution.
 The bands represent the dependence on the variables $a$ and $s$.
 }
 \label{fig:ghg_diags}
\end{figure}

Results for the ghost-gluon vertex are shown in \fref{fig:ghg}.
Due to the comparatively large angular dependence of this vertex, it is worthwhile to compare different kinematic configurations.
Again, a selection of the family of solutions is shown.
At low momenta, the vertex dressing function becomes constant.
However, while it becomes one for all decoupling solutions, it settles at a value larger than one for the scaling solution.
This difference can be understood analytically from the IR behavior of the propagators and vertices in the integrals.
It is known that with the boundary condition of a divergent ghost dressing function \cite{Zwanziger:2001kw,Lerche:2002ep}, vertex dressing functions behave like $(p^2)^{(-m+n/2)\kappa}$ in the IR \cite{Alkofer:2004it}, where $m/n$ is the number of gluon/ghost legs.
This behavior is induced by diagrams with a bare ghost-gluon vertex in the DSE \cite{Huber:2007kc,Huber:2009wh} and, consequently, there is an IR constant contribution from these diagrams that change the total IR value away from one given by the tree level.
For a decoupling solution, on the other hand, one can show that all diagrams are IR suppressed \cite{Alkofer:2008jy} and thus the tree-level value is not altered.
Numerically, this behavior was already confirmed for scaling \cite{Schleifenbaum:2004id}\footnote{The deviation from one is negative in \cite{Schleifenbaum:2004id} due to a sign error in the kernels.} and decoupling \cite{Fister:2011uw,Huber:2012kd}.
It should be noted that for the scaling solution the IR limit of the ghost-gluon vertex depends on the direction from which one approaches zero momentum.

When comparing the ghost-gluon vertex results to lattice results, it becomes evident that there is a mismatch in the scale.
Most likely, this is not due to comparing to $SU(2)$ lattice data, since the differences between $SU(2)$ and $SU(3)$ are only weak \cite{Sternbeck:2007ug,Cucchieri:2007zm}.
However, the position of the bump is found at lower momenta in all continuum results \cite{Huber:2012kd,Aguilar:2013xqa,Pelaez:2013cpa,Mintz:2017qri,Aguilar:2018csq}.
As can be seen, the scaling solution is the solution with the bump at lowest momenta.
Moving away from the scaling solution, the position of the bump moves to higher momenta.
In the bottom right plot, the full angular dependence for the lowest decoupling solution, corresponding to the black continuous lines in the other plots, is shown together with lattice data for various kinematics.
For a meaningful comparison, the data is plotted over the variable $S_0$, see \eref{eq:S3_vars}.
For this single solution, the agreement with the lattice data is better than for the other solutions, but there is room for improvement.
From all solutions, though, this is the one closest to minimal Landau gauge as employed on the lattice.
This hypothesis is supported by similar observations for the propagators; see Sec.~\ref{sec:props}.

Fig.~\ref{fig:ghg_diags} shows the contributions from the individual diagrams.
Both of them exhibit an angular dependence.
The non-Abelian diagram is clearly larger than the Abelian one.

\subsection{Three-gluon vertex}

\begin{figure*}[tb]
 \includegraphics[width=0.48\textwidth]{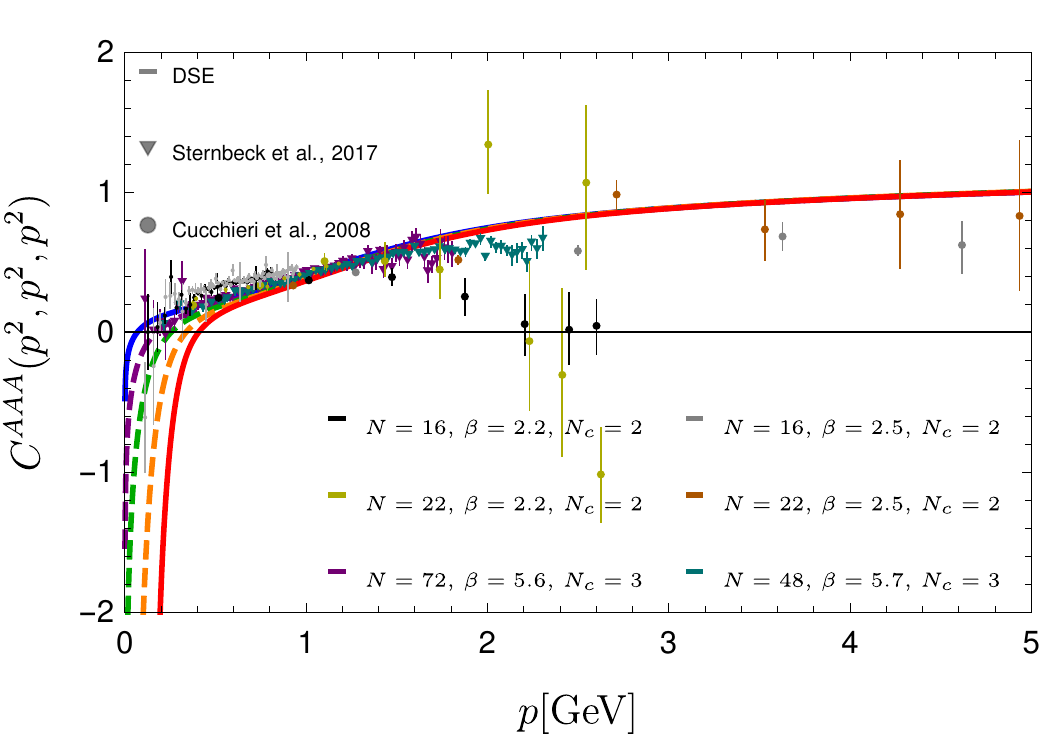}\hfill
 \includegraphics[width=0.48\textwidth]{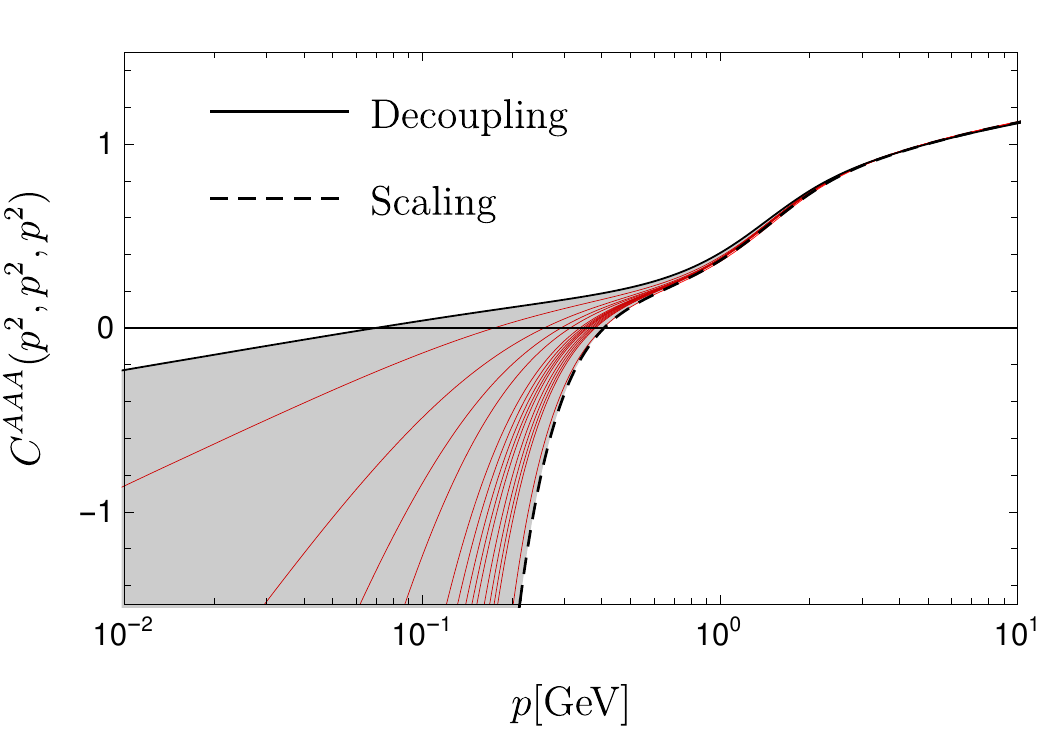}
 \caption{Left: Solutions for the three-gluon vertex dressing function at the symmetric point in comparison to lattice data \cite{Cucchieri:2008qm,Sternbeck:2017ntv}).
 Right: Several solutions over a logarithmic momentum scale.
 Data is renormalized to $1$ at $5\,\text{GeV}$.}
 \label{fig:tg}
\end{figure*}

\begin{figure}[tb]
 \includegraphics[width=0.48\textwidth]{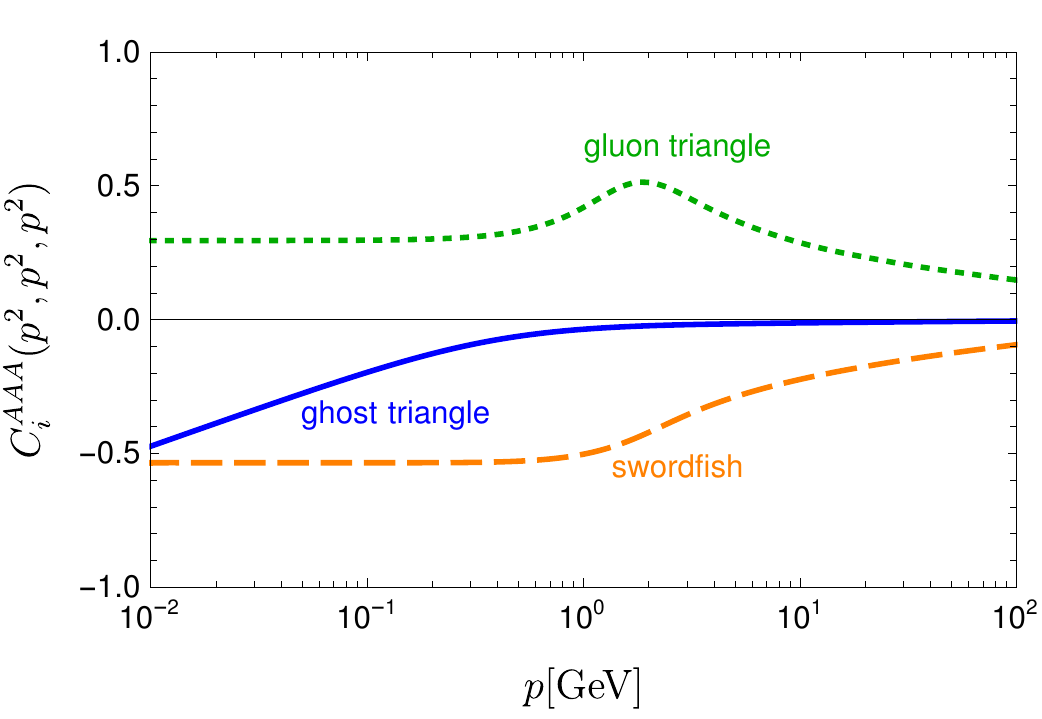}
 \caption{Contributions of individual diagrams in the three-gluon vertex equation for the lowest decoupling solution.
 Shown are results for the symmetric point, where all swordfish diagrams are identical and only the contribution of one is shown.}
 \label{fig:tg_diags}
\end{figure}

The three-gluon vertex shows a remarkable small angular dependence as illustrated previously in \fref{fig:tg_angle_truncs} where the dependence on the momentum scale $S_0$ is shown and the small band represents the variation of the dressing function with the variables $a$ and $s$.
As a consequence, it is sufficient to compare with other results for one momentum configuration only.
For the sake of comparison, all results were renormalized to one at $5\,\text{GeV}$.
Fig.~\ref{fig:tg} shows the family of solutions and a comparison to various lattice results for the symmetric configuration.
All solutions cross zero and diverge; logarithmically for decoupling solutions and like $(p^2)^{-3\kappa}$ for the scaling solution.
The position of the zero crossing correlates with the position of the maximum of the gluon propagator as argued for in Ref.~\cite{Aguilar:2013vaa}.
Extrapolating the position of the zero crossing for the case of a gluon propagator with no maximum is compatible with the vanishing of the zero crossing.
Again we find that the decoupling solution close to the Higgs branch agrees best with lattice results.
It should be noted that the three-gluon vertex has a non-trivial IR structure in the scaling case \cite{Alkofer:2008dt}, which, however, is not visible here due to the employed projection onto the tree-level tensor.

The contributions of individual diagrams are shown in \fref{fig:tg_diags}.
The IR divergence is created only by the ghost triangle, while the gluonic diagrams are IR finite but with opposite signs.
Due to the logarithmic running, the amount of cancellations between the gluonic diagrams is not immediately obvious.
However, in three dimensions, it was found that they almost cancel completely, making the ghost triangle the dominant nonperturbative contribution \cite{Huber:2016tvc}.
The figure shows the symmetric point where all swordfish diagrams are identical.
It should be noted that the individual swordfish diagrams do have an angular dependence which, however, is canceled in the symmetrization.

\subsection{Four-gluon vertex}

\begin{figure*}
 \begin{center}
  \includegraphics[width=0.48\textwidth]{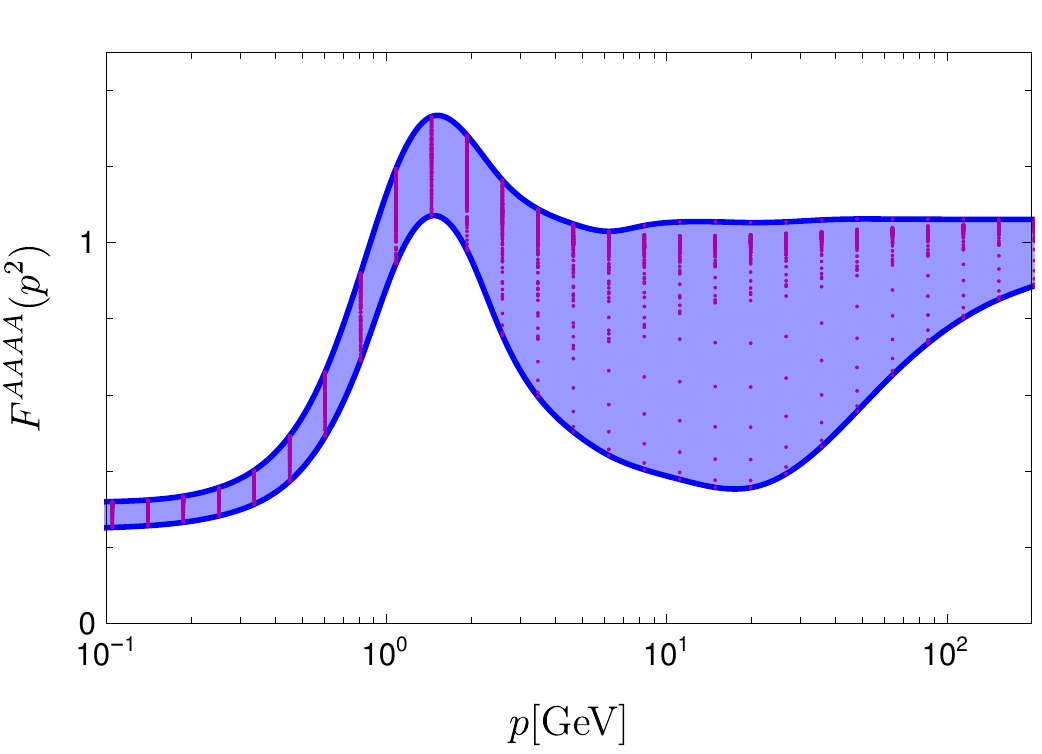}\hfill
  \includegraphics[width=0.48\textwidth]{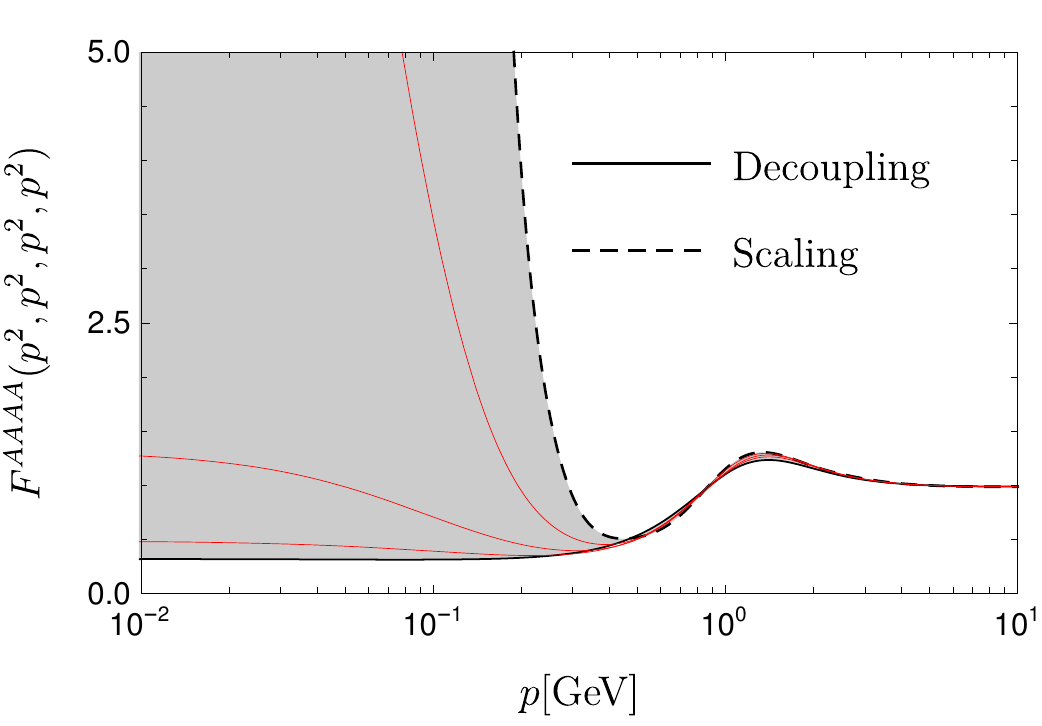}
 \end{center}
 \caption{Four-gluon vertex dressing function from its DSE.
 Left: Dressing function for the decoupling solution corresponding to the black line in the right plot as a function of $p=\sqrt{S_0}$ with the band indicating the angular dependence.
 Calculated points are shown as dots.
 Right: Family of solutions at the symmetric point renormalized to $1$ at $5\,\text{GeV}$.}
 \label{fig:fg}
\end{figure*}

\begin{figure}[tb]
 \includegraphics[width=0.48\textwidth]{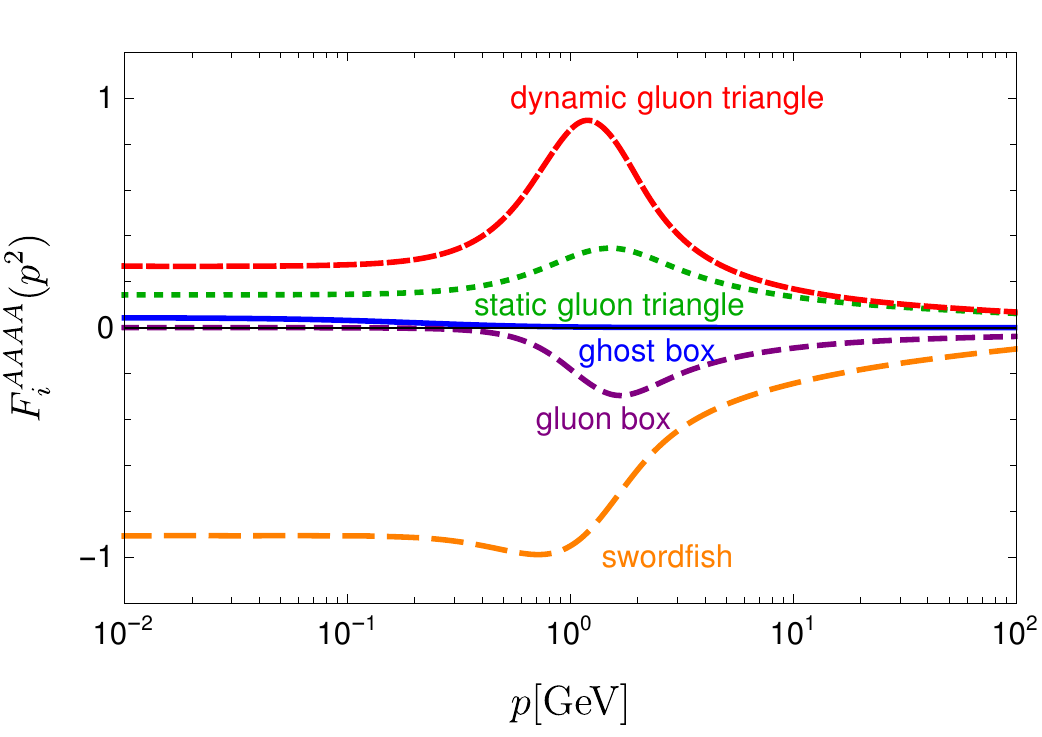}
 \caption{Contributions of individual diagrams in the four-gluon vertex equation for the lowest decoupling solution.
 Shown are results for the symmetric point, where all diagrams of the same class are identical.}
 \label{fig:fg_diags}
\end{figure}

The four-gluon vertex is calculated with the singlet-doublet approximation for the kinematics and only one dressing function as discussed in Sec.~\ref{sec:corrFuncs}.
The result is shown in \fref{fig:fg} where one can see that the angular dependence is stronger than for the three-gluon vertex.
However, the breadth of the dressing function around $10\,\text{GeV}$ comes from a few outlying points whereas the majority clusters in a much smaller band.
This is indicated by plotting besides the band also individual points in \fref{fig:fg}.
The qualitative behavior is in agreement with previous calculations \cite{Kellermann:2008iw,Binosi:2014kka,Cyrol:2014kca,Cyrol:2016tym,Huber:2019wxx}.
In the IR, the vertex diverges like $(p^2)^{-4\kappa}$ for the scaling solution.
For the decoupling solution, no logarithmic IR divergence is found for the tree-level dressing function as found previously \cite{Binosi:2014kka,Cyrol:2014kca}.
It should be noted that the individual ghost box diagrams are logarithmically divergent, see \fref{fig:fg_diags}, but the symmetrization of the equation cancels this effect.
A small number of other tensors has already been explored \cite{Binosi:2014kka,Cyrol:2014kca}.
Interestingly, in that case logarithmic divergences were found \cite{Binosi:2014kka,Cyrol:2014kca}, but the corresponding dressing functions are smaller than the tree-level dressing function.
There are $5\times41$ tensors in total, and the question of their quantitative importance remains open.

A remarkable feature of the four-gluon vertex is that the zero momentum limit depends on the kinematic configuration for which it is approached, as can be seen by the finite width of the band in the IR in the left plot of \fref{fig:fg}.
This is similar to the ghost-gluon vertex for the scaling solution.
For the three-gluon vertex, any such dependence is at best too small to allow any conclusions.
Such an IR irregularity is interesting insofar, as it was argued in Ref.~\cite{Cyrol:2016tym} that the creation of the gluon mass gap for decoupling solutions requires vertices that show an irregular IR behavior.
One should be careful, though, before drawing any final conclusions, because currently the four-gluon vertex is calculated with the most severe approximations of all quantities.

Individual contributions to the four-gluon vertex are shown in \fref{fig:fg_diags}.
Similar to the three-gluon vertex, the ghost diagrams are very small in the midmomentum regime.
Due to the cancellation from the symmetrization mentioned above, the ghost diagrams are thus not relevant for the four-gluon vertex.
The gluonic diagrams have opposite signs which makes the overall deviation from the tree-level small.

\subsection{Self-tests}
\label{sec:selftests}

Assessing the reliability of results from functional equations is one of the main challenges of this approach.
Beyond some trivial self-consistency checks, as, for example, the correct asymptotic IR and UV behaviors, a standard test is to compare to other results, typically from lattice calculations.
However, this has two drawbacks:
First, one relies on the availability of other results.
Second, even more importantly, such comparisons should not necessarily be interpreted too literally on a quantitative level.
For instance, the renormalization schemes are different for lattice and functional methods.
Actually, also within the latter category, care has to be taken when comparing results from different calculations.
While for DSE or $n$PI calculations typically a MOM scheme is employed, the renormalization scheme is implicit for FRG calculations.
Furthermore, the correspondence of members of the family of solutions between different methods is not clear due to the different ways they are realized for different methods, e.g., by changing the ghost dressing function at zero momentum, by choosing a UV mass parameter, or by a selection algorithm of Gribov copies.

In this situation, any way of checking the results without requiring external input is welcome.
Here, two possibilities are explored.
The first relies on the different vertex couplings and the requirement that they should agree perturbatively.
The second self-test checks the irrelevance of an unphysical parameter.
In addition, the application of the present results to the calculation of a gauge-invariant object is discussed.

\subsubsection{Couplings}
\label{sec:couplings}

\begin{figure*}
 \begin{center}
  \includegraphics[width=0.48\textwidth]{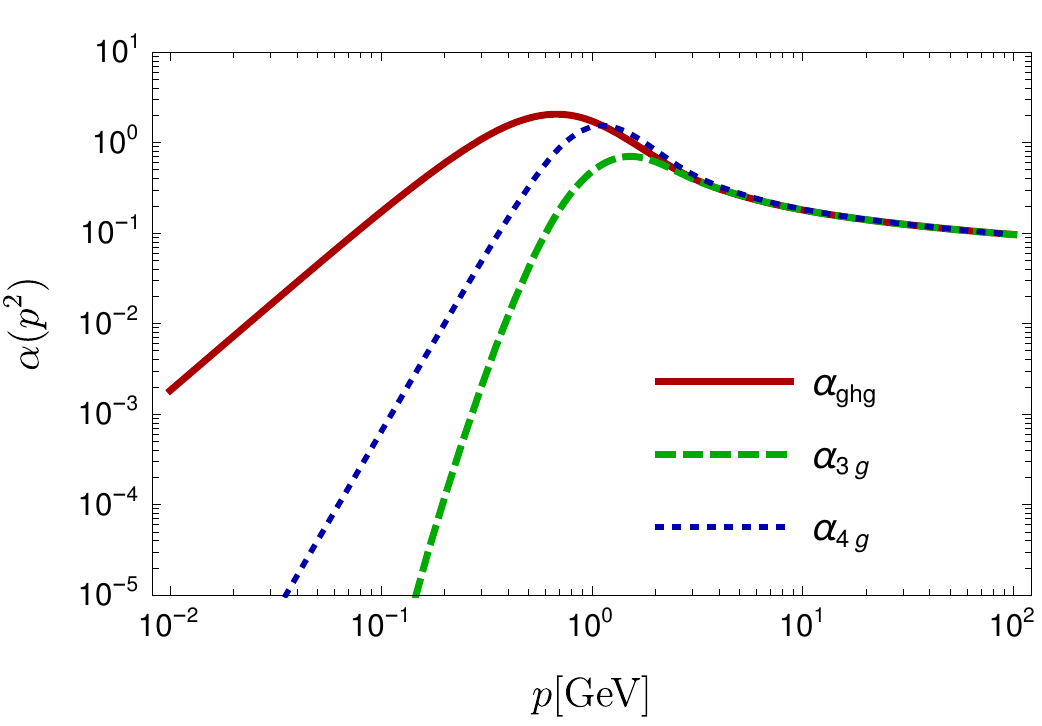}\hfill
  \includegraphics[width=0.48\textwidth]{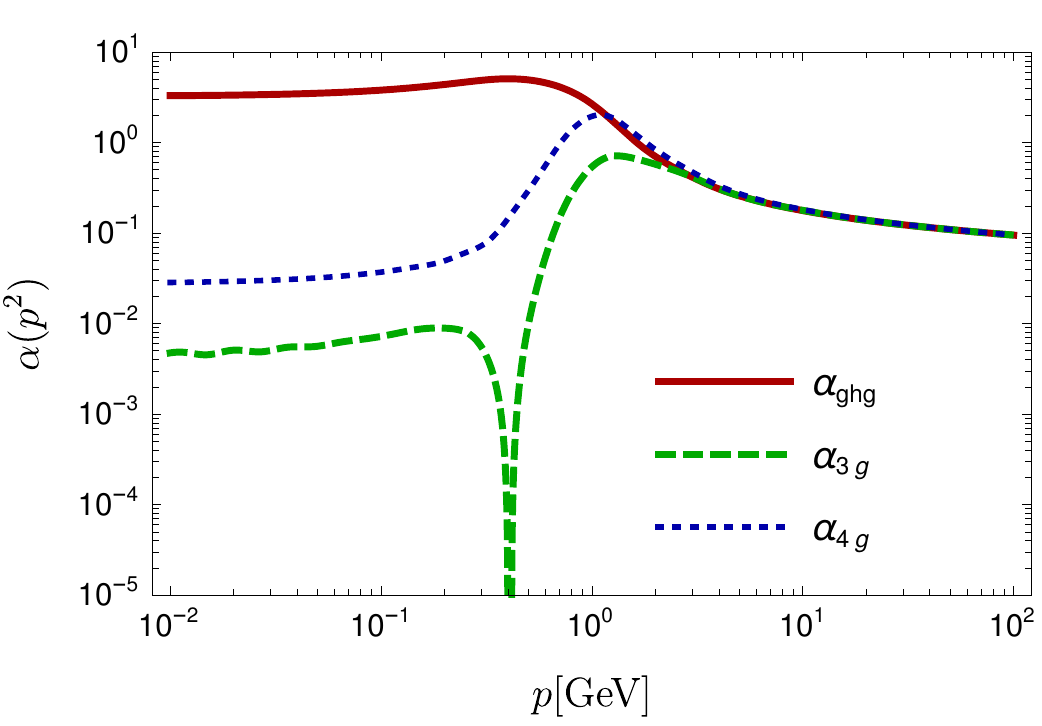}
 \end{center}
 \caption{The couplings from the ghost-gluon, three-gluon and four-gluon vertices for the decoupling (left) and scaling (right) solutions corresponding to the black continuous and dashed lines in Figs.~\ref{fig:gl}--\ref{fig:fg}, respectively.}
 \label{fig:couplings}
\end{figure*}

As discussed in Sec.~\ref{sec:renormalization}, the couplings extracted from the different vertices should agree in the perturbative regime.
This turns out to be a nontrivial requirement and can only be achieved when dynamically including all required quantities.
Such a computation was performed in Ref.~\cite{Cyrol:2016tym} and the couplings were found to agree down to a few GeV.
There, one can also find a comparison between different previous calculations which calculated different subsets of the system considered here.
The comparison reveals that such isolated calculations are insufficient as deviations between the couplings were found.
In this work, for the first time, all these quantities are calculated from their equations of motion jointly and self-consistently which leads to a good agreement between the various couplings.

The couplings from the ghost-gluon, three-gluon and four-gluon vertices, defined in \eref{eq:couplings}, are shown in \fref{fig:couplings}.
They show good agreement down to $3$~GeV.
The four-gluon vertex coupling deviates earlier from the ghost-gluon vertex coupling than that of the three-gluon vertex.
In Sec.~\ref{sec:comparison}, the couplings are also compared to results from the FRG.
It has to be stressed that the employed renormalization conditions for the vertex, see Sec.~\ref{sec:renormalization}, are not sufficient to obtain this degree of agreement, because they only force the couplings to agree at one specific point and do not fix their running.
Thus, the good agreement over several orders of magnitude comes from the dynamics of the equations.

The couplings have two convenient properties:
They are independent of the renormalization point and their runnings relate directly to the scale of Yang-Mills theory $\Lambda_\text{YM}$.
The former property makes the couplings a useful quantity for comparisons between different calculations.
The latter property could be used to carry over the scale from lattice simulations.
However, in the perturbative regime, for which dedicated lattice simulations exist to calculate the scale parameter of Yang-Mills theory $\Lambda_\text{YM}$ \cite{Boucaud:2008gn,Sternbeck:2012qs}, the slow logarithmic running of the coupling does not allow a precise determination of the scale for our purposes.
Thus, the scale fixing is done in the nonperturbative regime where it is most convenient to fix the scale via the position of the bump of the gluon dressing function for which $p_0=0.97\,\text{GeV}$ was chosen.
Even though the different solutions differ already in this regime, this procedure is accurate enough as can be seen by the good agreement of the different solutions above $1\,\text{GeV}$.

Having fixed the scale like this, another nontrivial check is to compare the couplings in the perturbative regime.
For this, the \textit{MiniMOM} coupling defined in \eref{eq:couplingMM} is most convenient as we can compare directly to high momentum results of Ref.~\cite{Sternbeck:2012qs}.
As reference points, we take the values $p^2=178\,\text{GeV}^2$ and $1785\,\text{GeV}^2$, which correspond to $1,000$ and $10,000$ in units of the Sommer scale $r_0$.
The results for the coupling from the lattice calculation are roughly $0.142$ and $0.107$, whereas the DSE calculation yields $0.144$ and $0.107$, respectively.
This good agreement further supports the quantitative reliability of the results and it should be stressed that there is no parameter to tune these values.

The use of a hard UV cutoff entails that gauge covariance is broken and the STIs need to be modified.
As a consequence, the momentum independent STIs given in \eref{eq:STIs} are not fulfilled exactly.
However, observing the good agreement between different couplings is a manifestation of the restoration of gauge covariance.
This can also be tested by comparing the values for the vertex renormalization constants calculated in their renormalization as discussed in Sec.~\ref{sec:renormalization} with the values obtained via the STIs given in \eref{eq:STIs} from the propagator renormalization constants.
The differences are approximately $1\,\%$ for $Z_1$ and $0.2\,\%$ for $Z_4$ which is an improvement over previous calculations where these differences were at the order of a few percent.

\subsubsection{Quadratic divergences}
\label{sec:glPAt0}

Another consequence of the hard UV cutoff is the appearance of quadratic divergences; see the discussion in Sec.~\ref{sec:renormalization}.
To deal with them, a second renormalization condition $D(x_m)$ was introduced which is given by the value of gluon propagator at a given IR scale $x_m$.
One could expect that this is similar to how the problem of quadratic divergences is handled in the flow equation of the gluon propagator where the value of the bare gluon mass is fine-tuned and leads to different solutions in the IR \cite{Cyrol:2016tym}.
Indeed, a first test for three-dimensional Yang-Mills theory using a simple truncation showed that the gluon mass renormalization can be used to obtain a family of solutions \cite{Huber:2016hns}.

\begin{figure*}
 \begin{center}
  \includegraphics[width=0.48\textwidth]{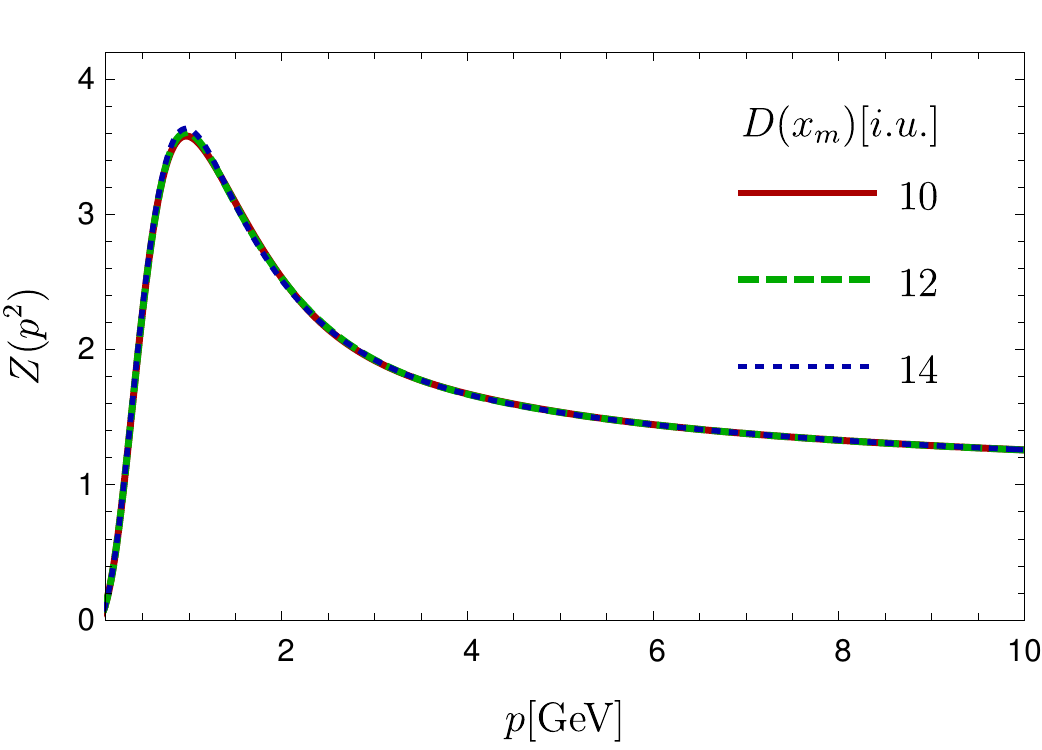}\hfill
  \includegraphics[width=0.48\textwidth]{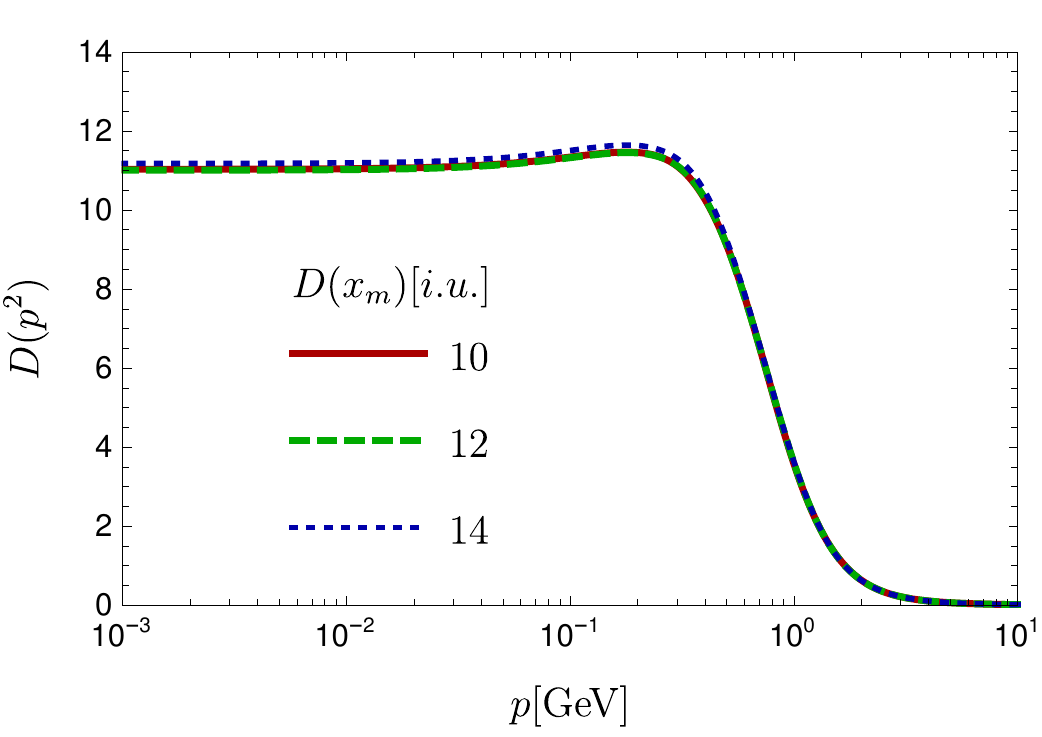}
 \end{center}
 \caption{Left/Right: Gluon dressing function/propagator for different renormalization conditions $D(x_m)$.}
 \label{fig:glPAt0}
\end{figure*}

However, the more sophisticated truncation of the present work sheds new light on this method to renormalize the gluon propagator DSE.
As it turns out, the results are basically independent of the new renormalization parameter $D(x_m)$.
Hence, it does not seem to be related to different decoupling solutions.
In the calculations, the value of $D(x_m)$ is given in internal units.
The conversion factor for physical units differs for calculations using different $D(x_m)$, because it is a dimensionful quantity.
However, in physical units, the solutions are equal, as can be seen by comparing any of the running couplings, which are RG invariant.
The propagators and vertices themselves are merely rescaled as required by multiplicative renormalizability.
Taking this finite renormalization into account, quantities can be compared for different values of $D(x_m)$.
For illustration, the gluon propagator and its dressing function are shown in \fref{fig:glPAt0} for the three values $D(x_m)=10$, $12$, and $14$ (in internal units).
The plots show the data in physical units renormalized at $6\,\text{GeV}$.
The small differences below $2\,\text{GeV}$ could be either numeric or truncation artifacts.
Since quadratic divergences are a manifestation of broken gauge invariance, the fact that they can be unambiguously subtracted seems naturally related to the good agreement of the different couplings which is a consequence of gauge invariance itself.
In turn, the small remaining differences might reflect the small deviations of the couplings discussed in Sec.~\ref{sec:couplings}.
However, quantitatively they are so small that they do not matter for physical applications like the calculation of glueballs discussed in Sec.~\ref{sec:glueballs}.
The differences in the other dressing functions are even smaller than for the gluon propagator.

The absence of a parameter related to the removal of quadratic divergences is a novel feature encountered for the first time within a DSE calculation.
The original idea that both the FRG and DSEs can create a family of solutions by varying a mass parameter related to the removal of quadratic divergences might have seemed appealing \cite{Huber:2016hns}, but the fact that these subtraction methods are not equivalent settles the problem of having a second parameter for DSEs, the ghost renormalization condition, that fulfills the same role.
Thus, we are finally in the position to subtract quadratic divergences unambiguously which is required to provide quantitatively reliable results.

\subsubsection{Glueballs}
\label{sec:glueballs}

An ultimate test of the quantitative correctness of the gauge-dependent propagators and vertices obtained here is employing them for the calculation of gauge-invariant quantities.
The obvious candidates for Yang-Mills theory are glueballs, i.e., bound states of gluons.
The simplest case are scalar and pseudoscalar glueballs which can be calculated from Bethe-Salpeter equations as two-gluon bound states \cite{Meyers:2012ka,Sanchis-Alepuz:2015hma,Souza:2019ylx,Huber:2020ngt}.
In Ref.~\cite{Huber:2020ngt}, the kernels of these Bethe-Salpeter equations were constructed from the same truncation of the 3PI effective action as here.
This provides a fully self-consistent setup.
The masses of the ground state and two excited states were then calculated using the propagators and vertices obtained here as input \cite{Huber:2020ngt}.
For the scalar glueball, the ground state mass was calculated as $1.75\pm0.12\,\text{GeV}$ and for the pseudoscalar glueball as $2.44\pm0.17\,\text{GeV}$.
These results agree well with corresponding lattice results of $1.73\pm0.13\,\text{GeV}$ and $2.59\pm0.17\,\text{GeV}$, respectively \cite{Morningstar:1999rf}.
Also the first excited states agree well: The bound state calculation yields $2.43\pm0.2\,\text{GeV}$ and $3.65\pm0.11\,\text{GeV}$ for the scalar and pseudoscalar masses, respectively, while on the lattice $2.67\pm0.31\,\text{GeV}$ and $3.64\pm\,0.24\,\text{GeV}$ are found.
Second excited states in each channel were also calculated.

The calculation of gauge-invariant quantities conveniently also provides the means to test if different solutions from the family of solutions are indeed physically equivalent.
To this end, the calculation of the glueball masses was repeated with various members of the family of solutions for the propagators and vertices. 
As no difference in the bound state masses was found, this provides another piece of evidence that different solutions just represent different nonperturbative completions of the perturbative Landau gauge.
Note that if the glueball masses varied with the input, one could not decide whether the employed truncation is insufficient or if the solutions really differ physically.

To test the impact of the self-consistency of the input, also mixed solutions were used, viz., propagators and vertices were taken from different solutions.
It should be noted that care has to be taken with regard to the employed units when mixing solutions like this.
All calculations are done in internal units and the conversion factor to physical units can be different for each calculation.
If this is not taken into account properly, the different (internal) units of the propagators and vertices destroy the consistency automatically.
For such mixed input, the bound state masses maintain the correct hierarchy but deviate when the input solutions are taken from members of the family of solutions which are close to each other.
If the mixed solutions are too far apart, though, the hierarchy of the masses gets lost and no sensible solutions can be obtained anymore.

As a final test, the solution from the system with the ghost-gluon vertex DSE instead of its EOM, discussed in Appendix~\ref{sec:ghg_eqs}, was used.
The glueball masses were equal within the given errors.

\begin{figure}[tb]
 \includegraphics[width=0.48\textwidth]{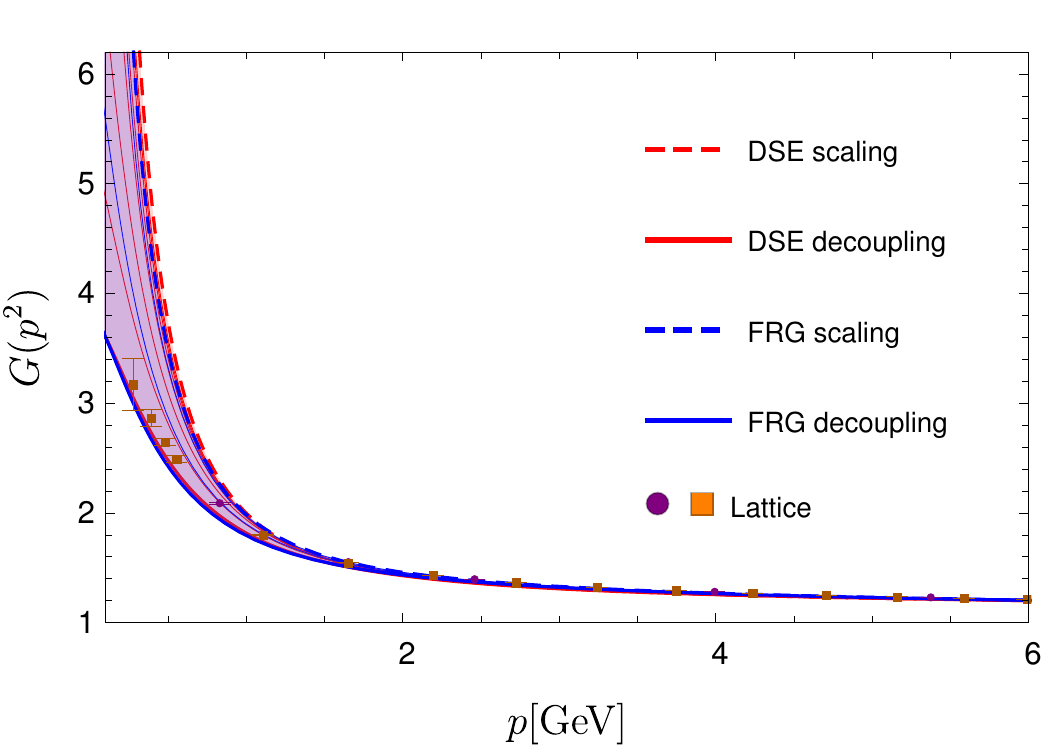}
 \caption{Ghost dressing function $G(p^2)$ in comparison to FRG \cite{Cyrol:2016tym} and lattice results \cite{Sternbeck:2006rd}.}
 \label{fig:gh_comp}
\end{figure}

\begin{figure*}[tb]
 \includegraphics[width=0.48\textwidth]{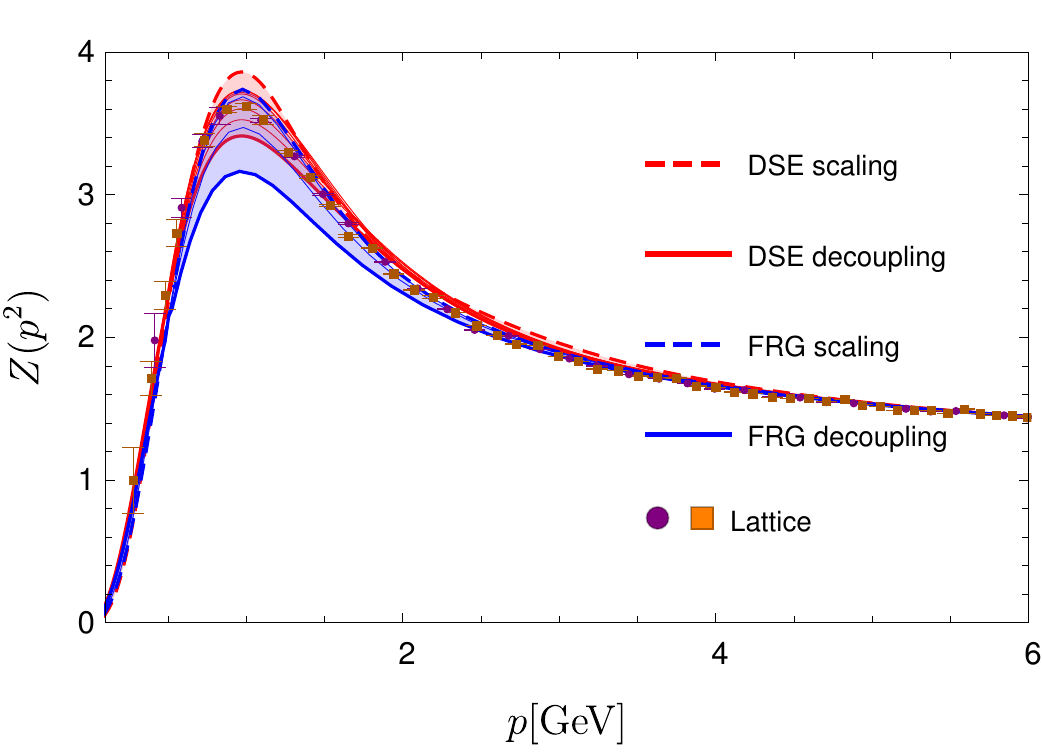}\hfill
 \includegraphics[width=0.48\textwidth]{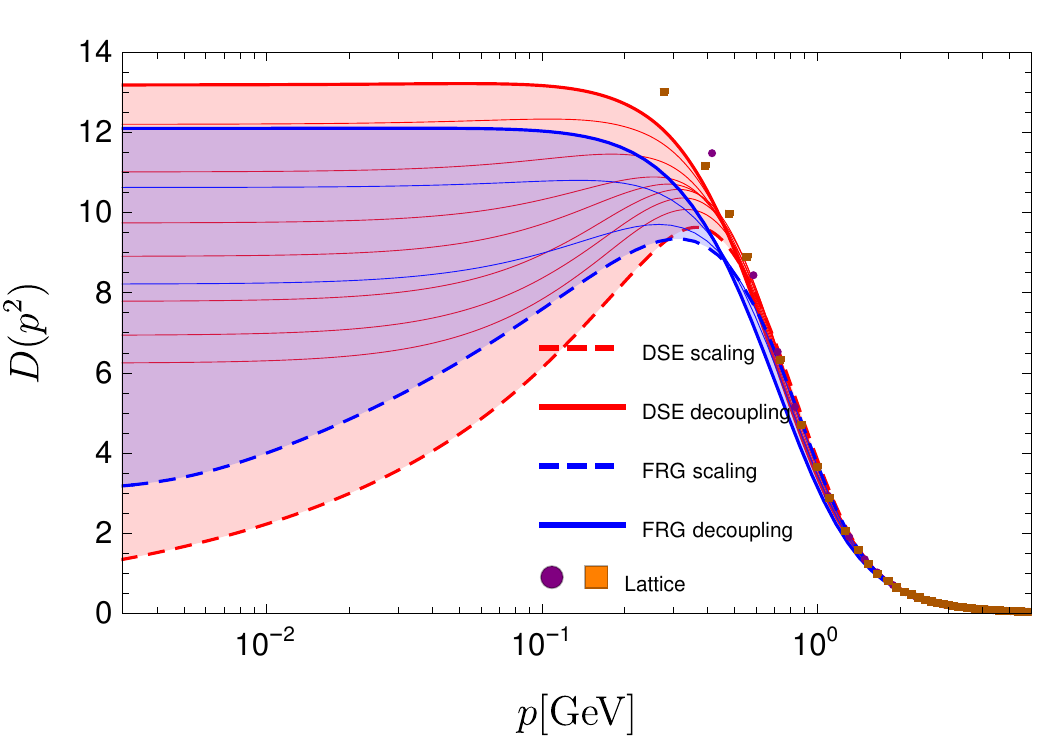}
 \caption{Gluon dressing function $Z(p^2)$ (left) and gluon propagator $D(p^2)$ (right) in comparison to FRG \cite{Cyrol:2016tym} and lattice results \cite{Sternbeck:2006rd}.}
 \label{fig:gl_comp}
\end{figure*}

\section{Comparison with other results}
\label{sec:comparison}

The results for all correlation functions are compared in this section to results from the lattice and the FRG.
For the former, various sources are used and for the latter results from Ref.~\cite{Cyrol:2016tym}.
In general, good agreement is found.
The purpose here is to identify and highlight the remaining differences.

The system of flow equations is formally truncated in the same way as here by including all primitively divergent correlation functions and neglecting the others.
Due to the different structure of the equations, the truncations are not equal, though.
For example, the two-ghost-two-gluon and the four-ghost vertices appear in the propagator flow equations.
The corresponding diagrams were dropped in Ref.~\cite{Cyrol:2016tym} but investigated for three dimensions \cite{Corell:2018yil}.
The FRG results shown here correspond to the '1D momentum-dependent' approximation of Ref.~\cite{Cyrol:2016tym}, i.e., the three-point functions are calculated only at the symmetric point.
As is clear from the small angular dependence of the three-gluon vertex, this is a good approximation in this case.
For the ghost-gluon vertex it is more severe, but the impact is still small \cite{Cyrol:2016tym}.
The four-gluon vertex is solved for the FRG for two kinematic configurations.

In the plots the DSE and FRG results are distinguished by colors.
The scaling solution is represented by a dashed line and the decoupling solutions by continuous ones.
For the comparison, all results were renormalized such that they agree in the region around $5\,\text{GeV}$.

The propagator results are compared in Figs.~\ref{fig:gh_comp} and \ref{fig:gl_comp}.
Lattice results were taken from Ref.~\cite{Sternbeck:2006rd}.
Similar results can be found, e.g., in Refs.~\cite{Cucchieri:2007md,Cucchieri:2008fc,Sternbeck:2007ug,Bogolubsky:2009dc,Maas:2011se,Oliveira:2012eh,Duarte:2016iko,Boucaud:2018xup}.
From the ghost dressing function and the gluon propagator one can see that for both functional methods the agreement with lattice data is best for a decoupling solution with a small gluon mass gap.
It should be noted that the gluon dressing function is slightly affected by the 1D approximation of the FRG calculation, but for the purpose of the comparison here this effect is small enough.
In general, it is not possible to find a solution for which both propagator dressing functions agree between the FRG and DSE results.
The cleanest comparison can be done for the scaling solution, because there is only one instance.
However, it cannot be expected that there is a clean one-to-one correspondence of single members of the family of solutions, because the similarities of the truncations are only superficial.
For example, the role of the tadpole diagram in the FRG calculation is partially played by the gluon loop, the sunset, and the squint diagrams in the DSE.
Thus, the treatment of the four-gluon vertex affects both systems differently.

\begin{figure}[tb]
 \includegraphics[width=0.48\textwidth]{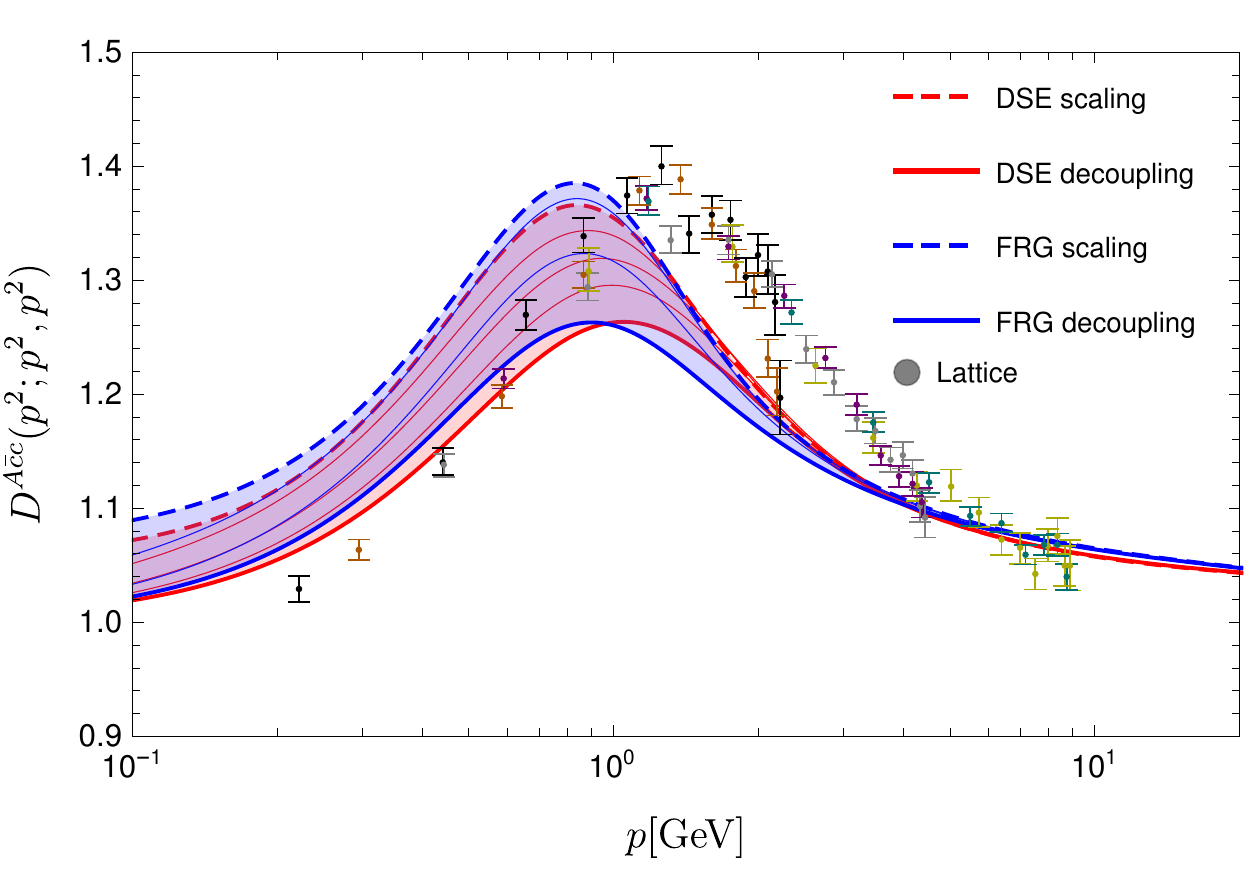}\hfill
 \caption{Ghost-gluon vertex dressing function at the symmetric point in comparison to FRG \cite{Cyrol:2016tym} and $SU(2)$ lattice data \cite{Maas:2019ggf} (details as explained in \fref{fig:ghg}).
 }
 \label{fig:ghg_comp}
\end{figure}

For the ghost-gluon vertex, the comparison is shown in \fref{fig:ghg_comp}.
Lattice results are from Ref.~\cite{Maas:2019ggf}.
Further results can be found, e.g., in Refs.~\cite{Ilgenfritz:2006he,Cucchieri:2008qm}.
Both functional results are very similar, and one should keep in mind that the '1D' approximation is most severe for this quantity.
In particular, the FRG results show a similar position of the maximum as the DSE results which differ from the lattice results.
As mentioned above, this is a generally observed difference between lattice and continuum results.
In addition, the shift of the position of the maximum from lower to higher momenta from the scaling solution through the decoupling solutions is similar for both functional methods.

\begin{figure}[tb]
\includegraphics[width=0.48\textwidth]{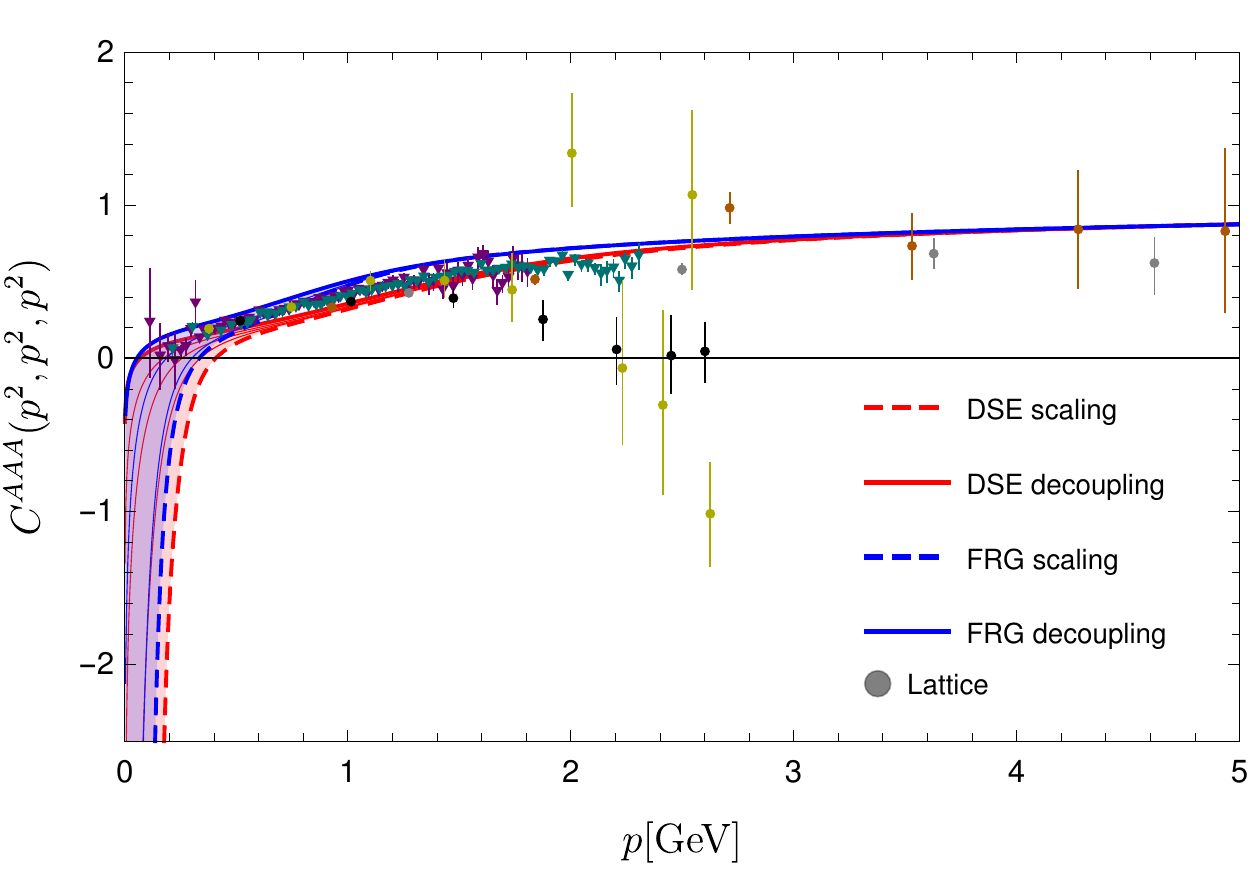}
 \caption{Three-gluon vertex dressing function at the symmetric point in comparison to FRG \cite{Cyrol:2016tym} and lattice data \cite{Cucchieri:2008qm,Sternbeck:2017ntv} (details as explained in \fref{fig:tg}).}
 \label{fig:tg_comp}
\end{figure}

The comparison for the three-gluon vertex is shown in \fref{fig:tg_comp} with lattice results taken from Refs.~\cite{Cucchieri:2008qm,Sternbeck:2017ntv}.
Further lattice results can be found, e.g., in Refs.~\cite{Athenodorou:2016oyh,Boucaud:2017obn}.
In this case, one finds an interesting difference between the DSE and FRG results.
The data is renormalized at $5\,\text{GeV}$.
As can be seen, the results deviate in the region below $2\,\text{GeV}$.
The FRG results have larger dressing functions at lower momenta.
If one tries to match the results to the lattice data by adjusting the normalization, the DSE results show a larger deviation than the FRG results.
These differences are most likely due to the different truncations.
As was discussed in Sec.~\ref{sec:equations}, there are also differences between a two-loop DSE and three-loop truncated 3PI EOM.
They are small, but the two-loop DSE results are also larger at low momenta.

\begin{figure}
 \begin{center}
  \includegraphics[width=0.48\textwidth]{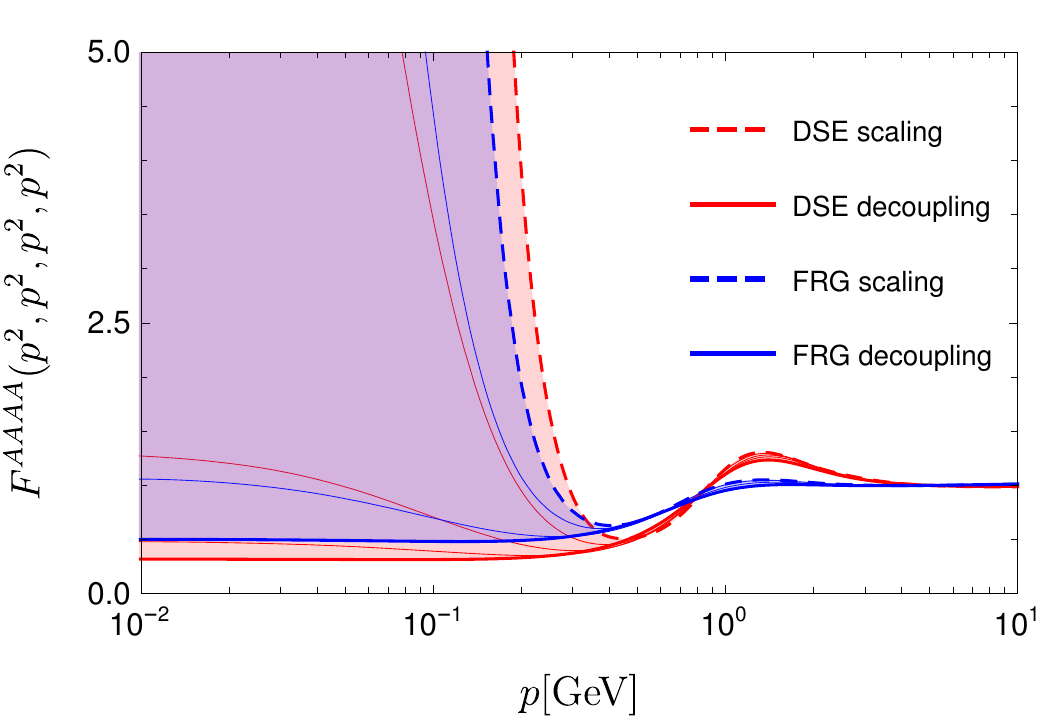}
 \end{center}
 \caption{Four-gluon vertex dressing function at the symmetric point in comparison to FRG results \cite{Cyrol:2016tym}.
 }
 \label{fig:fg_comp}
\end{figure}

Also for the four-gluon vertex deviations are observed, especially below $3\,\text{GeV}$; see \fref{fig:fg_comp}.
It can be seen in the present (see \fref{fig:fg}) and previous DSE calculations \cite{Cyrol:2014kca} that there is a large angular dependence compared to the three-gluon vertex.
Not taking this into account most likely affects the FRG calculation and also here only three out of six kinematic variables were taken into account.
This may explain at least a part of the quantitative differences.

\begin{figure*}[tb]
 \includegraphics[width=0.48\textwidth]{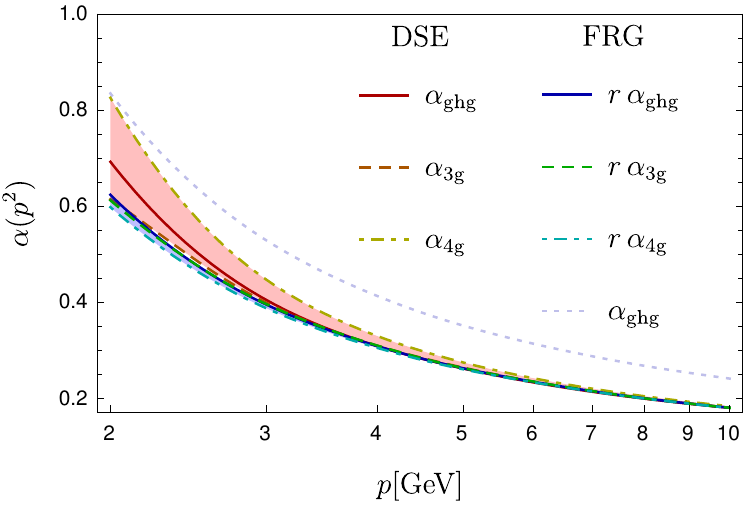}\hfill
 \includegraphics[width=0.48\textwidth]{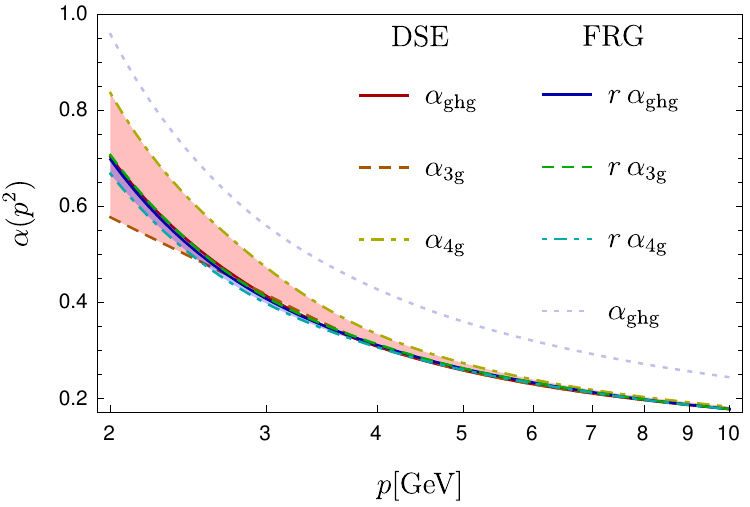}\hfill
 \caption{The vertex couplings from \eref{eq:couplings} for the DSE and FRG solutions \cite{Cyrol:2016tym}.
 The FRG couplings were rescaled by $r=0.75$ for the left plot (decoupling) and by $r=0.73$ for the right plot (scaling).
 }
 \label{fig:coup_comp}
\end{figure*}

As final quantities, the vertex couplings given in \eref{eq:couplings} are compared in \fref{fig:coup_comp}.
The running of the couplings agree quite well, but it is found that they deviate by an overall factor.
This is taken into account by rescaling the FRG couplings such that they agree with the DSE results at $10\,\text{GeV}$.
For reference, the ghost-gluon vertex coupling is also shown without rescaling.
Whether this difference is entirely due to differences in the renormalization schemes or due to other sources still has to be investigated.

The largest difference is seen in the four-gluon vertex coupling.
This is most likely a direct effect of the different approximations.
It should be noted that the four-gluon vertex is here calculated from its DSE and not an EOM, since at least the 4PI effective action is required for an EOM.
It is also confirmed that the effect of kinematic approximations cannot be solely responsible for this deviation, as a calculation with the vertex restricted to the symmetric point yields a very similar coupling.
The ghost-gluon and three-gluon vertex couplings agree very well down to $4\,\text{GeV}$, for the FRG even further to approximately $2.5\,\text{GeV}$.

In general, the agreement between the vertex couplings is a desirable feature of any calculation of correlation functions which can only be realized in self-consistent truncations.
In many previous studies, this is not found and the runnings of the couplings disagree already in the perturbative regime.
To what extent the agreement needs to be realized is not fully clear yet and might depend on the specific problem.
For example, for the calculation of glueball masses, the present degree of agreement was found to be sufficient \cite{Huber:2020ngt}.

\section{Discussion}
\label{sec:discussion}

The present truncation includes all primitively divergent correlation functions of Yang-Mills theory.
While the propagators and the ghost-gluon vertex are calculated entirely, for the three- and four-gluon vertices approximations of their full forms were made.
The approximation for the three-gluon vertex consists of taking into account only one out of four transverse dressing functions.
It was found previously that the basis can be chosen such that the dressing function of the tree-level tensor dominates \cite{Eichmann:2014xya}.
Indirectly, the results here corroborate that, but it remains to be seen what effect additional, albeit small contributions of the three-gluon vertex can have.
In particular, the gluon propagator DSE is quite sensitive to changes in the midmomentum regime, simply because the gluon propagator is the inverse of a sum of diagrams.
Small changes around the peak of the gluon dressing function can thus have a large effect, which is one of the reasons why the subtraction of quadratic divergences needs to be done with high enough precision.

For the four-gluon vertex, the same approximation was made with regard to the tensor basis.
In this case, no calculation using the full tensor basis exists.
Only projections onto some tensors were tested and found to be smaller than the dressing function of the tree-level tensor \cite{Binosi:2014kka,Cyrol:2014kca}.
The four-gluon vertex enters via the sunset diagram in the gluon propagator DSE.
This softens its possible effect on the gluon propagator, because this diagram is subleading compared to the gluon loop diagram and the squint diagram through which the three-gluon vertex enters \cite{Huber:2016tvc}.
However, as discussed in Sec.~\ref{sec:eoms}, the presence of the dressed four-gluon vertex has within the present truncation a stabilizing effect.
A second approximation for the four-gluon vertex is made with respect to its kinematic dependence.
Instead of six kinematic variables, only three are taken into account to reduce the computational costs.
A qualitative comparison with the results of Ref.~\cite{Cyrol:2014kca}, where the full kinematic dependence was considered, shows a similarly large angular dependence.
For a detailed estimation of the induced error, though, a dedicated calculation with all six kinematic variables is required.

To assess the truncation error, several tests of the results were performed.
One crucial test is the good agreement of the vertex coupling down to a few GeV.
This was also found in FRG calculations \cite{Cyrol:2016tym}, for which the agreement is even better than observed here.
The reason is most likely that here the vertices are not treated on the same footing: The three-point functions are calculated from their EOMs while for the four-gluon vertex its DSE is used.
Thus, the agreement between the ghost-gluon and three-gluon vertex couplings is better than that of the four-gluon vertex coupling with either of them.
However, the difference is small enough to not affect the calculation of glueballs and the handling of quadratic divergences.
It is an open question in what cases the found agreement is sufficient and when an improvement will be required.

A second test concerns the problem of quadratic divergences.
The treatment of these divergences in previous calculations of the gluon propagator DSE should be considered part of the truncation.
Here, on the other hand, the solution was shown to be independent of the parameter introduced by the subtraction of the divergences.
Strictly speaking, it cannot be excluded that the disappearance of this dependence is a lucky coincidence.
However, together with the other findings, it seems likely that the problem is solved.
Closely related to this is the good agreement of the vertex couplings mentioned above, as both issues are related to the breaking of gauge invariance.
A successful solution to both problems hints at an effective restoration of it.
However, in both cases, there is still room for improvement as illustrated by the remaining tiny dependence on the subtraction parameter, illustrated in \fref{fig:glPAt0}, and the not yet perfect agreement of the vertex couplings down to scales of $2-3\,\text{GeV}$.

It has to be stressed that, given the good quantitative agreement with lattice results, the errors have reached a new qualitative level compared to previous results from equations of motion.
As a consequence, the application of these results to other problems is promising.
The calculation of glueballs discussed in Sec.~\ref{sec:glueballs} is one example where this was successfully realized.

An important issue for all functional calculations is the stability of an employed truncation.
Naturally, one cannot exclude the possibility that two or more individual effects somehow cancel each other.
Hence, adding only one of them would lead to wrong conclusions.
An example of such an effect was observed in three-dimensional Yang-Mills theory where it was found that the effect of including a certain correlation function into a system of flow equations was counteracted by the inclusion of another one \cite{Corell:2018yil}.
For the truncation employed here, the situation with regard to stability of the truncation can be summarized as follows.
The impact of neglected higher correlation functions was partially tested already in the past.
This concerns diagrams with the two-ghost-two-gluon and four-ghost vertices \cite{Huber:2017txg}, while correlation functions beyond four-point functions are currently untested.
For the three-gluon vertex, it was shown that a one-loop truncation of its DSE leads to a qualitative defect and thus this truncation can be considered insufficient.
On the other hand, the EOM from the three-loop truncated 3PI effective action and a DSE including most two-loop diagrams yield very similar results, which are in addition close to lattice and FRG results.
This indicates that the three-gluon vertex is described quite well, but some small differences remain which could be corrected by extensions of the present setup.
The natural candidate would be the EOM from a 4PI effective action to include a dressed four-gluon vertex.
For the ghost-gluon vertex, the situation is different.
Using EOM or the DSE yields different results, but at the same time the other correlation functions remain basically unaffected; see Appendix~\ref{sec:ghg_eqs}.
This could reflect a general stability of the system of equations.

It should be stressed that this is by no means a trivial observation, as correlation functions not only react quantitatively to changes in other ones but often the convergence of the system itself is jeopardized by shortcomings in the truncation.
Two concrete examples are the gluon propagator and the three-gluon vertex.
It is found that the dressing function of the former cannot have a much larger maximum, as then the DSE does not converge anymore.
In a sense, the gluon propagator solution exists close the border of possible solutions.
For the three-gluon vertex, a balance between the gluon triangle and the other gluonic diagrams is required, as a too strong gluon triangle, as caused by a too large gluon propagator, prevents the equation from converging.

\section{Conclusions}
\label{sec:conclusions}

The parameter free calculation of the correlation functions of QCD is a necessary step for functional equations to make the transition from providing effective descriptions based on models fixed by phenomenology to establishing them as a first principles method.
Solving a given set of equations is one challenge in this regard.
Proving that the results are indeed sufficiently close to the correct solution is a second.
In this work, all primitively divergent correlation functions of Yang-Mills theory were solved for the first time simultaneously from a coupled system of equations of motion.
To address the second challenge, several tests were performed that show the qualitative and quantitative improvements compared to previous calculations.

In general, good quantitative agreement with results from other methods, namely lattice simulations and the FRG, is found which provides additional reassurance about the reliability of the results.
At the same time, with this level of precision one can now focus on the remaining differences.
Albeit being small, they might play an important role in the further development of truncations.
In particular, one should learn which deviations are relevant and which are not.

Several extensions of the this work present itself.
Naturally, it would be interesting to bring calculations with other gauges to the same level of truncation.
In fact, in some cases, like the maximally Abelian gauge, this seems like a requirement for numeric calculations anyway, because no simpler working truncation was found yet; see Refs.~\cite{Huber:2009wh,Huber:2018ned,Alkofer:2011di} for details.
In other cases, though, solutions of simple truncations exist.
An example are linear covariant gauges \cite{Aguilar:2015nqa,Huber:2015ria}, for which, however, the treatment of the longitudinal sector will be an additional challenge.
Another example is the Coulomb gauge, for which the Hamiltonian approach provides an alternative functional approach that was already successfully employed in the past, e.g., \cite{Epple:2006hv,Campagnari:2010wc,Campagnari:2011bk,Huber:2014isa,Vastag:2015qjd,Campagnari:2016wlt,Campagnari:2018flz}.

For Yang-Mills theory in the Landau gauge, it would be interesting to consider the 4PI effective action and its equations of motion to put all correlation functions on the same footing.
This should then further improve the agreement of the four-gluon vertex coupling  with the other couplings.
Physically, the interesting extension is the inclusion of quarks.
The quark sector was already calculated with the 3PI effective action, but the Yang-Mills propagators were used as fixed input \cite{Williams:2015cvx}.
From the delicate balancing of the equations in the Yang-Mills sector found here and from FRG calculations \cite{Cyrol:2017ewj}, it can by no means be expected to be trivial to include quarks.
Nevertheless, such calculations will be necessary to describe full QCD in a self-contained way with this method.

A successful inclusion of fermions would also pave the way for addressing additional interesting questions with functional methods in a qualitatively new way.
For example, technicolor scenarios could be probed or the effects of temperature and density in QCD with a largely reduced or even eliminated need for modeling.
Such questions are difficult to answer with any method, but the recent progress with functional methods leads to new perspectives.

\section*{Acknowledgments}

I would like to thank Anton K. Cyrol, Jan M. Pawlowski, Mario Mitter, H\`elios Sanchis Alepuz, and Richard Williams for useful discussions.
HPC Clusters at the Universities of Giessen and Graz were used for the numerical computations.
Support by the FWF (Austrian Science Fund) under Contract No. P27380-N27, the BMBF under Contract No. 05P18RGFP1, and the DFG (German Research Foundation) under Contract No. Fi970/11-1 is gratefully acknowledged.
Feynman diagrams were created with Jaxodraw \cite{Binosi:2003yf}.

\appendix

\section{The ghost-gluon vertex from different equations}
\label{sec:ghg_eqs}

\begin{figure*}[tb]
 \includegraphics[width=0.48\textwidth]{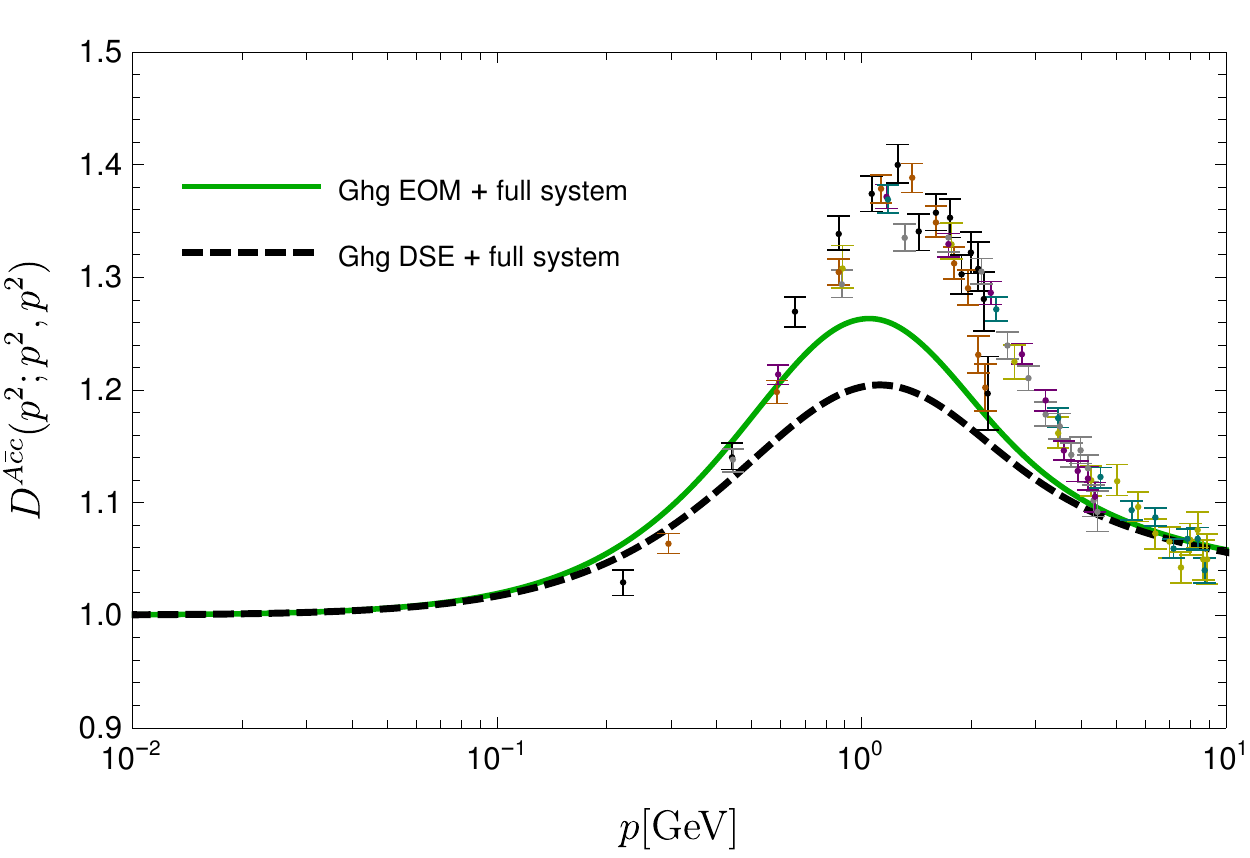}\hfill
 \includegraphics[width=0.48\textwidth]{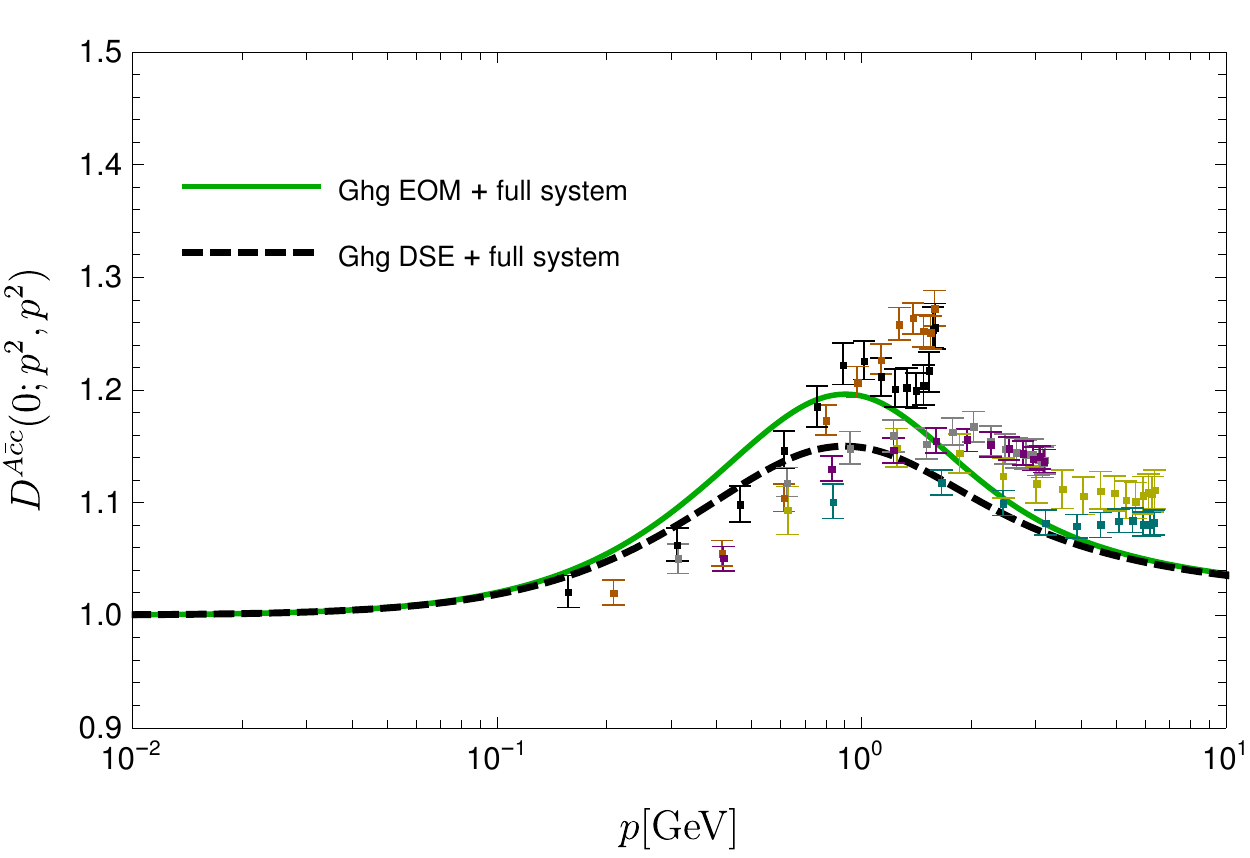}\\
 \includegraphics[width=0.48\textwidth]{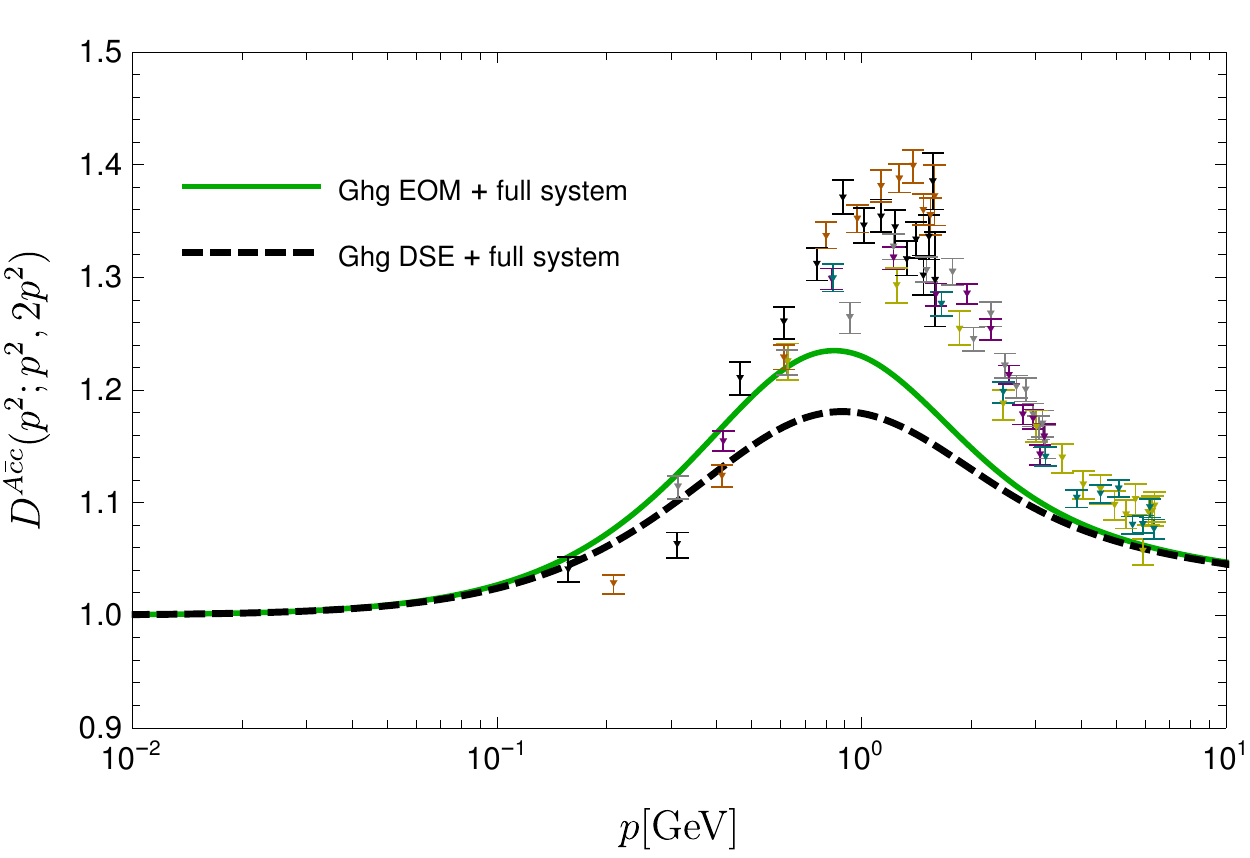}\hfill
 \includegraphics[width=0.48\textwidth]{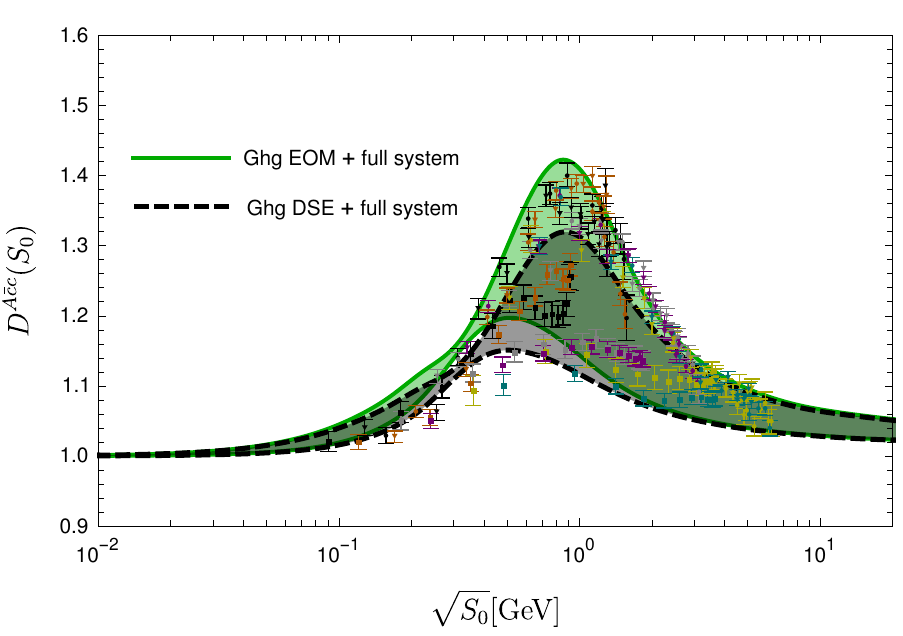}
 \caption{Ghost-gluon vertex dressing function from the full system using the DSE (black, dashed) or EOM (green, continuous) for the vertex.
 Lattice results as explained in \fref{fig:ghg}.
 The bottom right panel shows the full angular dependence, while the other panels show individual kinematic configurations: symmetric point (top left), soft gluon limit (top right), orthogonal ghost and gluon momenta with equal magnitude (bottom left).
 }
 \label{fig:ghg_DSEvsEOM} 
\end{figure*}

\begin{figure*}[tb]
 \includegraphics[width=0.48\textwidth]{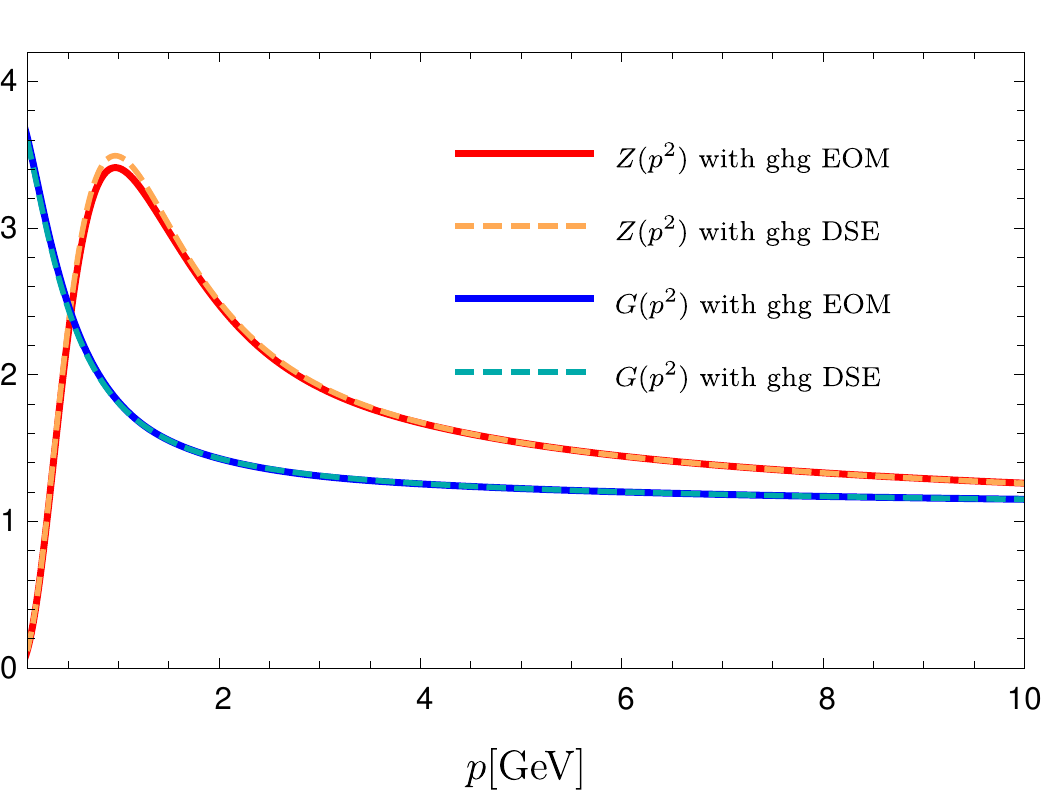}\hfill
 \includegraphics[width=0.48\textwidth]{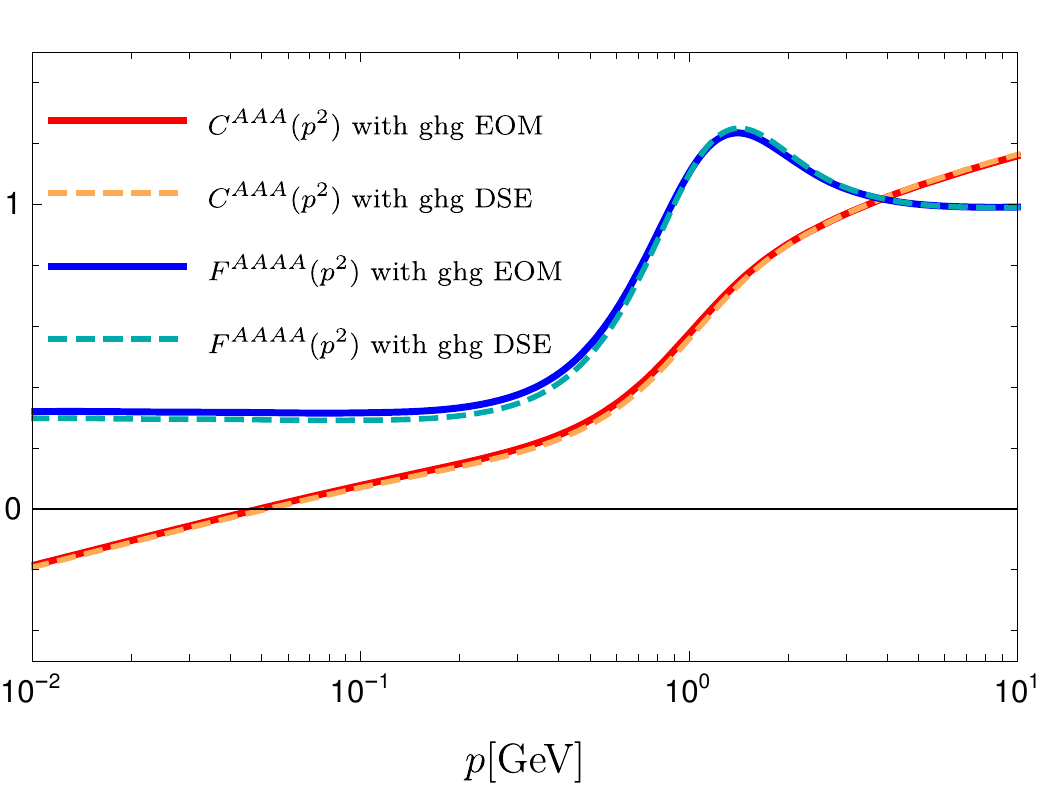}
 \caption{Propagator (left) and three- and four-gluon vertex dressing functions (right) from the full calculation when using the EOM (continuous) or DSE (dashed) for the ghost-gluon vertex.
 The vertices are shown at the symmetric points.}
 \label{fig:props_tg_fg_ghg_DSEvsEOM}
\end{figure*}

As discussed in Sec.~\ref{sec:equations}, the EOM from the three-loop truncated 3PI effective action is used for the ghost-gluon vertex.
Since its DSE looks very similar, one might wonder if this equation provides an equally good description.
To this end, the same system of equations was solved but with the ghost-gluon vertex EOM replaced by its DSE.
A comparison between the DSE and EOM solutions is shown in \fref{fig:ghg_DSEvsEOM} where it can be seen that the strength of the vertex dressing function changes roughly by $10\,\%$.
From the lower right plot, one might even think that the DSE solution agrees better with the lattice results than the EOM solution.
However, as can be seen from specific kinematic configurations, as shown in the other panels, this is not the case.

An interesting finding is that the dressing functions of the other correction functions, which are shown in \fref{fig:props_tg_fg_ghg_DSEvsEOM}, do not change despite the differences in the ghost-gluon vertex.
This independence of all quantities from the ghost-gluon vertex is rather unexpected.
For example, the ghost propagator is known to react to the used ghost-gluon vertex input, e.g., \cite{Dudal:2012zx,Huber:2012kd,Aguilar:2013xqa}.
The absence of such a dependence could be another indication of the stability of the presented solution for the full system.

\section{Subtraction of quadratic divergences for the scaling solution}
\label{sec:aud_divs_scal}

The ghost and gluon dressing functions behave like $G(x)=d_\text{gh} x^{\de_\text{gh}}$ and $Z(x)=d_\text{gl} x^{\de_\text{gl}}$, respectively.
Since $2\de_\text{gh}+\de_\text{gl}=0$, the coupling approaches a fixed point $\alpha_\text{MM}(0)=\alpha_0$.
Note that this is not changed by the dressed ghost-gluon vertex under some regularity assumptions \cite{Lerche:2002ep,Huber:2012zj}.
From $\alpha_0$, one can calculate the coefficient $d_\text{gl}$ of the gluon dressing function if the ghost dressing function is known.
However, at the numerically accessible values of $p^2$, the deviation of the coupling from $\alpha_0$ is large enough to be relevant for calculating $C_\text{sub}$.
Thus, one needs to take into account the momentum dependence by tracing back the derivation of the IR fixed point value $\alpha_0$.
The IR leading contributions in the ghost and gluon DSEs can be written as follows:
\begin{subequations}
\begin{align}
\label{eq:GIR}
 d^{-1}_\text{gh}x^{-\de_\text{gh}}&=g^2\,N_c\,d_\text{gh}d_\text{gl}x^{\de_\text{gh}+\de_\text{gl}}I_\text{gh},\\
 \label{eq:ZIR}
 d^{-1}_\text{gl}x^{-\de_\text{gl}}&=g^2\,N_c\,d^2_\text{gh}x^{2\de_\text{gh}}I_\text{gl}.
\end{align}
\end{subequations}
$I_\text{gh}$ and $I_\text{gh}$ are functions of $\de_\text{gh}$ and $\de_\text{gl}$ which can be calculated analytically for the IR limit.
From the two equations, one can infer that $I_\text{gh}=I_\text{gl}$ which allows to calculate values for $\de_\text{gh}$ and $\de_\text{gl}$.
We use this equality in \eref{eq:ZIR},
\begin{align}
 d^{-1}_\text{gl}=g^2 \,N_c\,d^2_\text{gh}I_\text{gh},
\end{align}
to calculate $d^{-1}_\text{gl}$.
It remains to rewrite the equation in terms of the ghost self-energy $\Sigma_G(x)=g^2\,N_c\,d_\text{gh}d_\text{gl}x^{\de_\text{gh}+\de_\text{gl}}I_\text{gh}$ and the dressing functions,
\begin{align}
\label{eq:dgl_update}
 d^{-1}_\text{gl}=
 \frac{d_\text{gh}\,\Sigma_G}{ d_\text{gl} x^{\de_\text{gh}+\de_\text{gl}}}
 =\frac{d^2_\text{gh}\Sigma_G}{G(x)Z(x)}.
\end{align}
Note that $d^{-1}_\text{gl}$ also appears on the right-hand side.
The equation is trivially true for the solution of the system of equations, but not during the iteration process.
Hence, to update $d_\text{gl}$, \eref{eq:dgl_update} is used and $D(x_m)$ is calculated from it.
Technically, it might be worthwhile to point out that in this case starting guesses for the dressing functions were used for the iteration that were obtained using the IR fixed point value and a very small relaxation parameter for the gluon.
This led to a stable solution and no further study of the stability of this procedure for different starting guesses was performed.

\section{Numerical implementation}
\label{sec:numerics}

This section describes some numerical details of the calculations.
As basis, \textit{CrasyDSE} \cite{Huber:2011xc} was used that provides basic \textit{C++} routines for integration and interpolation.

\renewcommand{\theenumi}{(\roman{enumi})}
\begin{enumerate}
 \item The three-dimensional interpolation for the vertices uses cubic spline interpolation for $S_0$, linear interpolation for $\rho$, and trigonometric interpolation for $\eta$.
 \item Starting conditions: For most calculations, results from previous calculations were used, as this turned out to be more convenient than using generic starting ans\"atze.
 \item Iterative process: The calculation used three levels of iteration.
 In the innermost loop, equations were iterated by themselves.
 However, typically, only for the ghost propagator more than one iteration step was done.
 All equations were then iterated consecutively (meta iteration), starting with the propagators.
 After at most 10 iterations, the renormalization constants were updated and the meta iteration started again (super iteration).
 Normally, only the ghost propagator equation required relaxation.
 \item Integration: Standard Gauss-Legendre integration was used for all integrals.
 The integration intervals for the propagators were split at the external points, as explained, e.g., in Ref.~\cite{Hopfer:2014szm}.
 For the vertices, also appropriate splittings for the intervals were performed.
 \item Extrapolation: The propagators were extrapolated in the IR and UV by the known analytic forms.
 For the ghost-gluon vertex, the boundary values were employed.
 The three- and four-gluon vertices were extrapolated in the IR by their boundary values.
 It was tested that this does not affect the results.
 In the UV, STI motivated extrapolation functions proportional to $Z(x)/G(x)$ and $Z(x)/G(x)^2$, respectively, were used.
 \item Symmetrization: As discussed in Sec.~\ref{sec:equations}, the three- and four-gluon vertex were symmetrized to smooth numeric artifacts.
 For the DSEs, this also has the effect of restoring Bose symmetry which is broken in the truncated equations.
 Using the permutation group variables, symmetrization can be easily realized by averaging over the corresponding angle $\eta$ of the doublet.
 For the three-gluon vertex, this is
 \begin{widetext}
 \begin{align}
  C^{AAA}(S_0, \rho, \eta) \rightarrow \left(C^{AAA}(S_0, \rho, \eta)+C^{AAA}(S_0, \rho, \eta+2\pi /3)+C^{AAA}(S_0, \rho, \eta-2\pi/3)\right)/3,
 \end{align}
 and for the four-gluon vertex one can work out that symmetrization corresponds to
 \begin{align}
   F^{AAAA}(S_0, \rho,\eta)\rightarrow& \big(F^{AAAA}(S_0, \rho, \eta)+F^{AAAA}(S_0, \rho, -\eta)
  +F^{AAAA}(S_0, \rho, \eta+2\pi/3)\nnnl
  &+F^{AAAA}(S_0, \rho, -\eta+2\pi/3)
  +F^{AAAA}(S_0, \rho, \eta-2\pi/3)+F^{AAAA}(S_0, \rho, -\eta-2\pi/3)\big)/6.
 \end{align}
 Here, $F^{AAAA}(S_0, \rho,\eta)$ corresponds to the calculation of only one diagram of each type.
 The symmetry factors of the diagrams need to be adjusted appropriately, see Ref.~\cite{Cyrol:2014kca} for details.
 \item Scale setting: The coupling in the calculation was set to $\alpha(\mu^2)=0.05$.
 This determines the scale, which is determined a posteriori.
 However, instead of determining the physical value of $\mu$ from the coupling $\alpha(\mu^2)=0.05$, the scale is fixed by putting the maximum of the gluon dressing function at $p_0=0.97\,\text{GeV}$.
 \end{widetext}
\end{enumerate}

\bibliographystyle{utphys_mod}
\bibliography{literature_YM_fullSystem}

\end{document}